\documentclass[10pt,graphicx,subfigure,axodraw]{article}
\usepackage{bm}
\usepackage{multirow}
\usepackage{amsfonts}
\usepackage{amssymb}
\usepackage[normalem]{ulem}
\usepackage{epsfig}
\usepackage{float}
\usepackage{amsmath}
\usepackage{amsthm}
\usepackage{slashed}
\usepackage[title]{appendix}
\restylefloat{table}
\usepackage{tabulary}
\usepackage{hyperref}
\usepackage{cleveref}
\usepackage[usenames,dvipsnames]{color}
\usepackage[parfill]{parskip}
\usepackage[numbers,sort&compress]{natbib}
\usepackage{kbordermatrix}
\newcommand{\code}[1]{\texttt{#1}}
\newcommand{\beqn}{\begin{eqnarray}}
\newcommand{\eeqn}{\end{eqnarray}}

\renewcommand{\kbldelim}{(}
\renewcommand{\kbrdelim}{)}

\begin{document}

\begin{titlepage}

\author{Amin Aboubrahim$^{a,b}$\footnote{Email: aabouibr@uni-muenster.de}~, Pran Nath$^b$\footnote{Email: p.nath@northeastern.edu} ~and Raza M. Syed$^{b,c}$\footnote{
Email: rsyed@aus.edu}\\~\\
$^{a}$\textit{\normalsize Institut f\"ur Theoretische Physik, Westf\"alische Wilhelms-Universit\"at M\"unster,} \\ 
\textit{\normalsize Wilhelm-Klemm-Stra{\ss}e 9, 48149 M\"unster, Germany}\\
$^{b}$\textit{\normalsize Department of Physics, Northeastern University, Boston, MA 02115-5000, USA} \\
$^{c}$\textit{\normalsize Department of Physics, American University of Sharjah, P.O. Box 26666, Sharjah, UAE\footnote{Permanent address}} \\}

\title{Corrections to Yukawa couplings from higher dimensional operators in a natural SUSY $\mathsf{SO(10)}$ and {HL-LHC} implications}

\maketitle

\begin{abstract}
\noindent We consider a class of unified models based on the gauge group $\mathsf{SO(10)}$
which with appropriate choice of Higgs representations generate in a natural way
 a pair of light Higgs doublets needed to accomplish electroweak symmetry breaking.
In this class of models higher dimensional operators of the form   matter-matter-Higgs-Higgs
 in the  superpotential after spontaneous breaking of the GUT symmetry generate
 contributions to Yukawa couplings which are comparable to the ones from cubic interactions.
Specifically we consider an $\mathsf{SO(10)}$ model  with a sector  consisting of
$\mathsf{126+\overline{126} + 210}$ of heavy Higgs which breaks the GUT symmetry down to the standard model gauge group and a sector
consisting of $2\times \mathsf{10+120}$ of light Higgs fields. In this model we compute the corrections from the quartic interactions
to the Yukawa couplings for  the top and the bottom quarks and for the tau lepton. It is then shown that
inclusion of these corrections to the GUT scale Yukawas
 allows for consistency of the top, bottom and tau masses with
experiment for low $\tan\beta$ with a value as low as $\tan\beta$ of 5$-$10.
We compute the sparticle spectrum
for a set of benchmarks and find that satisfaction of the relic density  is achieved via a compressed spectrum
and coannihilation and
three sets of coannihilations appear: chargino-neutralino, stop-neutralino and stau-neutralino.
We investigate the chargino-neutralino coannihilation in detail
for the possibility of observation of the light
chargino at  the high luminosity LHC (HL-LHC) and at the
 high energy LHC (HE-LHC) which is a possible future 27 TeV hadron collider. It is shown that all benchmark models but one can be discovered at HL-LHC and all  would be discoverable at HE-LHC. The ones discoverable at both machines require a much shorter time scale
 and a lower integrated luminosity at HE-LHC.
\end{abstract}

\end{titlepage}

{\small

\section{Introduction}\label{sec1}
Grand unified models based on $\mathsf{SO(10)}$ \cite{georgi,Fritzsch:1974nn}
are the most desirable of grand unified models as they provide unification of the standard model gauge group
and  a unification of one generation of matter consisting of quarks and leptons in a single irreducible representation.
The Higgs sector of $\mathsf{SO(10)}$ models is very rich  consisting of several possible representations which can be used to break
the grand unified symmetry down to the standard model gauge group. Some of these consist of $\mathsf{16+\overline{16}}$, $\mathsf{45}$, $\mathsf{54}$, $\mathsf{126+\overline{126}}$, $\mathsf{210}$ among others. In this work we will focus on large Higgs representations to break the grand unified theory (GUT) symmetry
 for reasons explained below. Large representations have been used in the literature for quite some time,
 a small sample of which are~\cite{Clark:1982ai,Aulakh:1982sw,Babu:1992ia} and  for some more recent works
see, e.g.,~\cite{Nath:2001uw,Nath:2001yj,Nath:2003rc,Aulakh:2003kg,Bajc:2004xe,Aulakh:2004hm,Aulakh:2008sn} and the references therein. However, in grand unified models  with small as well as with large Higgs represenations
the Higgs doublets lie in irreducible representations of the
unified gauge group along with other components which carry color, such as color triplets. The super-partners of these
enter in proton decay (for a review see~\cite{Nath:2006ut})
and they must be very heavy, i.e., of the GUT scale size, which makes the Higgs doublets also superheavy and thus unsuitable for electroweak symmetry breaking. One can, of course,  manufacture
a light Higgs doublet pair  by  fine tuning which, however, is rather large.

It is more appealing to have models where some higher symmetry,
  a group theoretic constraint, or a vacuum selection constraint
    leads to a pair of light Higgs doublets.
Such unified models may be viewed as natural, and GUT models which exhibit this property may be viewed as natural GUTs.
  In string theory examples of such  models exist,
 see e.g., \cite{Braun:2005nv,Bouchard:2006dn,Anderson:2009mh}
 and for natural GUTs see \cite{Bouchard:2005ag}.
  Natural GUT models may also be realized in the  framework of field theory.
Thus the Dimopoulos-Wilczek mechanism allows for
  generation of light Higgs doublets in $\mathsf{SO(10)}$~\cite{DW}.
  Another possibility  to generate a  light vector-like Higgs doublet
 is by a combination of Higgs representations.  In $\mathsf{SU(5)}$ one finds~\cite{Masiero:1982fe,Grinstein:1982um}
  that a combination  of $\mathsf{5+\bar 5}$, $\mathsf{50+\overline{50}}$ and $\mathsf{75}$ of Higgs  conspire to make the color Higgs triplets all
 heavy but leaves one pair of  Higgs doublets light. A similar phenomenon occurs in $\mathsf{SO(10)}$~\cite{Babu:2006nf,Babu:2011tw} where a pair of light Higgs doublets can arise purely by a proper combination of heavy and light Higgs multiplets. Models of this type are referred to as missing partner models
 and they belong to
 the larger class of natural models as defined above. This last class of models involve  Higgs fields in large tensor and spinor
 representations~\footnote{An example of a natural GUT model with spinor Higgs representations
is the case when the heavy Higgs consists of $\mathsf{560+\overline{560}}$ and the light Higgs consists of
$2\times\mathsf{10+320}$~\cite{Babu:2011tw}.}.

 The mechanism that operates in natural  field theoretic GUT models is the following:  Suppose the GUT model consists of two types of Higgs fields, where one set is heavy and the other set is light.
 Let us further suppose
  the heavy sector possesses $n^H_D$ number of Higgs doublet pairs, and  the light sector
possesses $n^L_D$ number of Higgs doublet  pairs and $n^L_D> n^H_D$. In this case if the light and the heavy sectors mix, $n^H_D$ number of
light Higgs doublet pairs will become heavy leaving  $n^L_D-n^H_D$ number of  Higgs doublet pairs light.
 In the class of models we
consider $n^L_D-n^H_D=1$ and thus one naturally produces one pair of light Higgs  doublets which is desired for electroweak symmetry breaking.
At the same time we need to make sure that the number of color triplets/anti-triplets $n_T^H$
in the heavy sector  and the number of color
triplets/anti-triplets $n_T^L$ in the light sector match, i.e., $n_T^L-n_T^H=0$ which makes all the
color Higgs triplets/anti-triplets heavy when the light and the heavy sectors mix.

It is of interest to investigate physics implications of $\mathsf{SO(10)}$ models of this type. Thus
proton stability in these models has been discussed in~\cite{Du:2013nza}. Here we will discuss
 quark-charged lepton masses and the sparticle spectrum in a class of these models and also investigate the
   implications for supersymmetry (SUSY) discovery at the HL-LHC
and HE-LHC.
In this work we will consider one specific model where the heavy  sector consists of  $\mathsf{126+\overline{126} +210}$ of Higgs fields and the light  sector
consists of $2\times \mathsf{10+ 120}$ of Higgs fields.
In this case
 using the counting discussed above only one Higgs doublet pair remains light while all the color triplet/anti-triplet
pairs become heavy. An important result of our analysis is to show that
 in models of this type,
higher dimensional operators can generate contributions to  Yukawa couplings which are
comparable to the contributions from the cubic interactions. The reason for this is the following: the quartic interactions of the type matter-matter-light Higgs- heavy Higgs
suppressed by a heavy (cutoff) mass produce contributions comparable to those from the cubic interactions
  after spontaneous breaking of the GUT symmetry.
This is so because one of the heavy Higgs fields, i.e., the 210-plet, has  a large vacuum expectation value (VEV)
and thus can make a non-negligible contribution even when suppressed by the cutoff mass.

The outline of the rest of the paper is as follows: In section~\ref{sec2} we give a description of the model.
{In section \ref{sec3a} we compute the contribution to Yukawas for the third generation from the
cubic interactions.}
In section \ref{sec3} we give computations of the
Yukawa couplings which arise from the cubic matter-matter-Higgs interactions and from the quartic  matter-matter-Higgs-Higgs
interactions
where one
of the Higgs fields belongs either to the $\mathsf{10}$-plets or to the $\mathsf{120}$-plet while the other 
Higgs field is the 210-plet.
 After spontaneous symmetry breaking
at the GUT scale these quartic interactions contribute to the Yukawa couplings.
 We show that the
quartic superpotential corrections to the Yukawa couplings can be substantial and can modify the well known
constraint that for the $t-b-\tau$ unification one needs a large $\tan\beta$~\cite{Ananthanarayan:1992cd}.
In section \ref{sec4} we give a numerical estimate of the VEVs of the heavy fields which break the $\mathsf{SO(10)}$ symmetry down to the standard model gauge group.
This is done for a set of benchmarks for the parameters involving the heavy fields. In this section we also give the numerical computations of the contributions of the
quartic operators in the Yukawa couplings.  {Implications of the model at high luminosity LHC (HL-LHC)} regarding the possible observation of supersymmetry in this model are also discussed.
Conclusions are given in section~\ref{sec5}.
Several appendices are also included.
Thus
notation of the model is given in Appendix A where we also
give a decomposition of the relevant  irreducible representations of
$\mathsf{SO(10)}$ in irreducible representations of  $\mathsf{SU(5)}$. In Appendix B we discuss spontaneous breaking of the $\mathsf{SO(10)}$ symmetry by the heavy Higgs fields $\mathsf{126+\overline{126}}$ and
$\mathsf{210}$. In Appendix C we exhibit for completeness the $7\times 7$ Higgs doublet mass matrix as a result of the mixing of the heavy fields
$\mathsf{126+\overline{126}+210}$ with the light fields $2\times \mathsf{10+120}$ discussed in section \ref{sec2}.  In Appendix D details of the computation of the contributions of quartic interactions to
Yukawa couplings are given.

\section{The model \label{sec2}}
The heavy Higgs sector of our $\mathsf{S0(10)}$ model consists of $\mathsf{126}(\Delta)+\mathsf{\overline{126}}(\overline{\Delta})+\mathsf{210}(\Phi)$, while the light sector contains $\sum_{r=1}^2\mathsf{10_r}({^r}\Omega)+\mathsf{120}({\Sigma})$. This particular particle content in the Higgs sector gives
after mixing of the light and heavy sectors just a pair of light Higgs doublets~\cite{Babu:2011tw,Nath:2015kaa}. Finally, the Yukawa sector contains a single (third) generation of quarks and leptons that reside in the $\mathsf{16}(\Psi_{(+)})$ multiplet spinor representation.
The GUT symmetry is broken via the superpotential \cite{Nath:2015kaa}
\begin{align}
W_{\textsc{gut}}=&~ M^{126}\Delta_{\mu\nu\rho\sigma\lambda}\overline{\Delta}_{\mu\nu\rho\sigma\lambda}+M^{210}\Phi_{\mu\nu\rho\sigma}\Phi_{\mu\nu\rho\sigma}
+\eta\Phi_{\mu\nu\rho\sigma}\Delta_{\mu\nu\lambda\tau\xi}\overline{\Delta}_{\rho\sigma\lambda\tau\xi} \nonumber \\
&+\lambda \Phi_{\mu\nu\rho\sigma} \Phi_{\rho\sigma\lambda\tau} \Phi_{\lambda\tau\mu\nu}\,.
\label{gut-1}
\end{align}
Here the VEVs of the $\mathsf{126+\overline{126}}$ fields, i.e.,  $\mathcal V_{1_{126}}$ and $\mathcal V_{1_{\overline{126}}}$ and the VEVs of the $\mathsf{210}$-plet fields
$\mathcal V_{1_{210}}$, $\mathcal V_{24_{210}}$, $\mathcal V_{75_{210}}$ break the GUT symmetry down to the gauge group symmetry of $\mathsf{SO(10)}$.
Details of this breaking are given in Appendix B.
Next we discuss the generation of the light Higgs doublet.
The couplings appearing in the superpotential that generate a light Higgs doublet pair are~\cite{Nath:2015kaa}
\begin{align}
W_{\textsc{dt}}=~& a~{^1}\Omega_{\mu}\overline{\Delta}_{\mu\nu\rho\sigma\lambda}\Phi_{\nu\rho\sigma\lambda}+\sum_{r=1}^2 b_r~{^r}\Omega_{\mu}{\Delta}_{\mu\nu\rho\sigma\lambda}\Phi_{\nu\rho\sigma\lambda}+c~ \Sigma_{\mu\nu\rho}\Delta_{\nu\rho\sigma\lambda\tau}\Phi_{\mu\sigma\lambda\tau}\nonumber\\
&+
\overline{c}~ \Sigma_{\mu\nu\rho}\overline{\Delta}_{\nu\rho\sigma\lambda\tau}\Phi_{\mu\sigma\lambda\tau}\,.
\label{dtsuper}
\end{align}
{In writing Eqs.(\ref{gut-1}) and (\ref{dtsuper}) we have imposed the condition that there be a light sector consisting of
$2\times {\mathsf{10}+\mathsf{120}}$.
This condition requires absence of explicit masses for $2\times {\mathsf{10}}+{\mathsf{120}}$ as well as absence of couplings
${10\cdot120\cdot210}$ and ${120\cdot120\cdot210}$ which would otherwise give superheavy masses to them after the
 $210$ develops a heavy VEV.
These constraints which were given in Eqs. (11) and (12) of \cite{Nath:2015kaa} are needed to guarantee
the existence of a light sector  which is necessary for the missing partner mechanism to operate.}
The $\mathsf{SO(10)}$ heavy Higgs multiplets $\mathsf{126}+\mathsf{\overline{126}}+\mathsf{210}$
contain three heavy  $\mathsf{SU(2)}$ doublet pairs:
\begin{align}
\{{}^{({5}_{\overline{126}})}\!{\mathsf D}^{a}, {}^{(\overline{5}_{{126}})}\!{\mathsf D}_{a}\},~~ \{{}^{({45}_{{126}})}\!{\mathsf D}^{a},~~ {}^{(\overline{45}_{\overline{126}})}\!{\mathsf D}_{a}\},~~ \{{}^{({5}_{{210}})}\!{\mathsf D}^{a}, {}^{(\overline{5}_{{210}})}\!{\mathsf D}_{a}\}.
\label{3-heavy}
\end{align}
The $\mathsf{SO(10)}$ light  Higgs multiplets $2\times \mathsf{10}+\mathsf{120}$ contain four light  $\mathsf{SU(2)}$ doublet pairs:
\begin{align}
 \{{}^{({5}_{{10_1}})}\!{\mathsf D}^{a}, {}^{(\overline{5}_{{10_1}})}\!{\mathsf D}_{a}\},
~~\{{}^{({5}_{{10_2}})}\!{\mathsf D}^{a}, {}^{(\overline{5}_{{10_2}})}\!{\mathsf D}_{a}\}, ~~\{{}^{({5}_{{120}})}\!{\mathsf D}^{a}, {}^{(\overline{5}_{{120}})}\!{\mathsf D}_{a}\}, ~~\{{}^{({45}_{{120}})}\!{\mathsf D}^{a}, {}^{(\overline{45}_{{120}})}\!{\mathsf D}_{a}\}.
\label{4-light}
\end{align}
Because of the mixings of the heavy Higgs and light Higgs sectors via Eq.~(\ref{dtsuper}), three linear combinations of the four light Higgs doublet pairs in Eq.~(\ref{4-light})
mix with the three heavy Higgs  doublet pairs  of Eq.~(\ref{3-heavy})
and
become heavy leaving only one pair of Higgs doublets light. This light Higgs doublet is the one that enters the electroweak symmetry breaking.
The specific linear combination of the seven Higgs doublet pairs that yield a light Higgs doublet pair can be gotten by diagonalizing the $7\times 7$
Higgs doublet mass matrix given in Appendix C.
The doublet mass matrix is diagonalized by two unitary matrices $U$ and $V$ whose relevant elements are displayed in Eq.~(\ref{doublet mass eigenstates}),
\begin{eqnarray}\label{doublet mass eigenstates}
  \begin{pmatrix}{}^{(\overline{5}_{10_1})}\!{\mathsf D}_{a}\\{}^{(\overline{5}_{10_2})}\!{\mathsf D}_{a}\\{}^{(\overline{5}_{120})}\!{\mathsf D}_{a}\\{}^{(\overline{5}_{{126}})}\!{\mathsf D}_{a}\\{}^{(\overline{5}_{{210}})}\!{\mathsf D}_{a}\\{}^{(\overline{45}_{120})}\!{\mathsf D}_{a}\\{}^{(\overline{45}_{\overline{126}})}\!{\mathsf D}_{a}\end{pmatrix}
  =  \begin{pmatrix}
  V_{d_{11}} &  & \cdots &  \\
  V_{d_{21}} &  & \cdots & \\
   V_{d_{31}} &  & \cdots &  \\
    0 & & \cdots & \\
  0 &  & \cdots &  \\
    V_{d_{61}} &  & \cdots &  \\
   0 &  & \cdots &  \\
 \end{pmatrix}\begin{pmatrix}{\mathbf{H_d}}_a\\{}^{2}\!{\mathsf D}_{a}^{\prime}\\{}^{3}\!{\mathsf D}_{a}^{\prime}\\{}^{4}\!{\mathsf D}_{a}^{\prime}\\{}^{5}\!{\mathsf D}_{a}^{\prime}\cr{}^{6}\!{\mathsf D}_{a}^{\prime}\\{}^{7}\!{\mathsf D}_{a}^{\prime}\end{pmatrix}
 ;~~\begin{pmatrix}{}^{({5}_{10_1})}\!{\mathsf D}^{a}\cr {}^{({5}_{10_2})}\!{\mathsf D}^{a}\cr{}^{({5}_{120})}\!{\mathsf D}^{a}\cr{}^{({5}_{\overline{126}})}\!{\mathsf D}^{a} \cr{}^{({5}_{{210}})}\!{\mathsf D}^{a} \cr{}^{({45}_{120})}\!{\mathsf D}^{a} \cr{}^{({45}_{126})}\!{\mathsf D}^{a}
\end{pmatrix}= \begin{pmatrix}
  U_{d_{11}} & & \cdots &  \\
  U_{d_{21}} &  & \cdots & \\
   U_{d_{31}} &  & \cdots &  \\
    0 &  & \cdots &  \\
  0 &  & \cdots & \\
    U_{d_{61}} &  & \cdots &  \\
   0 &  & \cdots & \\
 \end{pmatrix}\begin{pmatrix}{\mathbf{H_u}}^a\cr {}^{2}\!{\mathsf D}^{\prime a}\cr{}^{3}\!{\mathsf D}^{\prime a}\cr{}^{4}\!{\mathsf D}^{\prime a} \cr{}^{5}\!{\mathsf D}^{\prime a} \cr{}^{6}\!{\mathsf D}^{\prime a} \cr{}^{7}\!{\mathsf D}^{\prime a}\end{pmatrix},
\end{eqnarray}
 where $\mathsf D$'s and ${\mathsf D}^{\prime}$'s represent the normalized kinetic energy basis and normalized kinetic and mass eigenbasis, respectively of the doublet mass matrix of Eq. (\ref{doublet mass matrix}). The pair of doublets $({\mathbf{H_d}}_a,{\mathbf{H_u}}^a)$ are identified to be light and are the
 normalized   electroweak Higgs doublets of the minimal supersymmetric standard model (MSSM). The matrix elements of $U$ and $V$ marked by dots do not contribute in the low energy theory. Numerical values of the non-zero matrix elements of  $U$ and $V$ are displayed in Tables \ref{tab2} and \ref{tab3} for benchmarks of Table \ref{tab1}.

{
\section{Yukawa couplings from cubic interactions \label{sec3a}}
}
Since the product
$\mathsf{16}\times \mathsf{16}=\mathsf{10}_s+\mathsf{120}_{a}+\mathsf{\overline{126}}_{s}$,
 the $\mathsf{16}$-plet of matter has couplings with
the $\mathsf{10}$-plet, $\mathsf{120}$-plet and  $\mathsf{\overline{126}}$-plet of Higgs. Here the subscripts $a$ and $s$ indicate if the Higgs tensor appears symmetrically or anti-symmetrically under the exchange of two $\mathsf{16}$'s. Since the
$\mathsf{126}$ field is  superheavy it does not contribute to the fermion cubic couplings
as can be seen from  Eq.~(\ref{doublet mass eigenstates}), where one has  $U_{d_{41}}=0$ and $V_{d_{71}}=0$.
The $\mathsf{120}$-plet  couplings are anti-symmetric in the generation space and so  they also do not contribute
 because we consider here only one  {, i.e., the third} generation.
{Thus  only the
$\mathsf{16-16-10}$  cubic couplings contribute to the  Yukawa
couplings and their computation follows from}
\begin{equation}
\label{w3}
W_3=\sum_{r=1}^2f^{10_{r}}~\langle\Psi_{(+)}^*|B \Gamma_{\mu}|\Psi_{(+)}\rangle~ {^r}{\Omega_{\mu}}\,.
\end{equation}
Here $B$ and $\Gamma$'s are the $\mathsf{SO(10)}$ charge conjugation and gamma matrices~\cite{Nath:2001uw}.
  The decomposition of an
$\mathsf{SO(10)}$ vertex in the $\mathsf{SU(5)}$ basis using the oscillators~\cite{Mohapatra:1979nn} and the techniques developed
in \cite{Nath:2001uw,Nath:2001yj,Nath:2003rc,Nath:2005bx}
allow us to compute particle content in the $\mathsf{SU(3)_C\times SU(2)\times U(1)_Y}$
basis. Thus for  $W_3$ in $\mathsf{SU(5)}$  decomposition we get
\begin{eqnarray}
W_3&=&i\sum_{r=1}^2f^{10_{r}}\left[2\sqrt{2}\mathsf{M}^{ij}\mathsf{M}_{i}\mathsf{H}^{(10_r)}_j+\frac{1}{2\sqrt{2}}
\epsilon_{ijklm}\mathsf{M}^{ij}\mathsf{M}^{kl}\mathsf{H}^{(10_r)m}+\cdots\right]\nonumber\\
&=&i\sum_{r=1}^2f^{10_{r}}\left[2\sqrt{2}\left(-\mathsf{M}^{a\alpha}\mathsf{M}_{\alpha}+\mathsf{M}^{ba}\mathsf{M}_{b}+\cdots\right)\mathsf{H}^{(10_r)}_a
+\frac{1}{2\sqrt{2}}
\left(-4\epsilon_{\alpha\beta\gamma ab}\mathsf{M}^{\alpha\beta}\mathsf{M}^{a\gamma}\mathsf{H}^{(10_r)b}+\cdots\right)+\cdots\right]\nonumber\\
&=&i\sum_{r=1}^2f^{10_{r}}\left[2\sqrt{2}\left(-\mathbf{{Q}}^{a\alpha}{\mathbf D}_{\alpha}^{\mathtt c}+\epsilon^{ab}{\mathbf E}^{\mathtt c}\mathbf{{L}}_{b}\right)\mathsf{H}^{(10_r)}_a-\frac{8}{2\sqrt{2}}\epsilon_{ab}{\mathbf U}_{\alpha}^{\mathtt c}\mathbf{{Q}}^{a\alpha}\mathsf{H}^{(10_r)b}+\cdots\right].
\label{cubic-coupling}
\end{eqnarray}

Yukawa Lagrangian of the MSSM is given by
\begin{eqnarray}
\mathcal{L}_{\textnormal{Yuk}}= +h^0_{\tau}~\epsilon^{ab}{\mathbf{H_d}}_a\mathbf{{L}}_{b}{\mathbf E}^{\mathtt c}
-h^0_{b}~{\mathbf{H_d}}_a\mathbf{{Q}}^{a\alpha}{\mathbf D}_{\alpha}^{\mathtt c}-h^0_{t}~\epsilon_{ab}{\mathbf{H_u}}^a\mathbf{{Q}}^{b\alpha}{\mathbf U}_{\alpha}^{\mathtt c} + \textnormal{h.c.},
\end{eqnarray}
while the third generation Yukawas arising from Eq.~(\ref{cubic-coupling})  are given by
\begin{eqnarray}\label{q&l masses from cubic coupling}
h^0_{\tau}&=&i2\sqrt{2}\sum_{r=1}^2f^{10_{r}}V_{d_{r1}},\nonumber\\
h^0_{b}&=&-i2\sqrt{2}\sum_{r=1}^2f^{10_{r}}V_{d_{r1}}, \nonumber\\
h^0_{t}&=&-i2\sqrt{2}\sum_{r=1}^2f^{10_{r}}U_{d_{r1}},
\end{eqnarray}
where $U_{dr1}$ and $V_{dr1}$ are defined by  Eq.~(\ref{doublet mass eigenstates}) and  evaluated numerically in Tables~(\ref{tab2}) and~(\ref{tab3}).

\section{Corrections to  Yukawa couplings from higher dimensional operators \label{sec3}}

In addition to the cubic interactions in this model, contributions to the quark and lepton masses arise from quartic interactions suppressed by
a heavy mass $M_c$ of the type
$(\text{matter})(\text{matter})(\text{light Higgs}\big)(\text{heavy Higgs})/M$ when the heavy Higgs fields develop a large VEV. We assume that such higher dimensional operators arise from a high scale above
the grand unification scale, possibly at the string scale, and are suppressed by a heavy mass $M_c$
larger than the grand unification mass.
In the following we consider contributions arising from the higher dimensional
operators $W_4$ involving matter fields and the Higgs fields consisting of the 10-plets, 120-plet and
the 210-plet of Higgs so that
\begin{align}
W_4=W^{(1)}_4+ W^{(2)}_4+ W^{(3)}_4,
\end{align}
where
\begin{eqnarray}
W^{(1)}_4&=&-\frac{f^{(1)}}{5!M_c}b_r\langle\Psi_{(+)}^*|B \Gamma_{[\lambda}\Gamma_\mu\Gamma_{\nu}\Gamma_{\rho}\Gamma_{\sigma]}|\Psi_{(+)}\rangle~
\left[{^r}\Omega_{\lambda}\Phi_{\mu\nu\rho\sigma}-{^r}\Omega_{\mu}\Phi_{\lambda\nu\rho\sigma}+{^r}\Omega_{\nu}\Phi_{\lambda\mu\rho\sigma}\right.\nonumber\\
&&\left.~~~~~~~~~~~~~~~~~~~~~~~~~~~~~~~~~~~~~~~~~~~~~~~~~~~-{^r}\Omega_{\rho}\Phi_{\lambda\mu\nu\sigma}+{^r}\Omega_{\sigma}\Phi_{\lambda\mu\nu\rho}\right]\,,
\label{w41}
\\
W_4^{(2)}&=&-\frac{f^{(2)}}{5!M_{c}}\langle\Psi_{(+)}^*|B \Gamma_{[\lambda}\Gamma_\mu\Gamma_{\nu}\Gamma_{\rho}\Gamma_{\sigma]}|\Psi_{(+)}\rangle~
\left[\Sigma_{\lambda\alpha\beta}\Phi_{\gamma\rho\sigma\lambda}-\Sigma_{\lambda\alpha\gamma}\Phi_{\beta\rho\sigma\lambda}
+\Sigma_{\lambda\alpha\rho}\Phi_{\beta\gamma\sigma\lambda}\right.\nonumber\\
&&\left.~~~~~~~~~~~~~~~~~~~~~~~~~~~~~~~~~~~~~~~~~~~~~~~~-\Sigma_{\lambda\alpha\sigma}\Phi_{\beta\gamma\rho\lambda}
-\Sigma_{\lambda\gamma\beta}\Phi_{\alpha\rho\sigma\lambda}
+\Sigma_{\lambda\rho\beta}\Phi_{\alpha\gamma\sigma\lambda}\right.\nonumber\\
&&\left.~~~~~~~~~~~~~~~~~~~~~~~~~~~~~~~~~~~~~~~~~~~~~~~~-\Sigma_{\lambda\sigma\beta}\Phi_{\alpha\gamma\rho\lambda}
-\Sigma_{\lambda\gamma\rho}\Phi_{\beta\alpha\sigma\lambda}
+\Sigma_{\lambda\gamma\sigma}\Phi_{\beta\alpha\rho\lambda}\right.\nonumber\\
&&\left.~~~~~~~~~~~~~~~~~~~~~~~~~~~~~~~~~~~~~~~~~~~~~~~~-\Sigma_{\lambda\rho\sigma}\Phi_{\beta\alpha\gamma\lambda}\right]\,, \label{w42}
\\
W_4^{(3)}&=&\frac{f^{(3)}}{M_{c}}\langle\Psi_{(+)}^*|B \Gamma_{\mu}|\Psi_{(+)}\rangle\Sigma_{\rho\sigma\lambda}\Phi_{\rho\sigma\lambda\mu}\,.
\label{w43}
\end{eqnarray}
The contribution from Eq.~(\ref{w41}) is computed in Appendix D and the
contribution from this term to the third generation Yukawas is given by
\begin{eqnarray}
\delta h^{(1)}_{t}&=& \frac{if^{(1)}}{60\sqrt{2}M_{c}}\left(\sum_{r=1}^2b_rU_{d_{r1}}\right)\left[\frac{5\sqrt{3}}{2}\mathcal V_{75_{_{{210}}}}-4\sqrt{15}\mathcal V_{24_{_{{210}}}}-8\sqrt{15}\mathcal V_{1_{_{{210}}}}\right], \label{q&l masses from quartic coupling 1a} \\
\delta h^{(1)}_{b}&=& \frac{if^{(1)}}{60\sqrt{2}M_{c}}\left(\sum_{r=1}^2b_rV_{d_{r1}}\right)\left[\frac{\sqrt{20}}{3}\mathcal V_{75_{_{{210}}}}-20\sqrt{\frac{5}{3}}\mathcal V_{24_{_{{210}}}}\right],   \label{q&l masses from quartic coupling 1b}\\
\delta h^{(1)}_{\tau}&=& \frac{if^{(1)}}{60\sqrt{2}M_{c}}\left(\sum_{r=1}^2b_rV_{d_{r1}}\right)\left[20\sqrt{3}\mathcal V_{75_{_{{210}}}}-20\sqrt{15}\mathcal V_{24_{_{{210}}}}\right],\label{q&l masses from quartic coupling 1c}
\end{eqnarray}
where the VEVs of the Standard Model singlets denoted by $\cal V$'s are defined in Eq. (\ref{VEVS}). We refer to  appendix D for further details of the computation.
The contribution from Eq.~(\ref{w42}) is computed in Appendix D and they produce the following contributions to the third generation Yukawas
\begin{align}
\delta h^{(2)}_{t}=&-\frac{i f^{(2)}}{120M_{c}} \left[\frac{10}{3}\sqrt{\frac{2}{3}}\mathcal V_{75_{_{{210}}}}U_{d_{61}}
+\frac{5}{3}\sqrt{\frac{10}{3}}\mathcal V_{24_{_{{210}}}}U_{d_{61}}+6\sqrt{5}\mathcal V_{24_{_{{210}}}}U_{d_{31}}-8\sqrt{5}\mathcal V_{1_{_{{210}}}}U_{d_{31}}\right],\label{q&l masses from quartic coupling 2a} \\
\delta h^{(2)}_{b}=&-\frac{i f^{(2)}}{120M_{c}} \Bigg[-\frac{20}{3}\sqrt{\frac{2}{3}}\mathcal V_{75_{_{{210}}}}V_{d_{61}}-\frac{20}{3}\mathcal V_{75_{_{{210}}}}V_{d_{31}}
-\frac{1}{3}\sqrt{\frac{10}{3}}\mathcal V_{24_{_{{210}}}}V_{d_{61}}-\frac{10\sqrt{5}}{3}\mathcal V_{24_{_{{210}}}}V_{d_{31}}-4\sqrt{\frac{10}{3}}\mathcal V_{1_{_{{210}}}}V_{d_{61}}\Bigg],\label{q&l masses from quartic coupling 2b} \\
\delta h^{(2)}_{\tau}=& -\frac{i f^{(2)}}{120M_{c}} \Bigg[-20\sqrt{\frac{2}{3}}\mathcal V_{75_{_{{210}}}}V_{d_{61}}-20\mathcal V_{75_{_{{210}}}}V_{d_{31}}
-\sqrt{\frac{10}{3}}\mathcal V_{24_{_{{210}}}}V_{d_{61}}-10\sqrt{5}\mathcal V_{24_{_{{210}}}}V_{d_{31}}-4\sqrt{30}\mathcal V_{1_{_{{210}}}}V_{d_{61}}\Bigg].\label{q&l masses from quartic coupling 2c}
\end{align}
Finally we compute the contribution arising from Eq.~(\ref{w43}). From the analysis of  Appendix D
we find
\begin{eqnarray}
\delta h_t^{(3)}&=& -\frac{3i}{8}\frac{f^{(3)}}{M_c}\left[\frac{2}{3}\sqrt{\frac{2}{3}}\mathcal V_{75_{210}}U_{d_{61}}
+\frac{1}{3}\sqrt{\frac{10}{3}}\mathcal V_{24_{210}}U_{d_{61}}-\frac{2}{\sqrt{5}}\mathcal V_{24_{210}}U_{d_{31}}+\frac{8}{3\sqrt{5}}\mathcal V_{1_{210}}U_{d_{31}}\right],~~~\label{q&l masses from quartic coupling 3a}\\
\delta h_b^{(3)}&=& -\frac{3i}{8}\frac{f^{(3)}}{M_c}\left[\frac{2}{3}\sqrt{\frac{2}{3}}\mathcal V_{75_{210}}V_{d_{61}}
+\frac{1}{3}\sqrt{\frac{10}{3}}\mathcal V_{24_{210}}V_{d_{61}}-\frac{2}{\sqrt{5}}\mathcal V_{24_{210}}V_{d_{31}}+\frac{8}{3\sqrt{5}}\mathcal V_{1_{210}}V_{d_{31}}\right],\label{q&l masses from quartic coupling 3b}\\
\delta h_{\tau}^{(3)}&=& \frac{3i}{8}\frac{f^{(3)}}{M_c}\left[\frac{2}{3}\sqrt{\frac{2}{3}}\mathcal V_{75_{210}}V_{d_{61}}
+\frac{1}{3}\sqrt{\frac{10}{3}}\mathcal V_{24_{210}}V_{d_{61}}-\frac{2}{\sqrt{5}}\mathcal V_{24_{210}}V_{d_{31}}+\frac{8}{3\sqrt{5}}\mathcal V_{1_{210}}V_{d_{31}}\right].\label{q&l masses from quartic coupling 3c}
\end{eqnarray}
The total  Yukawas are the sum of the contributions from the cubic and from the quartic terms at the GUT scale. Thus we have
\begin{align}
h_t=h^0_t +\delta  h_t^{(1)} + \delta h_t^{(2)}+\delta h_t^{(3)},~~h_b=h^0_b +\delta  h_b^{(1)} + \delta h_b^{(2)}+ \delta h_b^{(3)} ,~~
h_\tau=h^0_\tau +\delta  h_\tau^{(1)} + \delta h_\tau^{(2)}+\delta h_\tau^{(3)}\,.
\label{yuk-gut}
\end{align}
In the renormalization group (RG) evolution, Eq.~(\ref{yuk-gut}) acts as the boundary condition which produces the effective Yukawas at the electroweak scale $Q$ so that
at this scale the top, bottom, and tau lepton masses are related to the effective  Yukawa couplings
so that
\beqn\label{yukawa-1}
m_{t}(Q)=\frac{h_{t}(Q) v \sin\beta}{\sqrt 2}, ~~m_{b}(Q)=\frac{h_{b}(Q) v \cos\beta}{\sqrt 2},
~~m_{\tau}(Q)=\frac{h_{\tau}(Q) v \cos\beta}{\sqrt 2},
\eeqn
where we used the relations $\langle H_d\rangle=\frac{v}{\sqrt 2} \cos\beta$ and $\langle H_u\rangle=\frac{v}{\sqrt 2}\sin\beta$, and where $v=246$ GeV.

\section{Analysis of model implications \label{sec4}}

In this section we discuss the implications of the model discussed above. Here we will give numerical computations of the cubic and the quartic interactions to the Yukawa couplings
of the third generation of quarks and of the charged lepton and  show that significant deviations exist at the GUT scale from the universal value of the top, bottom, and the tau Yukawa couplings
predicted by a single $\mathsf{10}$-plet of $\mathsf{SO(10)}$ mode. These important corrections allow one to do two things: first unlike the case of  a single $\mathsf{10}$-plet of $\mathsf{SO(10)}$ the presence of two $\mathsf{10}$-plets
already give unequal Yukawas for the top and the bottom quarks. This already implies that a $\tan\beta$ as large as 50 is no longer
needed for consistency with the experimental data
on the top and bottom quark masses. In addition one finds that in this class of models the quartic couplings typically
contribute substantial amounts to the Yukawa couplings at the GUT scale
because  $\langle\Phi\rangle/M_c)$ is non-negligible 
and thus quartic interactions give significant contributions of size comparable to those of 
the cubic ones.
Further,  because of the experimental discovery that the Higgs boson mass at 125 GeV~\cite{Aad:2012tfa,Chatrchyan:2012ufa}
requires the size of weak scale supersymmetry to lie in the
TeV region, the sparticle spectrum for the scalars is typically  in the TeV region, and the current experimental limits on the gluino mass also lie in  the TeV  region.
The RG evolution of the Yukawas is sensitive to the sparticle spectrum and thus both the GUT boundary conditions and the sparticle spectrum enter in a significant way  in achieving consistency with
the data on the third generation masses for which currently the experimental limits are~\cite{Tanabashi:2018oca}
\begin{align}
m_t (\rm pole) &= 172.25\pm 0.08\pm 0.62 ~\rm GeV, \nonumber \\
\overline{m}_b(\overline{m}_b) &= 4.18^{+0.04}_{-0.03} ~\rm GeV, \nonumber \\
m_{\tau} (\rm pole) &= 1.77686\pm 0.00012 ~\rm GeV \,.
\label{bt-tau}
\end{align}
Thus in this analysis we  give a specific set of benchmarks where consistency with the data of Eq.~(\ref{bt-tau}) is achieved  with Yukawa couplings at the GUT scale including contributions from the cubic and the quartic terms in matter-Higgs interactions. We follow this up by a collider analysis of some of the benchmarks for some of the  sparticle spectrum that would be accessible at HL-LHC and HE-LHC.
Further details of the
analysis are as follows.  For the  high scale  parameters of the Higgs sector, i.e., $M_c, M^{210}, \eta$ and $\lambda$,   we take the ranges
$0.1\leq\eta,~\lambda\leq 2.0$, $2\times 10^{17}\leq M_c\leq 8.5\times 10^{17}$, 
and $1\times 10^{15}\leq M^{210}\leq 2.5\times 10^{16}$. 
We assume that the $SO(10)$ is broken near the scale $M^{126}=M_c$ to $SU(5)$ 
while $SU(5)$ is broken after the 210-plet of Higgs develops a VEV.
Ten representative benchmarks are chosen from this set. We then look at the spontaneous breaking of  the GUT symmetry which breaks the $\mathsf{SO(10)}$ gauge symmetry to the gauge symmetry of the
standard model. The VEVs that enter are $\mathcal V_{1_{_{{210}}}}$, $\mathcal V_{24_{_{{210}}}}$, $\mathcal V_{75_{_{{210}}}} $, and $\mathcal V_{1_{_{{126}}}} $.
Details of the spontaneous breaking of the GUT symmetry is given in appendix B.
 The numerical analysis of the VEVs for the  benchmarks is presented in Table \ref{tab1}.

\begin{table}[H]
\begin{center}
\resizebox{\linewidth}{!}{\begin{tabulary}{\linewidth}{l|cccccccc}
\hline\hline\rule{0pt}{3ex}
{Model}&{$\eta$}&{$\lambda$}& $M^{126}$ & $M^{210}$ &$\mathcal V_{1_{_{{210}}}}$&
 $\mathcal V_{24_{_{{210}}}}$&$\mathcal V_{75_{_{{210}}}} $&$\mathcal V_{1_{_{{126}}}} $ \\
\hline\rule{0pt}{3ex}
\!\!(a) & 4.33 & 2.39 & $8.07\times 10^{17}$ & $4.72\times 10^{15}$ & $1.44\times 10^{18}$ & $(-2.86-\imath0.67)\times 10^{18}$ & $(-6.51-\imath0.60)\times 10^{18}$ & $(3.49 +\imath0.24 i)\times 10^{17}$ \\
(b) & 1.57 & 2.75 & $2.74\times 10^{17}$ & $4.48\times 10^{15}$ & $1.36\times 10^{18}$ & $-2.64\times 10^{18}$ & $-1.44\times 10^{18}$ & $\imath 2.73\times 10^{18}$ \\
(c) & 1.54 & 0.33 & $6.55\times 10^{17}$ & $2.43\times 10^{16}$ & $3.30\times 10^{18}$ & $-3.69\times 10^{18}$ & $-1.30\times 10^{18}$ & $\imath 1.95\times 10^{18}$ \\
(d) & 3.39 & 4.46 & $7.87\times 10^{17}$ & $1.26\times 10^{16}$ & $1.80\times 10^{18}$ & $-3.47\times 10^{18}$ & $-1.88\times 10^{18}$ & $\imath 3.12\times 10^{18}$\\
(e) & 1.88 & 0.43 &$7.58\times 10^{16}$ & $2.27\times 10^{16}$ & $3.13\times 10^{17}$ & $3.52\times 10^{17}$ & $-2.89\times 10^{17}$ & $\imath 9.49\times 10^{16}$ \\
(f) & 2.59 & 1.52 & $4.15\times 10^{17}$ & $1.14\times 10^{16}$ & $1.24\times 10^{18}$ & $-2.15\times 10^{18}$ & $-1.07\times 10^{18}$ & $\imath 1.38\times 10^{18}$\\
(g) & 2.57 & 1.71 & $2.10\times 10^{17}$ & $3.35\times 10^{15}$ & $6.33\times 10^{17}$ & $(-1.23-\imath0.44)\times 10^{18}$ & $(-2.88-\imath0.41)\times 10^{18}$ & $(2.52 +\imath 0.25)\times 10^{17}$ \\
(h) & 2.77 & 2.95 &$5.84\times 10^{17}$ & $2.16\times 10^{16}$ & $1.63\times 10^{18}$ & $-2.94\times 10^{18}$ & $-1.51\times 10^{18}$ & $\imath 2.48\times 10^{18}$  \\
(i) & 2.98 & 2.33 &$6.66\times 10^{17}$ & $2.35\times 10^{16}$ & $1.73\times 10^{18}$ & $-3.02\times 10^{18}$ & $-1.51\times 10^{18}$ & $\imath 2.23\times 10^{18}$ \\
(j) & 1.40 & 0.12 &$2.31\times 10^{17}$ & $2.32\times 10^{16}$ & $1.28\times 10^{18}$ & $1.26\times 10^{18}$ & $-8.24\times 10^{17}$ & $\imath 2.84\times 10^{17}$ \\
\hline\hline
\end{tabulary}}
\caption{{\small A numerical estimate of the VEVs of the Standard Model singlets in $\mathsf{210}$, $\mathsf{126}$ and  $\mathsf{\overline{126}}$-plets {arising} in the spontaneous breaking of the $\mathsf{SO(10)}$ GUT gauge symmetry under the assumption $\mathcal V_{1_{_{{126}}}}=\mathcal V_{1_{_{\overline{126}}}}$. All VEVs and masses are in GeV.}}
\label{tab1}
\end{center}
\end{table}
To generate a pair of light Higgs doublets
needed for electroweak symmetry breaking, we use
the superpotential of Eq. (\ref{dtsuper}),
 and the results of  Eqs. (\ref{gut-2})$-$(\ref{gut-7}).
 Here  as discussed earlier the number of Higgs doublet pairs are seven  which produce a $7\times 7$ Higgs doublet mass matrix $M_d$  given in Appendix C which we diagonalize
 to recover a light Higgs doublet.
   The matrix $M_d$  is not symmetric
 and needs to be diagonalized by a biunitary transformation so that
 \begin{align}
U_d^\dagger M_d V_d=  M_d^{\text{diag}} =(0, m_{d_2}, m_{d_3}, \cdots, m_{d_7}).
\end{align}
 The massless mode is identified as the Higgs doublet  pair that enters in the electroweak symmetry breaking.
  The Higgs doublets in this pair do not involve components from  $\mathsf{126+\overline{126} + 210}$ heavy Higgs
  and have components only from $2\times\mathsf{10+120}$ light Higgs.
 For that reason the non-vanishing parts of $U_d$ are the components
 $U_{d_{11}}, U_{d_{21}}, U_{d_{31}}, U_{d_{61}}$ and similarly for $V_d$. These are recorded in Table~\ref{tab2} and Table~\ref{tab3}.
Here the parameters $a$, $b_{1,2}$, $c$ and $\bar{c}$ are as defined in appendix C and are taken to be in the range $0.1-2.0$.
{\small
\begin{table}[H]
\begin{center}
\resizebox{\linewidth}{!}{
\begin{tabulary}{\linewidth}{l|ccccccccc}
\hline\hline\rule{0pt}{3ex}
{Model}&{$a$}&{$b_{1}$}&{$b_{2}$}&{$c$}&{$\bar{c}$}& $U_{d{_{11}}}$& ${U_{d{_{21}}}}$& ${U_{d{_{31}}}}$& ${U_{d{_{61}}}}$ \\
\hline\rule{0pt}{3ex}
\!\!(a) & 4.09 & 3.06 & 2.92 & 0.04 & 3.89 & $-0.105-\imath 0.092$ & $0.114+\imath 0.093$ & $0.296-\imath0.346$ & $-0.669+\imath0.551$ \\
(b) & 0.39 & 1.07 & 0.21 & 0.99 & 1.89 & $-0.122+\imath0.006$  & $0.924-\imath0.049$  & $0.078-\imath0.004$  & $-0.349+\imath0.018$ \\
(c) & 1.19 & 0.59 & 0.60 & 0.45 & 1.59 & $0.185+\imath0.045$  & $-0.239-\imath0.058$  & $-0.093-\imath0.023$  & $0.919+\imath0.223$   \\
(d) & 3.63 & 2.55 & 3.86 & 0.09 & 3.65 & $-0.071-\imath0.009$  & $0.051+\imath0.006$  & $0.212+\imath0.026$  & $-0.966-\imath0.118$ \\
(e) & 0.54 & 1.38 & 0.12 & 0.34 & 0.86 & $0.022-\imath0.003$  & $-0.236+\imath0.036$  & $0.002-\imath0.000$ & $-0.960+\imath0.146$ \\
(f) & 2.72 & 1.26 & 0.55 & 0.32 & 0.60 & $-0.018+\imath0.001$  & $0.127-\imath0.006$ & $0.182-\imath0.008$ & $-0.974+\imath0.045$ \\
(g) & 2.99 & 2.23 & 1.82 & 0.49 & 2.43 & $-0.081-\imath0.141$  & $0.192+\imath0.139$ & $0.421-\imath0.155$ & $-0.843+\imath0.075$ \\
(h) & 0.88 & 0.16 & 2.68 & 0.90 & 1.16 & $-0.101+\imath0.004$  & $0.059-\imath0.003$  & $0.193-\imath0.008$  & $-0.973+\imath0.043$ \\
(i) & 1.64 & 2.59 & 0.88 & 0.35 & 0.43 & $-0.021+\imath0.000$  & $0.120-\imath0.002$  & $0.184-\imath0.003$ & $-0.975+\imath0.018$ \\
(j) & 0.20 & 1.29 & 1.06 & 0.18 & 1.41 & $-0.213-\imath0.269$  & $0.258+\imath0.326$  & $0.005+\imath0.006$  & $0.523+\imath0.660$ \\
\hline\hline
\end{tabulary}}
\caption{{\small A numerical estimate of the elements of the down Higgs zero mode eigenvector using the analysis of Table \ref{tab1}
and the couplings of Eq. (\ref{dtsuper}).}}
\label{tab2}
\end{center}
\end{table}

\begin{table}[H]
\begin{center}
\resizebox{\linewidth}{!}{
\begin{tabular}{l|ccccccccc}
\hline\hline\rule{0pt}{3ex}
{Model}&{$a$}&{$b_{1}$}&{$b_{2}$}&{$c$}&{$\bar{c}$}&$V_{d{_{11}}}$&${V_{d{_{21}}}}$&${V_{d{_{31}}}}$&${V_{d{_{61}}}}$ \\
\hline\rule{0pt}{3ex}
\!\!(a) & 4.09 & 3.06 & 2.92 & 0.04 & 3.89 & $-0.320+\imath0.000$ & $0.336-\imath0.002$ & $-0.412+\imath0.000$ & $0.772-\imath0.136$ \\
(b) & 0.39 & 1.07 & 0.21 & 0.99 & 1.89 & $-0.175$ & $0.963$ & $-0.045$ & $0.201$ \\
(c) & 1.19 & 0.59 & 0.60 & 0.45 & 1.59 & $-0.106$ & $0.215$ & $-0.098$ & $0.966$  \\
(d) & 3.63 & 2.55 & 3.86 & 0.09 & 3.65 & $0.172$ & $-0.115$ & $0.210$ & $-0.955$ \\
(e) & 0.54 & 1.38 & 0.12 & 0.34 & 0.86 & $-0.003$ & $-0.006$ & $-0.002$ & $1.000$ \\
(f) & 2.72 & 1.26 & 0.55 & 0.32 & 0.60 & $0.033$ & $-0.123$ & $0.183$ & $-0.975$ \\
(g) & 2.99 & 2.23 & 1.82 & 0.49 & 2.43 & $0.278+\imath0.000$ & $-0.350+\imath0.050$ & $0.419-\imath0.000$ & $-0.762+\imath0.205$ \\
(h) & 0.88 & 0.16 & 2.68 & 0.90 & 1.16 & $0.206$ & $-0.037$ & $0.190$ & $-0.959$ \\
(i) & 1.64 & 2.59 & 0.88 & 0.35 & 0.43 & $-0.039$ & $0.146$ & $-0.183$ & $0.971$  \\
(j) & 0.20 & 1.29 & 1.06 & 0.18 & 1.41 & $0.051$ & $-0.071$ & $0.009$ & $0.996$ \\
\hline\hline
\end{tabular}}
\caption{{\small A numerical estimate of the elements of the up Higgs zero mode eigenvector using the analysis of Table \ref{tab1} and the couplings of Eq. (\ref{dtsuper}).}}
\label{tab3}
\end{center}
\end{table}

\begin{table}[H]
\begin{center}
\begin{tabulary}{1.20\textwidth}{l|CCCC}
\hline\hline\rule{0pt}{3ex}
Model & $f^{(1)}$ & $f^{(2)}$ & $f^{(3)}$ & $f^{10_r}$\\
\hline\rule{0pt}{3ex}
\!\!(a) & 0.54 & 0.02 & 0.14 & (0.36, 0.78)\\
(b) & 0.08 & 0.08 & 0.09  & (1.49, 0.33) \\
(c) & 0.16 & 0.07 & 0.14  & (0.55, 0.71)\\
(d) & 0.08 & 0.01 & 0.18  & (0.76, 2.18)\\
(e) & 0.54 & 0.18 & 0.03  & (1.81, 0.89) \\
(f) & 0.08 & 0.03 & 0.08 & (1.96, 1.23) \\
(g) & 0.50 & 0.25 & 0.05  & (0.48, 0.62) \\
(h) & 0.27 & 0.24 & 0.07  & (0.24, 2.42)\\
(i) & 0.10 & 0.04 & 0.10  & (1.83, 1.16)\\
(j) & 0.09 & 0.02 & 0.03  & (0.84, 0.25) \\
\hline
\end{tabulary}\end{center}
\caption{{\small The GUT scale parameters in the cubic and quartic superpotentials $W_3$,  $W_4^{(1)}$, $W_4^{(2)}$ and  $W_4^{(3)}$
for the model points (a)$-$(j). The masses are in GeV.}}
\label{tab4}
\end{table}

\begin{table}[H]
\begin{center}
\begin{tabulary}{1.10\textwidth}{l|CCCCCCCCC}
\hline\hline\rule{0pt}{3ex}
Model & $h_t^0$ & $h_b^0$ & $h_\tau^0$ & $\delta h_t^{\rm GUT}$ &$\delta h_b^{\rm GUT}$ & $\delta h_{\tau}^{\rm GUT}$ & $h_t^{\rm GUT}$ & $h_b^{\rm GUT}$ & $h_{\tau}^{\rm GUT}$  \\
\hline\rule{0pt}{3ex}
\!\!(a) & 0.181 & 0.412 & 0.412 & 0.424 & 0.385 & 0.376 & 0.519 & 0.026 & 0.037 \\
(b) & 0.354 & 0.168 & 0.168 & 0.143 & 0.091 & 0.060 & 0.507 & 0.078 &  0.108 \\
(c) & 0.197 & 0.266 & 0.266 & 0.292 & 0.329 & 0.177 & 0.494 & 0.063 & 0.089 \\
(d) & 0.163 & 0.339 & 0.339 & 0.354 & 0.349 & 0.343 & 0.513 & 0.032 &  0.043 \\
(e) & 0.489 & 0.029 & 0.029 & 0.014 & 0.023 & 0.010 & 0.498 & 0.052 &  0.072 \\
(f) & 0.344 & 0.243 & 0.243 & 0.169 & 0.175 & 0.147 & 0.507 & 0.068 &  0.097 \\
(g) & 0.234 & 0.251 & 0.251 & 0.274 & 0.314 & 0.168 & 0.498 & 0.104 & 0.151 \\
(h) & 0.334 & 0.114 & 0.114 & 0.159 & 0.216 & 0.028 & 0.505 & 0.103 & 0.143 \\
(i) & 0.289 & 0.279 & 0.279 & 0.192 & 0.198 & 0.166 & 0.495 & 0.084 &  0.113 \\
(j) & 0.520 & 0.069 & 0.069 & 0.013 & 0.007 & 0.034 & 0.496 & 0.077 & 0.104 \\
\hline
\end{tabulary}\end{center}
\caption{{\small
The  magnitude of the contributions to the top, bottom, and tau Yukawa couplings from cubic interactions (columns 2-4),
 from quartic interactions (columns 5-7)
 and the magnitude of their complex  sum (columns 8-10)
 at the GUT scale for the parameter set  of Table \ref{tab4}.
The Yukawa couplings are in general complex and we add the contributions of the cubic and
quartic interactions as complex numbers and exhibit only their magnitudes in the table.
}}
\label{tab5}
\end{table}

Next we give a computation of the Yukawa couplings at the GUT scale. As discussed in section~\ref{sec3}, contributions to the Yukawa couplings arise  from cubic interactions of  Eq.~(\ref{cubic-coupling}) and  from quartic interactions of Eqs.~(\ref{w41})-(\ref{w43}).
The couplings that enter here are: $f^{10_1}, f^{10_2}$, 
 $f^{(1)}$, $f^{(2)}$ and $f^{(3)}$. Using the values of these parameters given in Table \ref{tab4},
we exhibit in
 Table~\ref{tab5}  the contribution to the Yukawa couplings from the cubic interactions, from the quartic interactions, and their sum.
 Table~\ref{tab5} defines the Yukawa couplings for the top and the bottom quarks,  and for the tau lepton at the GUT scale.
To evolve the Yukawas  from the GUT scale to the electroweak scale we use RG equations (RGE) within the supergravity (SUGRA) model~\cite{sugra-uni,Nath:2016qzm}. The running of the RGEs is implemented with the help of \code{SPheno-4.0.4}~\cite{Porod:2003um,Porod:2011nf} which uses two-loop MSSM RGEs and three-loop standard model (SM) RGEs and takes into account SUSY threshold effects at the one-loop level. The larger SUSY scale makes it necessary to employ a two-scale matching condition at the electroweak and SUSY scales~\cite{Staub:2017jnp} thereby improving the calculations of the Higgs boson mass and of the sparticle spectrum. The bottom quark mass and $\alpha_S$ (the fine structure constant for the $\mathsf{SU(3)_C}$)
are run up to the scale of the  $Z$ boson mass, $M_Z$,  using four-loop RGEs in the $\overline{\rm MS}$ scheme while for the top quark, the evolution starts at the pole mass and the $\overline{\rm MS}$ mass is computed by running down to the $M_Z$ scale including two-loop QCD corrections.

The tau mass is calculated at $M_Z$ including one-loop electroweak corrections. The calculation of the $\overline{\rm MS}$ Yukawas at the electroweak scale involve the first matching conditions to include SM thresholds. Those couplings are then run using 3-loop SM RGEs to $M_{\rm SUSY}$ where the second matching takes place to include SUSY thresholds at the one-loop level and a shift is made to the $\overline{\rm DR}$ scheme. The 2-loop MSSM RGEs of the $\overline{\rm DR}$ Yukawas and gauge couplings are then run to the GUT scale where the soft SUSY breaking boundary conditions are applied. Thus in addition to the GUT scale Yukawas we define the SUGRA parameters
 $m_0$, $A_0$, $m_1$, $m_2$, $m_3$ and  $\tan\beta$ where $m_0$ is the universal scalar mass, $A_0$ is the universal trilinear coupling, $m_1,m_2,m_3$ are the $\mathsf {U(1),  SU(2),  SU(3)}$ gaugino masses all at the GUT scale and $\tan\beta= \langle H_u\rangle/\langle H_d\rangle$ where
 $H_u$ gives mass to the up quarks and $H_d$ gives mass to the down quarks and the charged leptons.
 The choice of the SUGRA parameters is constrained by the dark matter relic density for which we take Eq.~(\ref{planck}) to be
 the upper limit,
 the Higgs boson mass constraint, and the experimental lower limits on sparticle masses.
The result of the RG analysis is shown in Table \ref{tab6} and here one finds that
consistency with the top, bottom and tau masses along with gauge coupling unification can be achieved for values of $\tan\beta$ as low as $\tan\beta\sim 5-10$. [A $b-t-\tau$ unification with low $\tan\beta$ also occurs in
unified Higgs models involving a $\mathsf{144+\overline{144}}$ of Higgs fields~\cite{Babu:2005gx,Nath:2007eg,Nath:2009nf,Ajaib:2013kka}.]\\

We note that in Table \ref{tab6}
 the non-universality of gaugino masses plays an important role in producing the compressed
 spectrum which is needed in part to satisfy the relic density constraint.
  In  general the smearing of
 gauge couplings will also occur.  However, it turns out that
 there is no direct relationship between the non-universality of the
 gaugino mass terms and corrections to the gauge coupling constants. This is so because
  the gaugino masses arise as soft terms and in gravity mediation such terms depend
 on the GUT sector, the Kahler metric and the gauge kinetic function. However,
 in supergravity models
  the corrections to the gauge couplings involve
a different parameter as discussed in~\cite{Dasgupta:1995js}
and there is no rigid connection between the two.
Because of the lack of any direct connection between the corrections in the two cases,
 we have not included such corrections to the evolution of the gauge couplings. Further since the
unification of gauge couplings works rather well in the standard  MSSM/SUGRA
(for a recent analysis see~\cite{Aboubrahim:2018bil}) one can only
infer  that the gravitational smearing corrections to the gauge couplings at the GUT scale are small.

    We discuss now briefly the mechanism at work for achieving $t-b-\tau$ unification with
    a low $\tan\beta$. Here there are two components at work. First we note that unlike
    the case of the standard model the light pair of Higgs doublets do not arise from
    a single 10-plet of Higgs which is the case for the standard $SO(10)$ model.
    Rather, the light pair of Higgs doublets  have components from
    the two 10-plets of Higgs and one 120-plet of Higgs. This leads to a splitting of the
    Yukawa couplings at the GUT scale  even without contributions from the higher
    dimensional operators
    as can be seen from columns 2, 3 and 4 of Table 5. Second, we have additional
    contributions from the higher dimensional operators which contribute to the Yukawas.
    Together, it is then possible to achieve $t-b-\tau$ unification at low values of $\tan\beta$
    in certain regions of the parameter space of supergravity models. We emphasize
    that the analysis does not predict the existence of a low $\tan\beta$ but only points
     to the  possibility to achieve such a unification within the missing partner $SO(10)$
     with a low $\tan\beta$. Specifically large values of $\tan\beta$ are not excluded
     but we focus in this work on low $\tan\beta$ values as those are difficult to achieve
     in the standard $SO(10)$ model if one wants $t-b-\tau$ unification.

\begin{table}[H]
\begin{center}
\begin{tabulary}{1.10\textwidth}{l|CCCCCC|CCC}
\hline\hline\rule{0pt}{3ex}
Model & $m_0$ & $A_0$ & $m_1$ & $m_2$ & $m_3$ & $\tan\beta$  &$m_t$ (pole) & $\overline{m}_b(\overline{m}_b)$ & $m_{\tau}$ (pole)  \\
\hline\rule{0pt}{3ex}
\!\!(a) & 3051 & -10193 & 838 & 492 & 3502 & 5  & 173.9 & 4.165 & 1.77682  \\
(b) & 1096 & 4572 & 925 & 562 & 4081 & 15  & 172.4 & 4.195 & 1.77682  \\
(c) & 4127 & 3359 & 1049 & 642 & 5498 & 12 & 174.0 & 4.210 & 1.77682  \\
(d) & 1150 & -5313 & 1177 & 676 & 3423 & 6  & 172.2 & 4.210& 1.77682  \\
(e) & 1865 & 805 & 1440 & 861 & 6929 & 10  & 174.0 & 4.150& 1.77682 \\
(f) & 3763 & 9793 & 1748 & 996 & 4048 & 13  & 173.1 & 4.180& 1.77682  \\
(g) & 4027 & -4880 & 1989 & 1093 & 4560 & 20  & 173.1 & 4.170 & 1.77682  \\
(h) & 1706 & -4508 & 2596 & 3219 & 1428 & 19  & 173.6 & 4.180 & 1.77682 \\
(i) & 12196 & -1035 & 3422 & 1817 & 1687 & 15  & 173.2 & 4.160 & 1.77682 \\
(j) & 1655 & -1418 & 4492 & 4807 & 2615 & 14  & 172.8 & 4.170 & 1.77682 \\
\hline
\end{tabulary}\end{center}
\caption{
The SUGRA parameters sets used for RG analysis where the boundary conditions for the Yukawas for the top, bottom, and the tau
are taken from Table \ref{tab5}. In the analysis the GUT scale ranges from $8.8\times 10^{15}$ GeV to $1.6\times 10^{16}$ GeV.}
\label{tab6}
\end{table}

\begin{table}[H]
\begin{center}
\begin{tabulary}{1.10\textwidth}{l|CCCCCCC}
\hline\hline\rule{0pt}{3ex}
Model & $h^0$  & $\tilde t$ & $\tilde g$ & $\tilde\tau$ & $\tilde\chi^0_1$ & $\tilde\chi^{\pm}_1$ & $\Omega h^2$ \\
\hline\rule{0pt}{3ex}
\!\!(a) & 125.3 & 4078 & 7189 & 3022 & 356.3 & 376.9 & 0.062 \\
(b) & 124.0 & 6159 & 8180 & 983 & 379.4 & 405.0 & 0.109 \\
(c) & 125.5 & 8477 & 10949 & 4069 & 435.1 & 461.8 & 0.104 \\
(d) & 124.4 & 4589 & 6937 & 1174 & 503.9 & 528.3 & 0.088 \\
(e) & 126.1 & 9929 & 13458 & 1854 & 606.5 & 633.6 & 0.103 \\
(f) & 123.5 & 6312 & 8212 & 3644 & 758.7 & 783.3 & 0.096 \\
(g) & 126.8 & 6727 & 9194 & 3868 & 881.4 & 888.4 & 0.048 \\
(h) & 125.5 & 1171 & 3118 & 1709 & 1162 & 2394 & 0.068 \\
(i) & 124.6 & 8210 & 3949 & 12011 & 1588 & 1591 & 0.109 \\
(j) & 124.0 & 3534 & 5407 & 2261 & 2032 & 2308 & 0.056 \\
\hline
\end{tabulary}\end{center}
\caption{Low scale SUSY mass spectrum showing the Higgs boson, the stop, the gluino, the stau and the light electroweakino masses and the LSP relic density for the benchmarks  of Table~\ref{tab6}.}
\label{tab7}
\end{table}

Some of the sparticle spectrum for each of the model points are  exhibited in Table \ref{tab7}. The sparticle spectrum of benchmarks (a)$-$(g) contains light electroweakinos, i.e., of mass less than 1 TeV while stops and gluinos are much heavier. Those points will be of interest in the next section where we discuss the LHC implications. The dark matter relic density is calculated using \code{micrOMEGAs-5.0.9}~\cite{Belanger:2014vza} and we use as an upper limit  the experimental value reported by the Planck collaboration~\cite{Aghanim:2018eyx}
\begin{equation}
(\Omega h^2)_{\rm PLANCK}=0.1198\pm 0.0012\,.
\label{planck}
\end{equation}
As seen from Table \ref{tab7} some model  points do not saturate the relic density and thus  these models can
 accommodate more than one dark matter component, e.g., a hidden sector Dirac fermion~\cite{Feldman:2010wy,Feldman:2011ms,Aboubrahim:2019mxn} or an axion~\cite{Baer:2018rhs,Halverson:2017deq}. We have checked that the spin-independent proton-neutralino cross-sections are very small for such model points and thus not yet excluded.
As noted earlier the  benchmarks of Tables~\ref{tab1}$-$\ref{tab7} are just a sample of a larger parameter space where consistency with Eq.~(\ref{bt-tau}) can be achieved with a $\tan\beta$ significantly smaller  than
50. This is exhibited  in the  right panel of
Fig.~\ref{fig1} which shows a large set of model points with $\tan\beta$ in the range 5$-$10 and all of the model points exhibited have $\tan\beta$ less than 20.
The   GUT scale splitting of the Yukawas and their evolution to the electroweak scale is exhibited graphically for models
(a), (e) and (i)  in Fig.~\ref{fig2}. Here the
 left panel shows the top and bottom Yukawas while the right panel shows the bottom and tau Yukawas.
The kink in the evolution of the Yukawas is due to sparticle mass threshold effects.
\begin{figure}[H]
 \centering
   \includegraphics[width=0.49\textwidth]{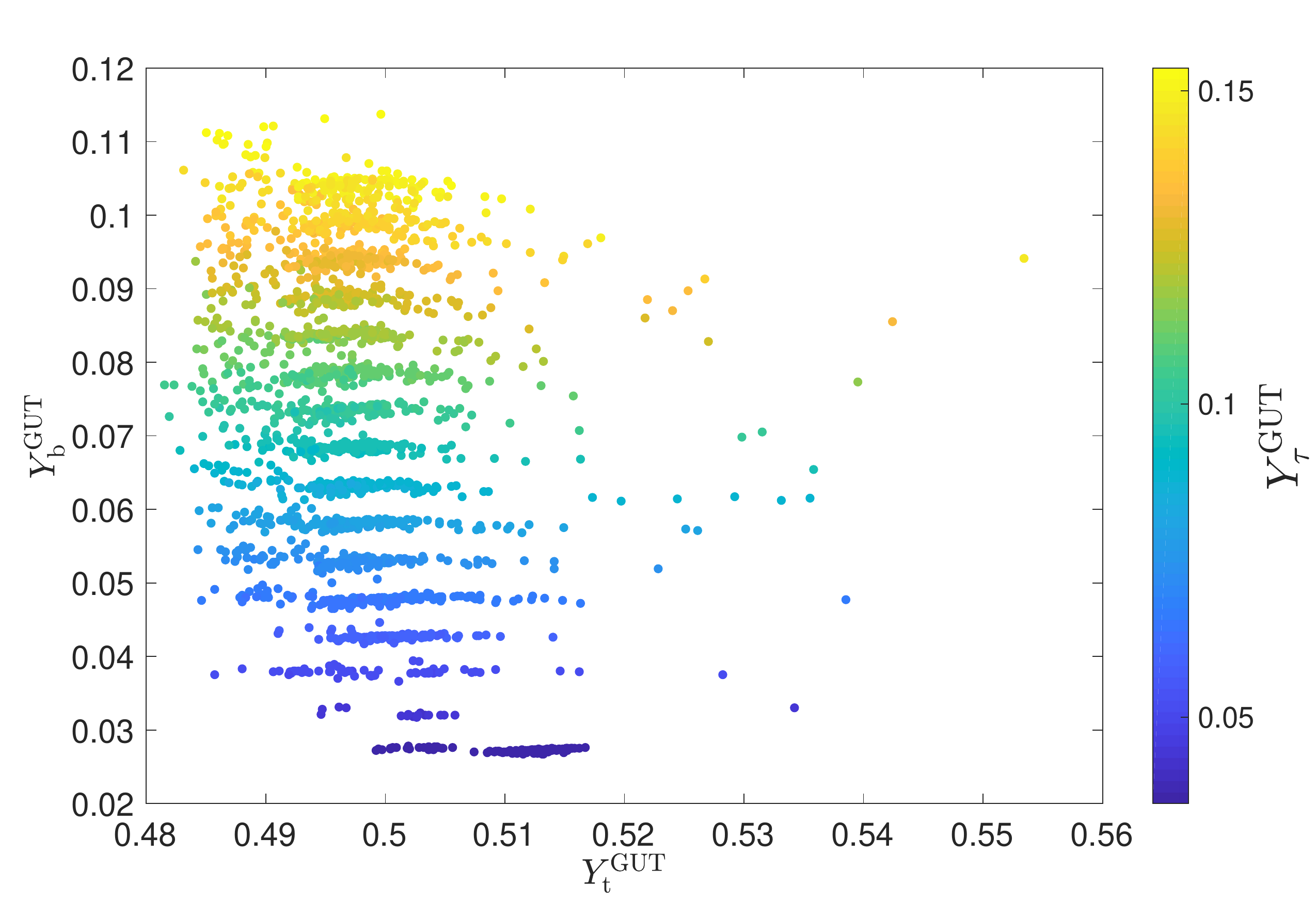}
   \includegraphics[width=0.49\textwidth]{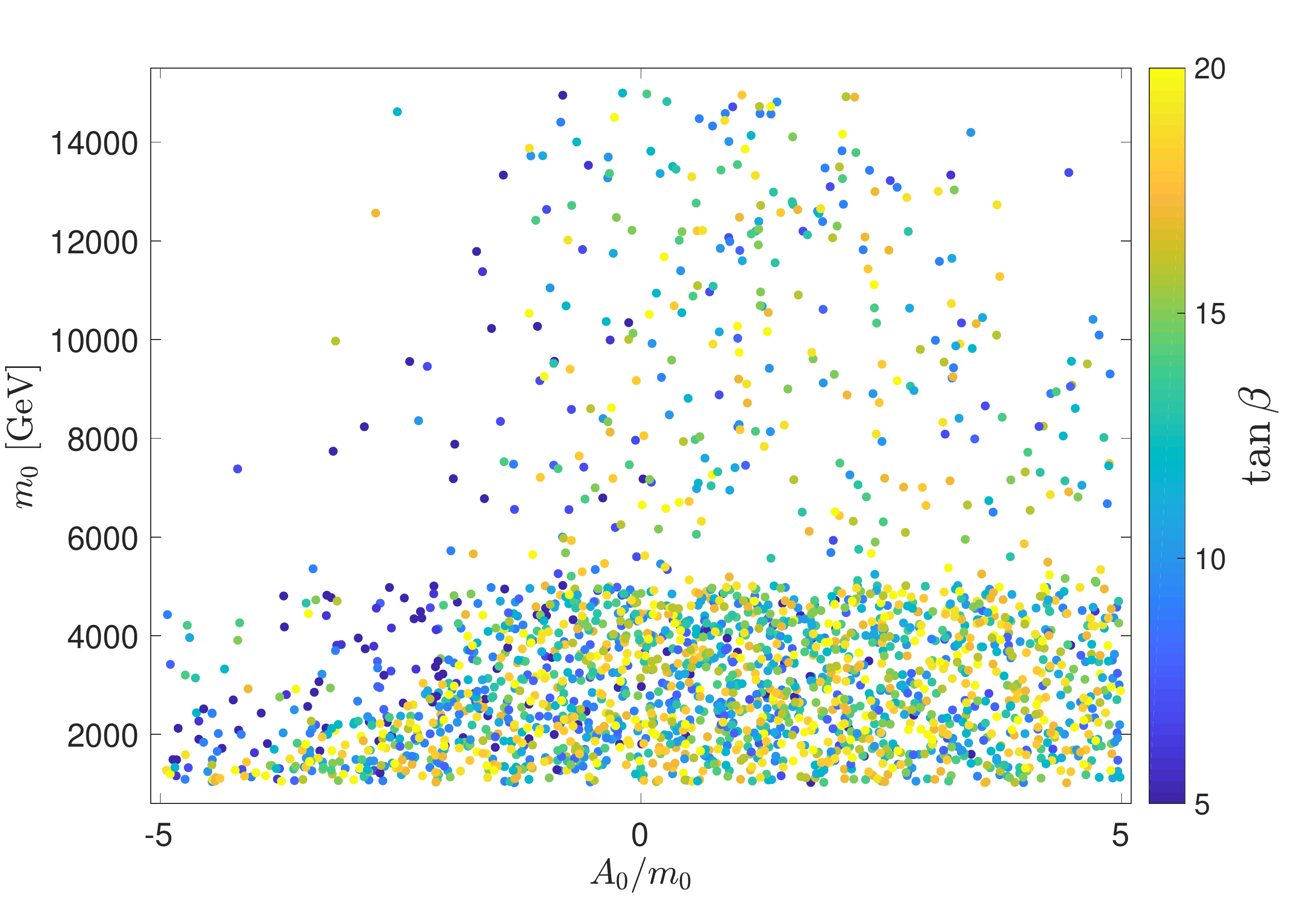}
   \caption{Left panel: A scatter plot of the top, bottom and tau GUT scale Yukawa couplings which produce the correct low scale top and bottom quark masses within a 2\% theoretical uncertainty and the exact tau mass. Right panel: a scatter plot in the $m_0$-$A_0/m_0$ plane with the color axis showing $\tan\beta$.}
	\label{fig1}
\end{figure}

\begin{figure}[H]
 \centering
 \includegraphics[width=0.49\textwidth]{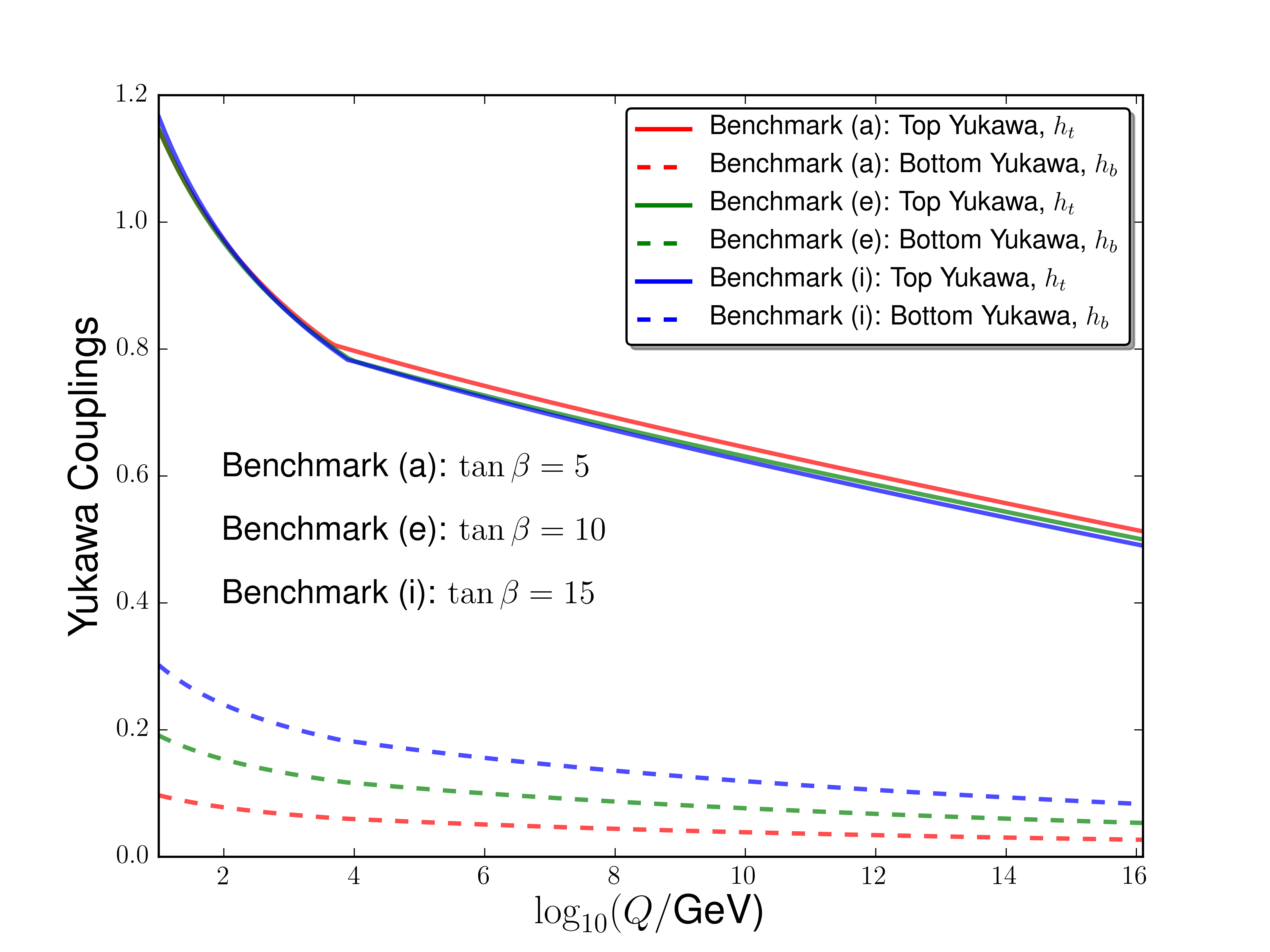}
 \includegraphics[width=0.49\textwidth]{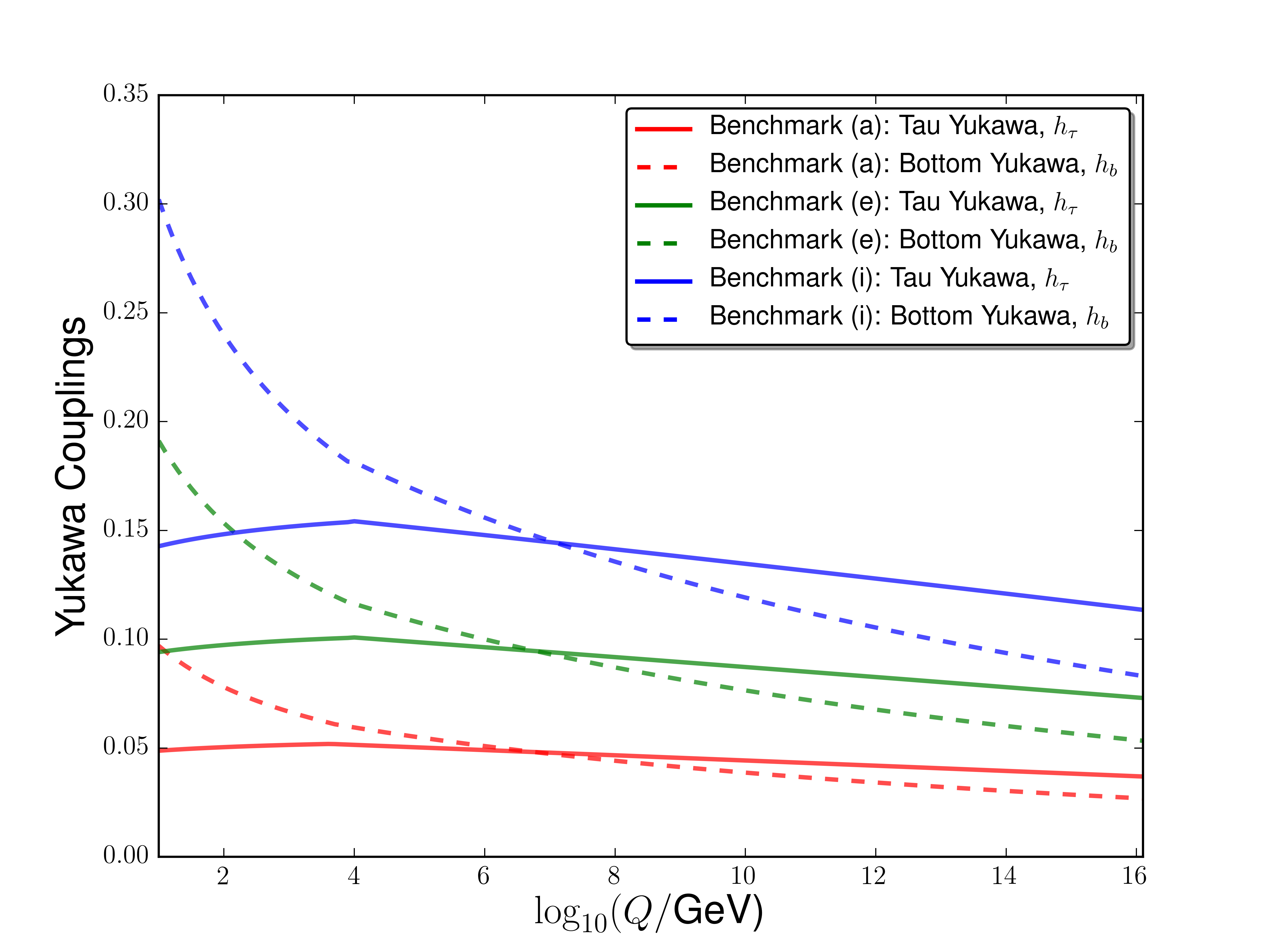}
   \caption{Left panel: The running of the top and bottom Yukawa couplings for benchmarks (a), (e) and (i). Right panel: The running of the bottom and tau Yukawas for the same benchmarks.}
	\label{fig2}
\end{figure}

\subsection{Electroweakino pair production at the LHC and their decay channels}

The low energy sparticle spectrum of the  benchmarks in Tables~\ref{tab6} and~\ref{tab7} contain light electroweakinos (charginos and neutralinos).
 In this section we investigate the potential of discovering light electroweakinos with small mass splittings at the LHC. According to Table~\ref{tab7}, points (a)$-$(f) possess the property of a small mass splitting between the lightest chargino and the lightest neutralino (LSP). Note that the second lightest neutralino has the same mass as the lightest chargino. Points (g) and (i) have very small mass splittings (less than 8 GeV) and require special treatment~\cite{Aboubrahim:2020wah}. Point (h) is an example of a stop coannihilation scenario where the stop lies close in mass to the LSP while point (j) points to a stau coannihilation region. We will not consider these scenarios here (for previous works involving stop and stau coannihilation, see, e.g.~\cite{Kaufman:2015nda,Aboubrahim:2017aen}) but focus on the chargino coannihilation, i.e. points (a)$-$(f). The electroweakino mass range under study is $\sim 350$ GeV to $\sim 800$ GeV with a
 chargino and neutralino
  mass splitting of $\sim 20-27$ GeV. It is worth noting that model point (h) with a stop mass of $\sim 1.2$ TeV is within the reach of HL-LHC. The possibility of detecting electroweakinos and gluinos at HL-LHC and HE-LHC has been studied in an earlier work~\cite{Aboubrahim:2018bil} as well as light charged and CP odd Higgs~\cite{Aboubrahim:2018tpf,Aboubrahim:2019vjl}.

Constraints on the electroweakino mass spectrum from the LHC have been taken into consideration when selecting the benchmarks under study. CMS has excluded charginos up to 230 GeV with a mass splitting of $\sim 20$ GeV while lighter masses were excluded for larger mass splitting (down to 100 GeV for 35 GeV splitting)~\cite{CMS:2017fij,Sirunyan:2018iwl}. More recent searches~\cite{Sirunyan:2019zfq} in the zero and one lepton channels excluded charginos up to 200 GeV for a larger range of mass splittings, up to 50 GeV. ATLAS has put more stringent constraints on charginos and neutralinos. For the small and intermediate mass splittings~\cite{Aad:2019vvi} chargino mass up to 345 GeV has been excluded and up to 200 GeV also ruled out for an almost degenerate spectrum. The limit on charginos reach a mass $\sim 1.1$ TeV associated with a massless neutralino~\cite{Aad:2015eda,Aaboud:2018jiw}. For chargino mass of more than 350 GeV, a mass splitting with the LSP of up to 50 GeV is still allowed and that mass gap increases for heavier spectra. The benchmarks (a)$-$(f) are in accordance with those constraints from ATLAS and CMS.

We consider electroweakino pair production, $\tilde\chi^0_2\,\tilde\chi^{\pm}_1$ and $\tilde\chi^+_1\,\tilde\chi^-_1$ in proton-proton collisions at 14 TeV (HL-LHC) and 27 TeV (HE-LHC). The NLO+NNLL production cross-sections for the benchmarks (a)$-$(f) are calculated with \code{Resummino-2.0.1}~\cite{Debove:2011xj,Fuks:2013vua} using the five-flavor NNPDF23NLO PDF set. The results are shown in Table~\ref{tab8} along with the branching ratios of $\tilde\chi^0_2$ and $\tilde\chi^{\pm}_1$ into the different final states of interest.

\begin{table}[H]
\begin{center}
\resizebox{\linewidth}{!}{\begin{tabulary}{\linewidth}{l|CC|CC|CCC}
\hline\hline\rule{0pt}{3ex}
Model  & \multicolumn{2}{c|}{$\sigma_{\rm NLO+NNLL}(pp\rightarrow \tilde\chi^0_2\,\tilde\chi^{\pm}_1)$} & \multicolumn{2}{c|}{$\sigma_{\rm NLO+NNLL}(pp\rightarrow \tilde\chi^+_1\,\tilde\chi^-_1)$}& \multicolumn{3}{c}{Branching ratios} \\
  &  &  &  &  &  &  &  \\
  & 14 TeV & 27 TeV & 14 TeV & 27 TeV & $\tilde\chi^{\pm}_1\rightarrow \tilde\chi^0_1q_i \bar{q}_j$ & $\tilde\chi^0_2\rightarrow \tilde\chi^0_1\ell^+\ell^-$ & $\tilde\chi^0_2\rightarrow \tilde\chi^0_1\tau^+\tau^-$  \\
\hline\rule{0pt}{3ex}
\!\!(a) & 174.3 & 540.5 & 84.9 & 270.3 & 0.67 & 0.28 & 0.14 \\
(b) & 129.5 & 414.3 & 62.8 & 206.8 & 0.64 & 0.29 & 0.21 \\
(c) & 75.8 & 258.6 & 36.5 & 128.5 & 0.67 & 0.12 & 0.07 \\
(d) & 40.6 & 148.8 & 19.4 & 73.7 & 0.66 & 0.32 & 0.17 \\
(e) & 18.5 & 76.7 & 8.7 & 37.7 & 0.66 & 0.33 & 0.17 \\
(f) & 6.2 & 30.8 & 2.9 & 15.0 & 0.67 & 0.07 & 0.04 \\
\hline
\end{tabulary}}
\end{center}
\caption{The NLO+NNLL production cross-sections, in fb, of electroweakinos: the second neutralino-chargino pair, $\tilde\chi^0_2\,\tilde\chi^{\pm}_1$ (second and third columns), and opposite sign chargino pair (fourth and fifth columns) at $\sqrt{s}=14$ TeV and at $\sqrt{s}=27$ TeV for benchmarks (a)$-$(f) of Table~\ref{tab6}. Also shown are the branching ratios to quarks and leptons for the electroweakinos of the same benchmarks. Note that $q\in\{u,d,c,s\}$ and $\ell\in\{e,\mu\}$.}
\label{tab8}
\end{table}

The second neutralino three-body decays into two light leptons (electrons and muons) proceed through an off-shell $Z$ and Higgs bosons. Light leptons may also come from the decay of taus. This three-body decay (shown in the last column of Table~\ref{tab8}) can also proceed via the exchange of a stau. We note that the branching ratio to two taus is particularly enhanced for benchmark (b) and this is because of a relatively light stau (983 GeV, see Table~\ref{tab7}). The three-body decay of a chargino into quarks is mediated by an off-shell $W$ boson and is the dominant decay channel as seen in Table~\ref{tab8}.

\subsection{Signal and background simulation and event selection}

The signal which consists of electroweakino pair production can be reconstructed based on specific final states of our choice. Here we look for a pair of same flavor and opposite sign (SFOF) light leptons (electron or muons), at least two jets and a large missing transverse energy (MET). The leptons are expected to be soft as a result of the small mass splitting between the LSP and the NLSP (chargino or second neutralino). However, the lepton and MET systems receive a kick in momentum as they recoil against a hard initial state radiation (ISR). This ISR-assisted topology is crucial in extracting the signal from the large standard model (SM) background. The signal region (SR) will be denoted as SR $2\ell$N$j$ with $N\geq 2$ as the number of jets required in the final state.  The dominant SM backgrounds come from diboson production, $Z/\gamma+$jets, dilepton production from off-shell vector bosons ($V^*\rightarrow\ell\ell$), $t\bar{t}$ and $t+W/Z$. The subdominant backgrounds are Higgs production via gluon fusion ($ggF$ H) and vector boson fusion (VBF).  The signal and SM backgrounds are simulated at LO at 14 TeV and 27 TeV with \code{MadGraph5\_aMC@NLO-2.6.3} interfaced to \code{LHAPDF}~\cite{Buckley:2014ana} using the NNPDF30LO PDF set. At the generator level, up to two partons are added to the main process to produce extra jets. The parton level events are passed on to \code{PYTHIA8}~\cite{Sjostrand:2014zea} for showering and hadronization using a five-flavor matching scheme in order to avoid double counting of jets. The matching/merging scale for the signal is set at 100/150 GeV. Additional jets from ISR and FSR (final state radiation) are allowed in order to boost the signal topology. Jets are clustered with \code{FastJet}~\cite{Cacciari:2011ma} using the anti-$k_t$ algorithm~\cite{Cacciari:2008gp} with jet radius $R=0.4$. \code{DELPHES-3.4.2}~\cite{deFavereau:2013fsa} is then employed for detector simulation and event reconstruction using the beta card for HL-LHC and HE-LHC studies. The cross-sections in the resulting files are then scaled to their NLO+NNLL values for the signal samples and NLO for the SM backgrounds. The corresponding ROOT files are then analyzed and analysis cuts are implemented with the help of \code{ROOT 6}~\cite{Antcheva:2011zz}.

The preselection criteria applied to the signal and background samples involve two SFOS leptons with the leading and subleading transverse momenta $p_T>15$ GeV for electrons and $p_T>10$ GeV for muons with $|\eta|<2.5$. Each event should contain at least two non-b-tagged jets with the leading $p_T>20$ GeV in the $|\eta|<2.4$ region. For the signal region analysis, we design a set of kinematic variables that are especially effective in reducing the SM background while retaining as much of the signal as possible. Since the signal is rich in missing transverse momentum, then a cut on $E^{\rm miss}_T$ is essential in reducing the background. The SFOS dilepton invariant mass, $m_{\ell\ell}$, is calculated using the leading and subleading leptons in an event. The total transverse momentum of the dilepton system is associated with the $Z$ boson and denoted by $p_T^Z$. The dijet system, consisting of the leading and subleading jets in an event, is reconstructed and associated to a $W$ boson which is closest in $\Delta\phi$ to the $Z\rightarrow\ell\ell$+MET system. The other jets are taken to be ISR, with $p_T^{\rm ISR}$ denoting the vector sum of all ISR transverse momenta in an event. From these observables we determine $E^{\rm miss}_T/p^Z_T$ and $\Delta\phi(\vec{p}^{\,\rm miss}_T,Z)$ which is the opening angle between the MET and $p^Z_T$. The normalized distributions in some of those variables are shown in Fig.~\ref{fig3}.

\begin{figure}[H]
 \centering
 \includegraphics[width=0.49\textwidth]{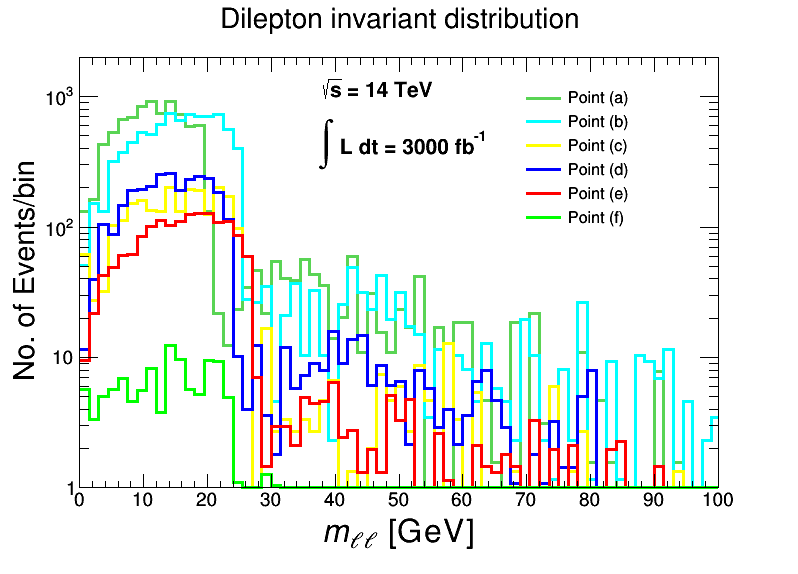}
  \includegraphics[width=0.49\textwidth]{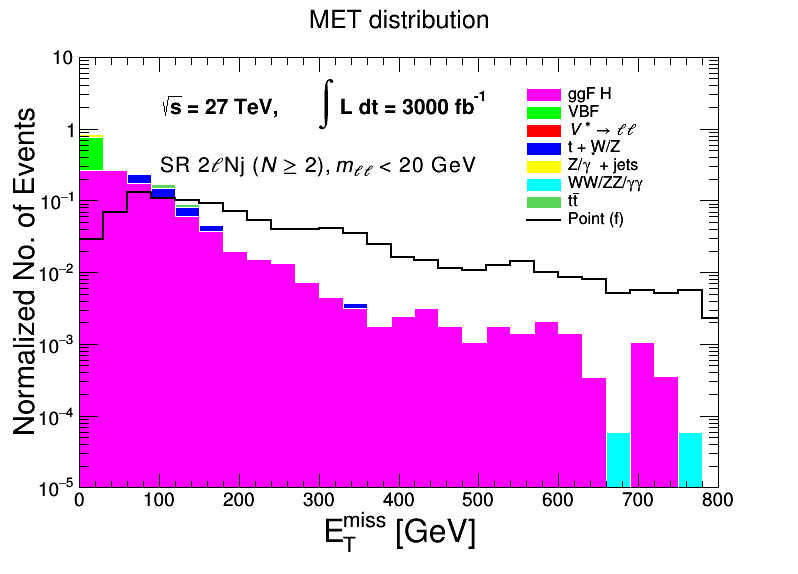} \\
 \includegraphics[width=0.49\textwidth]{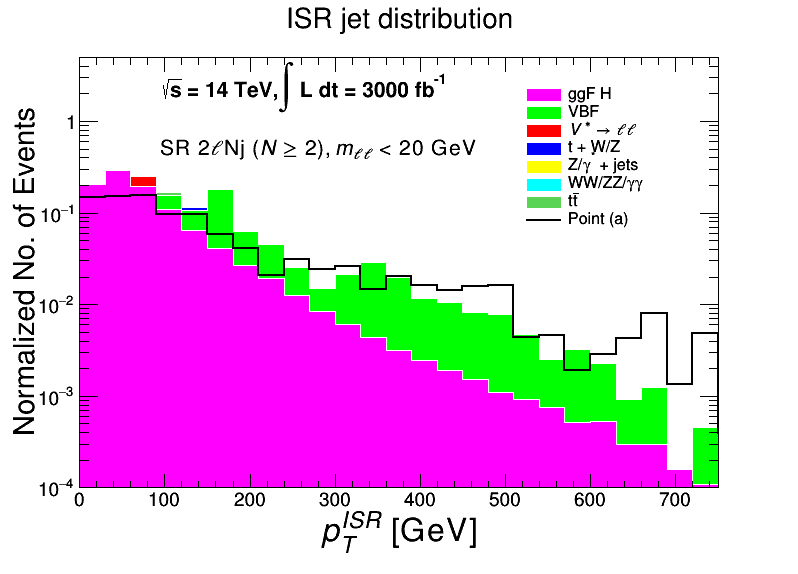}
 \includegraphics[width=0.49\textwidth]{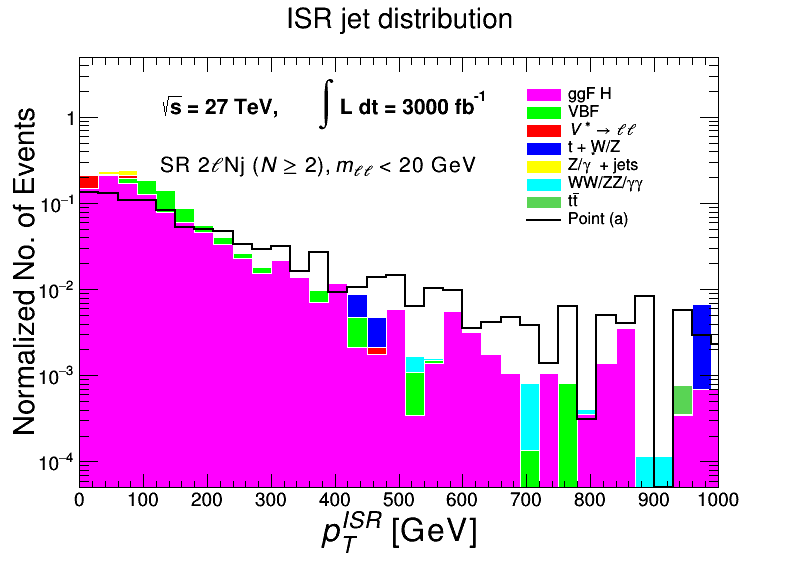}
 \caption{Top panels: an exhibition of the reconstructed dilepton invariant mass, $m_{\ell\ell}$, (left) for points (a)$-$(f) at 14 TeV and the distribution in MET (right) at 27 TeV for point (f). Bottom panels: an exhibition of the distributions in the ISR transverse momentum at 14 TeV (left) and 27 TeV (right) for benchmark (a).}
	\label{fig3}
\end{figure}

The dilepton invariant mass distribution at 14 TeV for the benchmarks (a)$-$(f) is shown in the upper left panel of Fig.~\ref{fig3}.  Here one finds that the distributions have a peak around 20 GeV for most points, consistent with the chargino (second neutralino)-LSP mass gap. The upper right panel shows the distribution in missing transverse energy for point (f) at 27 TeV for an integrated luminosity of 3000 fb$^{-1}$. In the bottom panels we show the distributions in the ISR jet transverse momentum for point (a) at 14 TeV (left) and at 27 TeV (right) both for an integrated luminosity of 3000 fb$^{-1}$.  Such distributions help design the selection criteria necessary to discriminate the signal from the SM backgrounds. The three distributions in MET and ISR jets are plotted after a selection cut on the dilepton invariant mass $m_{\ell\ell}$ where events with $m_{\ell\ell}>20$ GeV are rejected.  A cut around that value will remove most of the dominant backgrounds especially the $Z+$jets which has a peak around the $Z$ boson mass. A veto on b-tagged jets will reduce the $t\bar{t}$ background and further preselection criteria on MET will reduce the rest of the SM backgrounds. The dominant background remaining is  from dilepton production via off-shell vector bosons. More analysis cuts are required to reduce such a background. We summarize the preselection and selection criteria in Table~\ref{tab9}.

\begin{table}[H]
\begin{center}
\begin{tabular}{l|cc|cc}
\hline\hline\rule{0pt}{3ex}
\multirow{3}{*}{Observable} & SR-$2\ell$Nj-A & SR-$2\ell$Nj-B & SR-$2\ell$Nj-C & SR-$2\ell$Nj-D \\
\cline{2-5}
 & \multicolumn{2}{c|}{14 TeV} & \multicolumn{2}{c}{27 TeV} \\
 \cline{2-5}
 & \multicolumn{4}{c}{Preselection criteria} \\
\hline\rule{0pt}{3ex}
$N_{\ell}$ (SFOS) & \multicolumn{2}{c|}{$2$} & \multicolumn{2}{c}{$2$} \\
$N_{\rm jets}^{\rm non-b-tagged}$ & \multicolumn{2}{c|}{$\geq 2$} & \multicolumn{2}{c}{$\geq 2$} \\
$p^{\rm leading~jet}_T$ [GeV] & \multicolumn{2}{c|}{$>20$} & \multicolumn{2}{c}{$>20$} \\
$p_T^{\ell}$ (electron, muon) [GeV] & \multicolumn{2}{c|}{$>15$, $>10$} & \multicolumn{2}{c}{$>15$, $>10$} \\
$E^{\rm miss}_T$ [GeV] & \multicolumn{2}{c|}{$>90$} & \multicolumn{2}{c}{$>100$} \\
\cline{2-5}\rule{0pt}{3ex}
 & \multicolumn{4}{c}{Analysis cuts} \\
\cline{2-5}\rule{0pt}{3ex}
$p^{\rm ISR}_T$ [GeV] & $>100$ & $>100$ & $>120$ & $>120$ \\
$\Delta\phi(\vec{p}^{\,\rm miss}_T,Z)$ [rad] & $<1.2$ & $<1.2$ & $<1.2$ & $<1.2$ \\
$E^{\rm miss}_T/p^Z_T$ & $>12$ & $>15$ & $>12$ & $>25$ \\
$m_{\ell\ell}$ [GeV] & $<20$ & $<23$ & $<20$ & $<23$ \\
\hline
\end{tabular}
\end{center}
\caption{Preselection and analysis cuts (at 14 TeV and 27 TeV) applied to the signal and SM backgrounds for two signal regions targeting low and high electroweakino mass ranges.}
\label{tab9}
\end{table}

\subsection{Cut implementation and the estimated integrated luminosity}

Selection criteria are optimized per mass range and for each collider, i.e. for HL-LHC and HE-LHC. Starting with HL-LHC, the two signal regions we consider are SR $2\ell N$j-A and SR $2\ell N$j-B. They have the same preselection criteria but differ in terms of the analysis cuts on the variables $E^{\rm miss}_T$ and $m_{\ell\ell}$ as shown in Table~\ref{tab9}. Signal regions pertaining to HE-LHC are termed SR $2\ell N$j-C and SR $2\ell N$j-D and as HL-LHC, the only differences are in the same two variables mentioned before. For HE-LHC, harder cuts on $E^{\rm miss}_T$, $p^{\rm ISR}_T$ and $E^{\rm miss}_T/p^Z_T$ are applied. Another variable used in the analysis cuts is $\Delta\phi(\vec{p}^{\,\rm miss}_T,Z)$ which is the opening angle between the MET and $p^Z_T$ ensuring that no jets constructed from $W$ bosons fake the dilepton system. The variable $E^{\rm miss}_T/p^Z_T$ is a powerful discriminant since, unlike the backgrounds, the signal has the most MET and the softest of leptons so we expect the signal to have a larger value of this variable compared to the backgrounds. In order to design the optimal cuts on this variable, we plot the distributions in $E^{\rm miss}_T/p^Z_T$ for the lightest benchmark (a) and the heaviest (f) at 14 TeV and 27 TeV. The plots are shown in Fig.~\ref{fig4}.

\begin{figure}[H]
 \centering
 \includegraphics[width=0.49\textwidth]{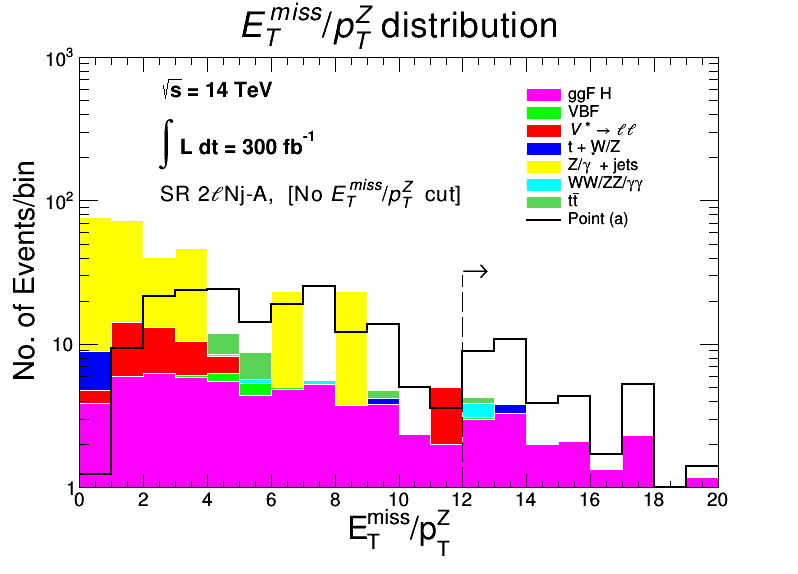}
 \includegraphics[width=0.49\textwidth]{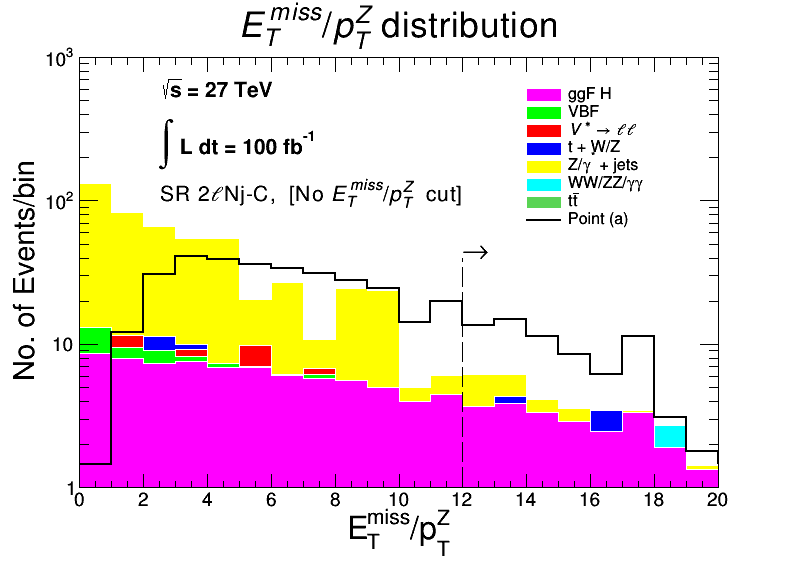} \\
 \includegraphics[width=0.49\textwidth]{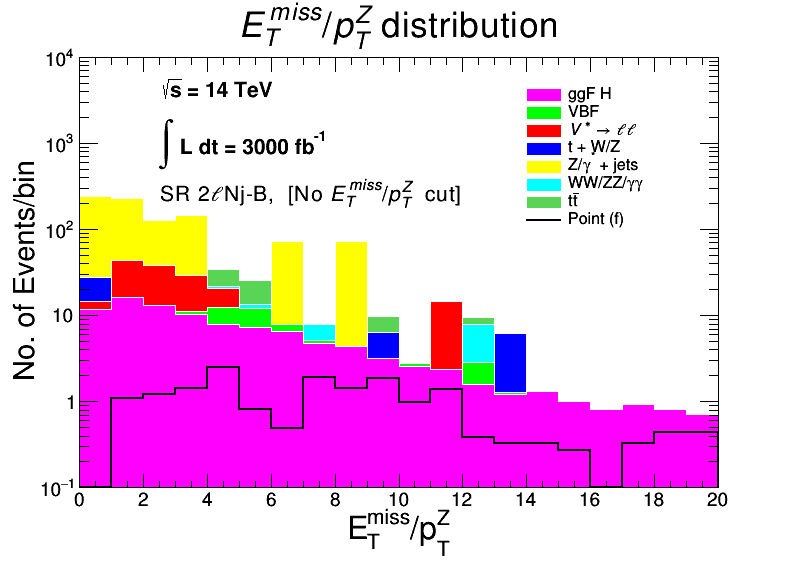}
 \includegraphics[width=0.49\textwidth]{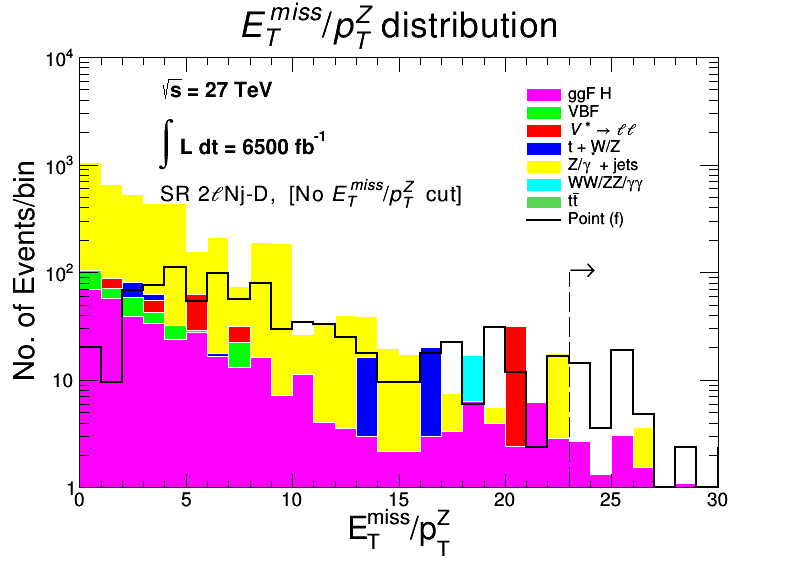}
\caption{An exhibition of distributions in $E^{\rm miss}_T/p^Z_T$ for benchmarks (a) and (f) at 14 TeV (left panels) and 27 TeV (right panels). The dashed line and arrow indicate the best cut on the variable.}
	\label{fig4}
\end{figure}

The top panel shows a comparison between HL-LHC (left) and HE-LHC (right) for point (a) where this benchmark can be visible at both colliders for 300 fb$^{-1}$ and 100 fb$^{-1}$ of integrated luminosity, respectively. The number of signal events in excess over the background are enough for a $5\sigma$ discovery if a cut on the variable $E^{\rm miss}_T/p^Z_T$ is made where the dashed line and arrow are located. This cut is shown in Table~\ref{tab9}. In contrast, point (f) cannot be discovered with $\mathcal{L}=3000$ fb$^{-1}$ at HL-LHC since the signal is completely below the background as seen in the bottom left panel of Fig.~\ref{fig4}. However, one can potentially discover this model point at HE-LHC with $\sim\mathcal{L}=6500$ fb$^{-1}$ at HE-LHC by applying a cut on $E^{\rm miss}_T/p^Z_T$ where the dashed line and arrow indicate. This cut is also shown in Table~\ref{tab9}. The estimated integrated luminosities for discovery of benchmarks (a)$-$(f) are shown in Fig.~\ref{fig5}.

\begin{figure}[H]
 \centering
 \includegraphics[width=0.8\textwidth]{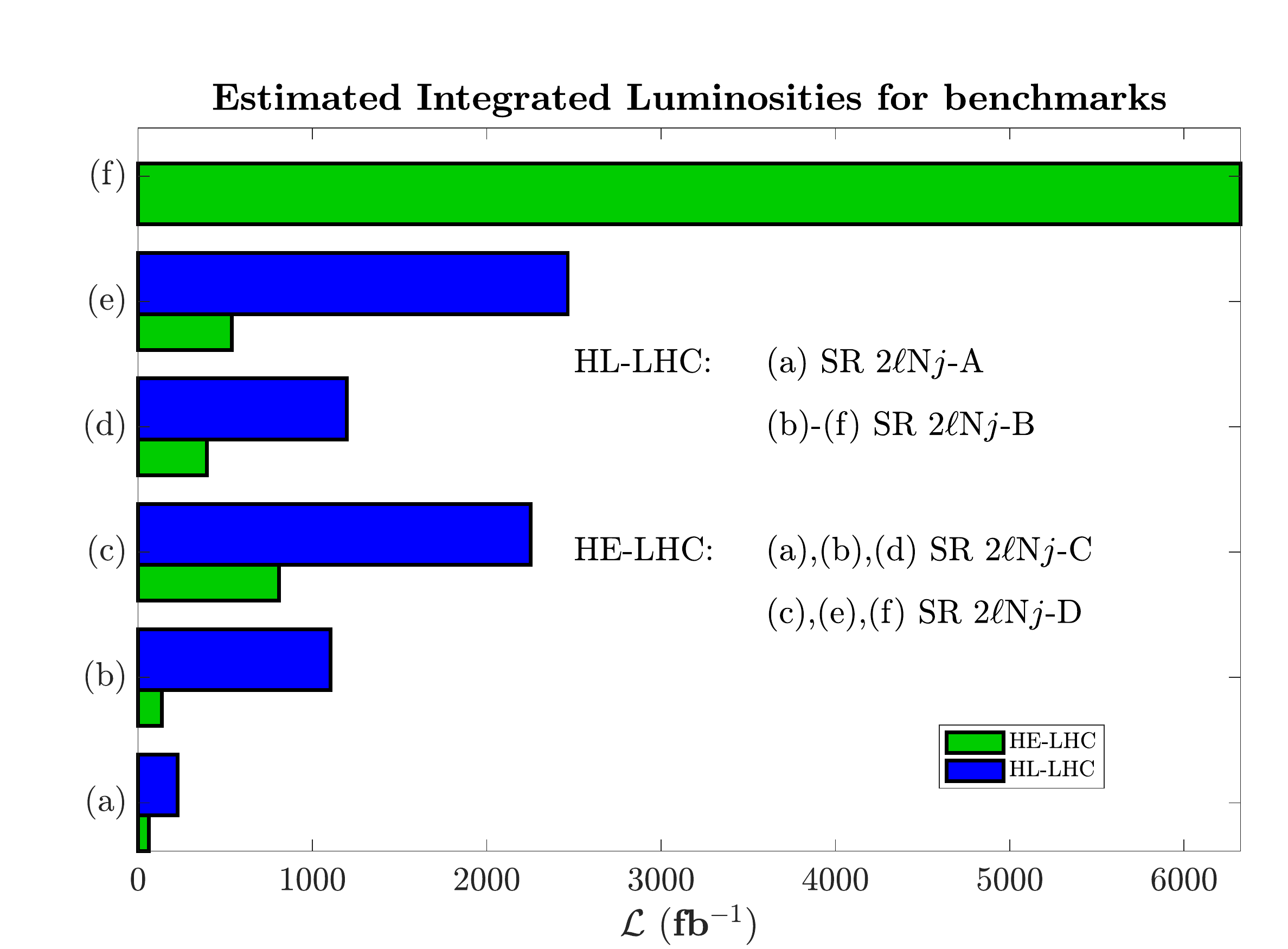}
 \caption{Estimated integrated luminosity for discovery of benchmarks (a)$-$(f) at HL-LHC and HE-LHC. All points except (f) are visible at HL-LHC while all points are discoverable at HE-LHC. }
	\label{fig5}
\end{figure}

The signal regions that give the optimal results for each of the benchmarks are shown in the plot per each collider. Starting with the lightest model point (a), a discovery at 14 TeV can be made with only $\sim 226$ fb$^{-1}$ which should be attainable in the upcoming Run 3. A much smaller integrated luminosity of $\sim 62$ fb$^{-1}$ is needed for discovery at HE-LHC. Model point (b) requires much more $\mathcal{L}$, around 1100 fb$^{-1}$ at HL-LHC while only 135 fb$^{-1}$ is needed at HE-LHC. Points (c), (d) and (e) require an integrated luminosity ranging between $\sim 1200$ fb$^{-1}$ to $\sim 2500$ fb$^{-1}$ for HL-LHC and $\sim 390$ fb$^{-1}$ to $\sim 800$ fb$^{-1}$ for HE-LHC. Point (f) which is only discoverable at HE-LHC require $\sim 6300$ fb$^{-1}$. Note that despite being heavier than point (c), point (d) requires less integrated luminosity for discovery. The reason is that point (c) has a small branching ratio to dileptons (see Table~\ref{tab8}) and so the overall cross-section to the required final states is smaller. Note also that the branching ratio to leptons for point (f) is very small ($7\%$) compounded with the fact that it is the heaviest makes it very difficult to detect and that is why even at HE-LHC, which could potentially collect around 15 ab$^{-1}$ of data~\cite{CidVidal:2018eel,Cepeda:2019klc}, the required integrated luminosity is large.

\section{Conclusion\label{sec5}}

In this work we consider a class of $\mathsf{SO(10)}$  models which lead to a  pair of light Higgs doublets without the necessity of a fine tuning
needed in generic grand unified models. In this class  we consider a model with $\mathsf{126+\overline{126}+210}$ of heavy Higgs
and a $2\times \mathsf{10 + 120}$ of light Higgs.
The focus of this work is to show that significant contributions from the higher dimensional operators to the Yukawa couplings arise from
matter-matter-Higgs-Higgs interactions in the superpotential
where one of the Higgs fields is light and the other heavy, even though the interactions are suppressed by a heavy mass.
This occurs because the  heavy field, after spontaneous symmetry breaking of the GUT symmetry,
develops  a VEV which is order the
GUT scale which  overcomes the suppression of the higher dimensional operator by the heavy mass.
 In this work we focused on computing  the corrections to the third generation Yukawas using quartic couplings of Eqs.~(\ref{w41})$-$(\ref{w43}).
 The analysis shows that the contribution of the quartic terms to the Yukawas can produce
substantial corrections to the GUT boundary conditions for the Yukawas.
The RG evolution using the modified boundary conditions shows
that a consistency with the third generation quarks and the charged leptons masses
can be achieved even with a low value of $\tan\beta$, i.e., a $\tan\beta$ as low as 5$-$10 consistent with gauge coupling unification.
 The sparticle spectrum for the models considered was investigated and it
 is found that the relic density as an upper limit constraint can be satisfied in three coannihilation regions
that arise in the models investigated, i.e., coannihilations involving chargino-neutralino, stau-neutralino,
and stop-neutralino.
 Further,  LHC implications for some of the chargino-neutralino coannihilation models  was
 carried out for the possibility of SUSY discovery via the detection of a light chargino at HL-LHC and at
 a possible future collider HE-LHC at 27 TeV. It is shown that most of the models
 investigated can be discovered at HL-LHC using up to its optimal integrated luminosity while
 all of the models are discoverable at HE-LHC with a significantly smaller integrated luminosity
 and on a much shorter time scale.
 Discovery of a chargino, a stau or a stop which appear as the lightest sparticles in the analysis
along with a determination of  $\tan\beta$ which indicates a low value for it would lend support to
this class of unified models. We note in passing that in the models of the type discussed the LSP
can both saturate the relic density or be only a fraction of it. This implies that
 dark matter could be either a  one component WIMP (neutralino) dark matter, or a
multicomponent  one where the WIMPs comprise only a fraction and the rest arises
from other sources such as axions or matter from the dark sector.\\

\textbf{Acknowledgments:} The analysis presented here was done using the resources of the
high-performance Cluster353 at the Advanced Scientific Computing Initiative (ASCI) and
the Discovery Cluster at Northeastern University. The research of AA was supported by the BMBF under contract 05H18PMCC1, while the research of PN was supported in part by the NSF Grant PHY-1913328.

\begin{appendices}

\section{Notation and decomposition of {\boldmath$\mathsf{SU(5)}$} fields in terms of {\boldmath$\mathsf{SU(3)_C}\times \mathsf{SU(2)_L}\times\mathsf{U(1)_Y}$}\label{appendix A}}
$\mathsf{SO(10)}$ spinor and Higgs fields of our model in terms of $\mathsf{SU(5)}$ fields are
\begin{eqnarray}
\mathsf{16}\left[\Psi_{(+)}\right]&=&\mathsf{1}({-5}) \left[\mathsf{M}\right] + \overline {\mathsf{5}}(3) \left[\mathsf{M}_{i}\right] + \mathsf{10} (-1) \left[\mathsf{M}^{ij}\right],\nonumber\\
\mathsf{10_r}\left[{^r}\Omega_{\mu}\right]&=&\mathsf{5}(2)\left[\mathsf{H}^{(10_r)i}\right] + \overline{\mathsf{5}}(-2)\left[\mathsf{H}^{(10_r)}_i\right],\nonumber\\
\mathsf{120}\left[\Sigma_{\mu\nu\rho}\right]&=&\mathsf{5}(2)\left[\mathsf{H}^{(120)i}\right] + \overline{\mathsf{5}}(-2)\left[\mathsf{H}^{(120)}_i\right]+\mathsf{10}(-6) + \overline{\mathsf{10}}(6)] +\mathsf{45}(2)\left[\mathsf{H}^{(120)ij}_k\right]+\overline{\mathsf{45}}(-2)\left[\mathsf{H}^{(120)k}_{ij}\right],\nonumber\\
\overline{\mathsf{126}}\left[\Delta_{\mu\nu\rho\sigma\lambda}\right]&=&\mathsf{1}(10)\left[\mathsf{H}^{(\overline{126})}\right] + \mathsf{5}(2)\left[\mathsf{H}^{(\overline{126})i}\right]+ \overline{\mathsf{10}}(6) +\mathsf{15}(-6)+\overline{\mathsf{45}}(-2)\left[\mathsf{H}^{(\overline{126})k}_{ij}\right]+ \mathsf{50}(2),\nonumber\\
\mathsf{210}\left[\Phi_{\mu\nu\rho\sigma}\right]&=&\mathsf{1}(0)\left[\mathsf{H}^{(210)}\right]+\mathsf{5}(-8)\left[\mathsf{H}^{(210)i}\right] + \overline{\mathsf{5}}(8)\left[\mathsf{H}^{(210)}_i\right]+\mathsf{10}(4) + \overline{\mathsf{10}}(-4)+\mathsf{24}(0)\left[\mathsf{H}^{(120)i}_j\right]\nonumber\\
&&+\mathsf{40}(-4)+\overline{\mathsf{40}}(4)
+\mathsf{75}(0)\left[\mathsf{H}^{(210)ij}_{kl}\right],
\end{eqnarray}
where  $\mu,~\nu,~\rho,~\sigma,~\lambda=1,...,10$ and $i,~j,~k,~l,~m,~n=1,...,5$ are $\mathsf{SO(10)}$ and  $\mathsf{SU(5)}$ indices and $r,~s=1,~2$ {count}
the number of $\mathsf{10}$ plet of $\mathsf{SO(10)}$.
The identification of  $ \mathsf{SU(2)}$  doublets, denoted by $\mathcal D$'s, contained in $\mathsf{5}$, $\mathsf{\overline{5}}$, $\mathsf{45}$, $\mathsf{\overline{45}}$ of  $\mathsf{SU(5)}$ are done through
\begin{eqnarray}\label{Extraction of doublets}
\mathsf{H}^{(\#)a}\equiv {}^{({5}_{\#})}\!{{\mathcal D}}^{a};&& \mathsf{H}^{(\#)}_a\equiv {}^{(\overline{5}_{\#})}\!{{\mathcal D}}_{a},\nonumber\\
\mathsf{H}^{(\#)\alpha a}_\beta=\frac{1}{3}\delta^{\alpha}_{\beta}~{}^{({45}_{\#})}\!{{\mathcal D}}^{a};&&\mathsf{H}^{(\#)ab }_c=\delta^{b}_{c}~{}^{({45}_{\#})}\!{{\mathcal D}}^{a}-\delta^{a}_{c}~{}^{({45}_{\#})}\!{{\mathcal D}}^{b},\nonumber\\
\mathsf{H}^{(\#)\beta}_{\alpha a}=\frac{1}{3}\delta^{\beta}_{\alpha}~{}^{({\overline{45}}_{\#})}\!{{\mathcal D}}_a;&&\mathsf{H}_{ab }^{(\#)c}=\delta^{c}_{b}~{}^{({\overline{45}}_{\#})}\!{{\mathcal D}}_{a}-\delta^{c}_{a}~{}^{({\overline{45}}_{\#})}\!{{\mathcal D}}_{b},
\end{eqnarray}
where $\alpha,~ \beta, ~\gamma=1,~2,~3$ are $\mathsf{SU(3)}$ color indices, while $a,~b=4,~5$ are $\mathsf{SU(2)}$ weak indices and $\#$ refers to the $\mathsf{10_r}, ~\mathsf{120},~ \mathsf{\mathsf{126}},~ \mathsf{210}$  fields of $\mathsf{SO(10)}$.  Note, however, that $\mathcal{D}$'s are un-normalized. To normalize the $\mathsf{SU(2)}$ doublets contained in the $\mathsf{SO(10)}$ tensors, we carry out the following field redefinition:
\begin{eqnarray}\label{Normalized doublets}
{}^{({5}_{{10_r}})}\!{\mathcal D}^{a}=\sqrt{2}~{}^{({5}_{{10_r}})}\!{\mathsf D}^{a};&&{}^{(\overline{5}_{{10_r}})}\!{\mathcal D}_{a}=\sqrt{2}~{}^{(\overline{5}_{{10_r}})}\!{\mathsf D}_{a},\nonumber\\
{}^{({5}_{{120}})}\!{\mathcal D}^{a}=\frac{4}{\sqrt{3}}~{}^{({5}_{{120}})}\!{\mathsf D}^{a};&&{}^{(\overline{5}_{{120}})}\!{\mathcal D}_{a}=\frac{4}{\sqrt{3}}~{}^{(\overline{5}_{{120}})}\!{\mathsf D}_{a},\nonumber\\
{}^{({5}_{{\overline{126}}})}\!{\mathcal D}^{a}=4\sqrt{\frac{2}{5}}~{}^{({5}_{{\overline{126}}})}\!{\mathsf D}^{a};&&{}^{(\overline{5}_{{126}})}\!{\mathcal D}_{a}=4\sqrt{\frac{2}{5}}~{}^{(\overline{5}_{{126}})}\!{\mathsf D}_{a},\nonumber\\
{}^{({5}_{{210}})}\!{\mathcal D}^{a}=8\sqrt{6}~{}^{({5}_{{210}})}\!{\mathsf D}^{a};&&{}^{(\overline{5}_{{210}})}\!{\mathcal D}_{a}=8\sqrt{6}~{}^{(\overline{5}_{{210}})}\!{\mathsf D}_{a},\nonumber\\
{}^{({45}_{{120}})}\!{\mathcal D}^{a}=\frac{1}{\sqrt{2}}~{}^{({45}_{{120}})}\!{\mathsf D}^{a};&&{}^{(\overline{45}_{{120}})}\!{\mathcal D}_{a}=\frac{1}{\sqrt{2}}~{}^{(\overline{45}_{{120}})}\!{\mathsf D}_{a},\nonumber\\
{}^{({45}_{{126}})}\!{\mathcal D}^{a}=\frac{1}{\sqrt{5}}~{}^{({45}_{{126}})}\!{\mathsf D}^{a};&&{}^{(\overline{45}_{\overline{126}})}\!{\mathcal D}_{a}=\frac{1}{\sqrt{5}}~{}^{(\overline{45}_{\overline{126}})}\!{\mathsf D}_{a},
\end{eqnarray}
where $\mathsf D$'s represent the doublet fields with canonically normalized kinetic energy terms.

The identification of various $\mathsf{SU(3)_C}\times \mathsf{SU(2)_L}\times\mathsf{U(1)_Y}$ singlets   are done through
 \begin{eqnarray}
\mathsf{H}^{(\overline{126})}=\mathcal S_{1_{_{\overline{126}}}};&&\mathsf{H}^{(210)} =\mathcal S_{1_{_{{210}}}},\nonumber\\
\mathsf{H}^{(210)a}_b=-\frac{1}{2}\delta^{a}_{b}~\mathcal S_{24_{_{{210}}}};&&\mathsf{H}^{(210)\alpha}_{\beta} =\frac{1}{3}\delta^{\alpha}_{\beta}~\mathcal S_{24_{_{{210}}}},\nonumber\\
\mathsf{H}^{(210)ab}_{cd}=\frac{1}{2}\left(\delta^{a}_{c}\delta^{b}_{d}-\delta^{a}_{d}\delta^{b}_{c}\right)~\mathcal S_{75_{_{{210}}}};&&\mathsf{H}^{(210)\alpha\beta}_{\gamma\tau} =\frac{1}{6}\left(\delta^{\alpha}_{\gamma}\delta^{\beta}_{\tau}-\delta^{\alpha}_{\tau}\delta^{\beta}_{\gamma}\right)~\mathcal S_{75_{_{{210}}}},\nonumber\\
\mathsf{H}^{(210)\alpha a}_{\beta b} =-\frac{1}{6}\delta^{a}_{b}\delta^{\alpha}_{\gamma}~\mathcal S_{75_{_{{210}}}}.
\end{eqnarray}
$\mathcal {S}$'s are un-normalized fields. To normalize them, we carry out the following field redefinition:
\begin{eqnarray}\label{Normalized singlets}
\mathcal S_{1_{_{{210}}}}=4\sqrt{\frac{5}{3}}~\mathbf S_{1_{210}};&&\mathcal S_{24_{_{{210}}}}=2\sqrt{\frac{3}{5}}~\mathbf S_{24_{210}},\nonumber\\
\mathcal S_{75_{_{{210}}}}=\frac{1}{\sqrt{3}}~\mathbf S_{75_{210}};&&\mathcal S_{1_{_{{126}}}}=\frac{2}{\sqrt{15}}~\mathbf S_{1_{126}},\nonumber\\
\mathcal S_{1_{_{\overline{126}}}}=\frac{2}{\sqrt{15}}~\mathbf S_{1_{\overline{126}}}.
\end{eqnarray}
We will denote the VEVs of the normalized fields by $\mathcal {V}$'s, so that
\begin{eqnarray}\label{VEVS}
\mathcal V_{1_{210}}\equiv\langle\mathbf S_{1_{210}}\rangle;~~\mathcal V_{24_{210}}\equiv\langle\mathbf S_{24_{210}}\rangle;~~\mathcal V_{75_{210}}\equiv\langle\mathbf S_{75_{210}}\rangle;~~\mathcal V_{1_{{126}}}\equiv \langle\mathbf S_{1_{126}}\rangle;~~\mathcal V_{1_{\overline{126}}}\equiv\langle\mathbf S_{1_{\overline{126}}}\rangle.
\end{eqnarray}

The $\mathsf{SU(5)}$ matter fields are
\begin{eqnarray}
\mathsf{{M}}={\bm\nu}^{\mathtt c};~~~~~~
\mathsf{{M}}_{\alpha}={\mathbf D}_{\alpha}^{\mathtt c};~~~~~~\mathsf{{M}}_{a}=\mathbf{{L}}_{a}={{{\mathbf E}}\choose {-{\bm\nu}}};\nonumber\\
\mathsf{{M}}^{a\alpha}=\mathbf{{Q}}^{a\alpha}={{{\mathbf U}}^{\alpha}\choose {{\mathbf D}}^{\alpha}};~~~~~~\mathsf{{M}}^{\alpha\beta}=\epsilon^{\alpha\beta\gamma}{\mathbf
U}_{\gamma}^{\mathtt c};~~~~~~{\mathsf M}^{ab}=\epsilon^{ab}{\mathbf E}^{\mathtt c}.
\end{eqnarray}
\section{Breaking the $\mathsf{SO(10)}$ gauge symmetry \label{Appendix B}}
In breaking the  GUT symmetry to the symmetry of the standard model gauge group the fields that enter are   $\mathcal S_{1_{_{{126}}}},~ \mathcal S_{1_{_{\overline{126}}}},~\mathcal S_{1_{_{{210}}}},~\mathcal S_{24_{_{{210}}}},~\mathcal S_{75_{_{{210}}}}$. Retaining these fields, Eq.~(\ref{gut-1}) takes the form
\begin{eqnarray}
W_{\textsc{gut}}&
=&{M^{126}}\left(\frac{15}{2}\mathcal S_{1_{_{{126}}}}                  \mathcal S_{1_{_{\overline{126}}}}     +\cdots\right)
+{M^{210}}\left(\frac{3}{4}\mathcal S_{75_{_{{210}}}}                  ^2+\frac{5}{12}\mathcal S_{24_{_{{210}}}}                  ^2+\frac{3}{80}\mathcal S_{1_{_{{210}}}}                  ^2+\cdots\right)\nonumber\\
&&+\eta\left(-\frac{3}{16}\mathcal S_{1_{_{{210}}}}                  \mathcal S_{1_{_{{126}}}}                  \mathcal S_{1_{_{\overline{126}}}}     +\cdots\right)+ \lambda\left({\frac{1}{18}}\mathcal S_{75_{_{{210}}}}                  ^3-{\frac{1}{18}}\mathcal S_{75_{_{{210}}}}                  ^2\mathcal S_{24_{_{{210}}}}                  +\frac{25}{864}\mathcal S_{75_{_{{210}}}}                  \mathcal S_{24_{_{{210}}}}                  ^2\right.\nonumber\\
  &&\left.+{\frac{3}{160}}\mathcal S_{75_{_{{210}}}}                  ^2\mathcal S_{1_{_{{210}}}}                  -\frac{35}{3888}\mathcal S_{24_{_{{210}}}}                  ^3-\frac{1}{192}\mathcal S_{24_{_{{210}}}}                  ^2\mathcal S_{1_{_{{210}}}}                  -\frac{3}{3200}\mathcal S_{1_{_{{210}}}}                  ^3+\cdots\right).
  \label{gut-2}
  \end{eqnarray}
 As noted above,  $\mathcal S_{1_{_{{126}}}},~ \mathcal S_{1_{_{\overline{126}}}},~\mathcal S_{1_{_{{210}}}},~\mathcal S_{24_{_{{210}}}},~\mathcal S_{75_{_{{210}}}}$  are the Standard Model singlets that acquire VEVs  after normalization through Eq. (\ref{Normalized singlets}). Eq. (\ref{gut-2}) corrects the coefficient of the term $\mathcal S_{75_{_{{210}}}}^2\mathcal S_{1_{_{{210}}}} $ that
appears in~\cite{Nath:2015kaa} by a factor of $3/4$ in agreement with the analysis of~\cite{Chen:2017quw}.
 The spontaneous symmetry breaking equations including this factor are
Eqs. (\ref{gut-3})$-$(\ref{gut-6}).
Vanishing of the $F$-terms leads to the immediate determination of $\mathcal V_{1_{_{{210}}}}$ and a cubic equation in $\mathcal V_{24_{_{{210}}}}$ through
\begin{eqnarray}
\label{gut-3}
\mathcal V_{1_{_{{210}}}}=2\sqrt{15}{\cal{M}}^{126},~~~~~~~~~~~~~~~~~~~~~~\\
\sqrt{15}\mathcal V_{24_{_{{210}}}}^3+{60}\mathcal V_{24_{_{{210}}}}^2\left(3{\cal{M}}^{126}-{{}5}{\cal M}^{210}\right)+ 240\sqrt{15}\mathcal V_{24_{_{{210}}}}                  \left[{{}{3}}({\cal{M}}^{126})^2-{{}5}{\cal{M}}^{126}{\cal M}^{210}+{{}{12}}({\cal M}^{210})^2\right]\nonumber\\
+{14400}\left({\cal{M}}^{126}
-2{\cal M}^{210}\right)\left({\cal{M}}^{126}+{{}}{\cal M}^{210}\right)^2=0,~~~~
\label{gut-4}
\end{eqnarray}
where the $\mathcal V$'s  are defined in Eq. (\ref{VEVS}) and
 $\displaystyle{\cal M}^{126}\equiv\frac{M^{126}}{\eta}, ~ {\cal M}^{210}\equiv\frac{M^{210}}{\lambda}$.
[For an early work on the appearance of a cubic equation in spontaneous breaking of the GUT symmetry see~\cite{Aulakh:2003kg}].
The remaining $\mathsf{SU(3)_C\times SU(2)_L\times U(1)_Y}$ singlet fields are functions of $\mathcal V_{24_{_{{210}}}}$  and are determined by
\begin{eqnarray}
\label{gut-5}
\mathcal V_{75_{_{{210}}}}&=&\frac{\sqrt{15}\mathcal V_{24_{210}}^2+60\mathcal V_{24_{210}}\left({\cal{M}}^{126}-2{\cal M}^{210}\right)}{\sqrt{3}\mathcal V_{24_{_{210}}}+12\sqrt{5}\left({\cal{M}}^{126}+{\cal M}^{210}\right)}, \\
\mathcal V_{1_{_{{126}}}}\cdot\mathcal V_{1_{_{\overline{126}}}}&=&\frac{3}{32\sqrt{5}}\left(\frac{\lambda}{\eta}\right)
\frac{1}{{\sqrt{3}\mathcal V_{24_{_{{210}}}}+{{}12\sqrt{5}}\left({\cal{M}}^{126}+{\cal M}^{210}\right)}}
\left\{{{}}\sqrt{15}\mathcal V_{24_{_{{210}}}}^3
-{{}120}\mathcal V_{24_{_{{210}}}}^2\left({{}}{\cal{M}}^{126}+{{}9}{\cal M}^{210}\right)\right.\nonumber\\
&&\left.-{{}80\sqrt{15}}\mathcal V_{24_{_{{210}}}}                  \displaystyle\left[{{}21}({\cal{M}}^{126})^2-{{}17}{\cal{M}}^{126}{\cal M}^{210}-{{}18}({\cal M}^{210})^2\right]\right.\nonumber\\
&&\left.
-{{}19200}{\cal{M}}^{126}\left(3{\cal{M}}^{126}-2{\cal M}^{210}\right)\left({{}}{\cal{M}}^{126}+{{}}{\cal M}^{210}\right)\displaystyle\right\}.
\label{gut-6}
\end{eqnarray}
Finally, setting $D$-terms to zero yields
\begin{eqnarray}
\mathcal V_{1_{_{{126}}}}=\mathcal V_{1_{_{\overline{126}}}}.
\label{gut-7}
\end{eqnarray}


\section{Higgs doublet mass matrix \label{appendix C}}
A computation of the $7\times 7$ doublet mass matrix was given in \cite{Nath:2015kaa}.
Here we record the matrix for completeness using the constraint of Eq.~(\ref{gut-7}).
We have
{\small
\begin{eqnarray}\label{doublet mass matrix}
M_d~~=~~
\kbordermatrix{
&{}^{(\overline{5}_{10_1})}\!{\mathsf D}_{a} &
{}^{(\overline{5}_{10_2})}\!{\mathsf D}_{a} &
{}^{(\overline{5}_{120})}\!{\mathsf D}_{a}&
 {}^{(\overline{5}_{{126}})}\!{\mathsf D}_{a} &
 {}^{(\overline{5}_{{210}})}\!{\mathsf D}_{a} &
 {}^{(\overline{45}_{120})}\!{\mathsf D}_{a} &
{}^{(\overline{45}_{\overline{126}})}\!{\mathsf D}_{a}\cr\cr
{}^{({5}_{10_1})}\!{\mathsf D}^{a} & 0& 0& 0&\displaystyle\left(\frac{b_1}{a}\right)\mathsf{d_2} &\displaystyle\left(\frac{b_1}{a}\right)\mathsf{d_1} &0 &\displaystyle\mathsf{d_3}\cr\cr
{}^{({5}_{10_2})}\!{\mathsf D}^{a} &0& 0&0 & \displaystyle\left(\frac{b_2}{a}\right)\mathsf{d_2}& \displaystyle\left(\frac{b_2}{a}\right)\mathsf{d_1}& 0&0 \cr\cr
{}^{({5}_{120})}\!{\mathsf D}^{a} & 0& 0& 0&\displaystyle\mathsf{d_5} &\displaystyle\mathsf{d_4} & 0&\displaystyle\left(\frac{\overline{c}}{c}\right)\mathsf{d_6} \cr\cr
{}^{({5}_{\overline{126}})}\!{\mathsf D}^{a} & \mathsf{d_2}&  0&\displaystyle\left(\frac{\overline{c}}{c}\right)\mathsf{d_5}& {\mathsf{d_9}}& {\mathsf{d_{11}}}&\displaystyle\left(\frac{\overline{c}}{c}\right)\mathsf{d_7} &0\cr\cr
{}^{({5}_{{210}})}\!{\mathsf D}^{a}& \displaystyle\mathsf{d_1}& 0&\displaystyle\left(\frac{\overline{c}}{c}\right)\mathsf{d_4}& {\displaystyle\mathsf{d_{11}}}& {\displaystyle\mathsf{d_{10}}}&0 &0\cr\cr
{}^{({45}_{120})}\!{\mathsf D}^{a} & 0& 0& 0&\displaystyle\mathsf{d_7} &0 & 0&\displaystyle\left(\frac{\overline{c}}{c}\right)\displaystyle\mathsf{d_8} \cr\cr
{}^{({45}_{126})}\!{\mathsf D}^{a}& \displaystyle\left(\frac{b_1}{a}\right)\mathsf{d_3}& \displaystyle\left(\frac{b_2}{a}\right)\mathsf{d_3}&\displaystyle \mathsf{d_6}& 0 &0 &\displaystyle\mathsf{d_8}&{\displaystyle\mathsf{d_{12}}}\cr\cr
},
\end{eqnarray}
}
\begin{eqnarray}\label{elements of doublet mass matrix}
\mathsf{d_1}&\equiv&\frac{a}{2\sqrt{5}}\mathcal V_{1_{_{{126}}}}                  ,\nonumber\\
\mathsf{d_2}&\equiv&-a\left[\frac{\sqrt{3}}{10}\mathcal V_{1_{_{{210}}}}                  +\frac{\sqrt{3}}{20}\mathcal V_{24_{_{{210}}}}                  \right],\nonumber\\
\mathsf{d_3}&\equiv&a\left[-\frac{1}{4\sqrt{6}}\mathcal V_{24_{_{{210}}}}                  +\frac{1}{4\sqrt{15}}\mathcal V_{75_{_{{210}}}}                  \right],\nonumber\\
{\mathsf{d_4}}&\equiv&{-\frac{c}{\sqrt{30}}}\mathcal V_{1_{_{{126}}}}                  ,\nonumber\\
\mathsf{d_5}&\equiv&c\left[-\frac{1}{10\sqrt{2}}\mathcal V_{1_{_{{210}}}}                  +\frac{3}{40\sqrt{2}}\mathcal V_{24_{_{{210}}}}                  \right],\nonumber\\
{\mathsf{d_6}}&\equiv&-c\left[{\frac{1}{48}}\mathcal V_{24_{_{{210}}}}                  +\frac{1}{12\sqrt{10}}\mathcal V_{75_{_{{210}}}}                  \right],\nonumber\\
\mathsf{d_7}&\equiv&c\left[\frac{1}{48\sqrt{3}}\mathcal V_{24_{_{{210}}}}                  +\frac{1}{12\sqrt{30}}\mathcal V_{75_{_{{210}}}}                  \right],\nonumber\\
\mathsf{d_8}&\equiv&-c\left[\frac{1}{20\sqrt{6}}\mathcal V_{1_{_{{210}}}}                  +\frac{1}{240\sqrt{6}}\mathcal V_{24_{_{{210}}}}                  +\frac{1}{12\sqrt{15}}\mathcal V_{75_{_{{210}}}} \right],\nonumber\\
\nonumber
{\mathsf{d_9}}&{\equiv}&{{2}M^{126}-\eta\left[{\frac{2}{5\sqrt{15}}}\mathcal V_{1_{_{{210}}}}+{\frac{3}{{20}}\sqrt{\frac{3}{5}}}\mathcal V_{24_{_{{210}}}}                  \right]},\nonumber\\
{\mathsf{d_{10}}}&{\equiv}&{2M^{210}-\lambda\left[{\frac{3}{10\sqrt{2}}}\mathcal V_{1_{_{{210}}}}+\frac{1}{2}\sqrt{\frac{3}{5}}\mathcal V_{24_{_{{210}}}}                  \right]},
\nonumber\\
{\mathsf{d_{11}}}&{\equiv}&{\frac{1}{5}\eta\mathcal V_{1_{_{{126}}}}},\nonumber\\
{\mathsf{d_{12}}}&{\equiv}&{M^{126}+\eta\left[-{\frac{1}{6\sqrt{15}}}\mathcal V_{24_{_{{210}}}} +{\frac{1}{15\sqrt{6}}}\mathcal V_{75_{_{{210}}}} \right]}.
\end{eqnarray}

\section{Details of  Yukawa couplings from quartic interactions\label{appendix D}}

In this appendix we give details of the computations of the Yukawa couplings arising from the quartic interactions of
$W_4^{(1)}, ~W_4^{(2)}$ and $W_4^{(3)}$ of the superpotential given in Eqs. (\ref{w41})$-$(\ref{w43}).
  We discuss their contribution in that order.

\begin{enumerate}
\item
We compute contriubtions from Eq. (\ref{w41}) first.
\begin{eqnarray}
W^{(1)}_4&=&\frac{f^{(1)}}{5!M_{c}}b_r\langle\Psi_{(+)}^*|B \Gamma_{[\lambda}\Gamma_\mu\Gamma_{\nu}\Gamma_{\rho}\Gamma_{\sigma]}|\Psi_{(+)}\rangle~
\left[-{^r}\Omega_{\lambda}\Phi_{\mu\nu\rho\sigma}+{^r}\Omega_{\mu}\Phi_{\lambda\nu\rho\sigma}
-{^r}\Omega_{\nu}\Phi_{\lambda\mu\rho\sigma}\right.\nonumber\\
&&\left.~~~~~~~~~~~~~~~~~~~~~~~~~~~~~~~~~~~~~~~~~~~~~~~~~+{^r}\Omega_{\rho}\Phi_{\lambda\mu\nu\sigma}
-{^r}\Omega_{\sigma}\Phi_{\lambda\mu\nu\rho}\right]
\nonumber\\
&=&-i\frac{f^{(1)}}{5!M_{c}}b_r\left[-15\epsilon_{ijklm}\mathsf{M}^{ij}\mathsf{M}^{np}\mathsf{H}^{(10_r)k}\mathsf{H}^{(210)lm}_{np}
-\frac{15}{2}\epsilon_{ijklm}\mathsf{M}^{ij}\mathsf{M}^{kn}\mathsf{H}^{(10_r)p}\mathsf{H}^{(210)lm}_{np}\right.\nonumber\\
&&\left.~~~~~~~~~~~~~~~+\frac{5}{2}\epsilon_{ijklm}\mathsf{M}^{ij}\mathsf{M}^{kl}\mathsf{H}^{(10_r)n}\mathsf{H}^{(210)m}_{n}
-\frac{3}{4}\epsilon_{ijklm}\mathsf{M}^{ij}\mathsf{M}^{kl}\mathsf{H}^{(10_r)m}\mathsf{H}^{(210)}\right.\nonumber\\
&&\left.~~~~~~~~~~~~~~~-60\mathsf{M}^{ij}\mathsf{M}_{k}\mathsf{H}^{(10_r)}_l\mathsf{H}^{(210)kl}_{ij}
-20\mathsf{M}^{ij}\mathsf{M}_{i}\mathsf{H}^{(10_r)}_k\mathsf{H}^{(210)k}_{j}\right.\nonumber\\
&&\left.~~~~~~~~~~~~~~~+80\mathsf{M}^{ij}\mathsf{M}_{k}\mathsf{H}^{(10_r)}_i\mathsf{H}^{(210)k}_{j}+\cdots\right].
\end{eqnarray}

\vskip 0.5cm
\begin{enumerate}
\item $\displaystyle-15\epsilon_{ijklm}\mathsf{M}^{ij}\mathsf{M}^{np}\mathsf{H}^{(10_r)k}\mathsf{H}^{(210)lm}_{np}=
    -60\epsilon_{\gamma\sigma\beta ac}\mathsf{M}^{\gamma\sigma}\mathsf{M}^{b\alpha}\mathsf{H}^{(10_r)a}\mathsf{H}^{(210)c\beta}_{b\alpha}\\
    ~~~~~~~~~~~~~~~~~~~~~~~~~~~~~~~~~~~~~~~~~~~~~~~~+15\epsilon_{ba\rho\gamma\sigma}\mathsf{M}^{\alpha\beta}\mathsf{M}^{b\rho}\mathsf{H}^{(10_r)a}
    \mathsf{H}^{(210)\gamma\sigma}_{\alpha\beta}+\cdots$
\begin{enumerate}
\item $\displaystyle-60\epsilon_{\gamma\sigma\beta ac}\mathsf{M}^{\gamma\sigma}\mathsf{M}^{b\alpha}\mathsf{H}^{(10_r)a}\mathsf{H}^{(210)c\beta}_{b\alpha}=
20\mathcal S_{75_{_{{210}}}}\epsilon_{ab}{\mathbf U}_{\alpha}^{\mathtt c}\mathbf{{Q}}^{b\alpha}\mathsf{H}^{(10_r)a}$
\item $\displaystyle15\epsilon_{ba\rho\gamma\sigma}\mathsf{M}^{\alpha\beta}\mathsf{M}^{b\rho}\mathsf{H}^{(10_r)a}
    \mathsf{H}^{(210)\gamma\sigma}_{\alpha\beta}=-10\mathcal S_{75_{_{{210}}}}\epsilon_{ab}{\mathbf U}_{\alpha}^{\mathtt c}\mathbf{{Q}}^{b\alpha}\mathsf{H}^{(10_r)a}$
\end{enumerate}

\item $\displaystyle-\frac{15}{2}\epsilon_{ijklm}\mathsf{M}^{ij}\mathsf{M}^{kn}\mathsf{H}^{(10_r)p}\mathsf{H}^{(210)lm}_{np}=
    \frac{15}{2}\epsilon_{\alpha\beta\gamma cd}\mathsf{M}^{\alpha\beta}\mathsf{M}^{b\gamma}\mathsf{H}^{(10_r)a}\mathsf{H}^{(210)cd}_{ba}\\
    ~~~~~~~~~~~~~~~~~~~~~~~~~~~~~~~~~~~~~~~~~~~~~~~~+15\epsilon_{\gamma\sigma\beta cb}\mathsf{M}^{\gamma\sigma}\mathsf{M}^{c\alpha}\mathsf{H}^{(10_r)a}
    \mathsf{H}^{(210)\beta b}_{\alpha a}\\
    ~~~~~~~~~~~~~~~~~~~~~~~~~~~~~~~~~~~~~~~~~~~~~~~~+30\epsilon_{\sigma\gamma\beta cb}\mathsf{M}^{\gamma\alpha}\mathsf{M}^{c\sigma}\mathsf{H}^{(10_r)a}
    \mathsf{H}^{(210)\beta b}_{\alpha a}+\cdots$
    \begin{enumerate}
\item $\displaystyle\frac{15}{2}\epsilon_{\alpha\beta\gamma cd}\mathsf{M}^{\alpha\beta}\mathsf{M}^{b\gamma}\mathsf{H}^{(10_r)a}\mathsf{H}^{(210)cd}_{ba}=
-15\mathcal S_{75_{_{{210}}}}\epsilon_{ab}{\mathbf U}_{\alpha}^{\mathtt c}\mathbf{{Q}}^{b\alpha}\mathsf{H}^{(10_r)a}$
\item $\displaystyle15\epsilon_{\gamma\sigma\beta cb}\mathsf{M}^{\gamma\sigma}\mathsf{M}^{c\alpha}\mathsf{H}^{(10_r)a}
    \mathsf{H}^{(210)\beta b}_{\alpha a}=\frac{5}{2}\mathcal S_{75_{_{{210}}}}\epsilon_{ab}{\mathbf U}_{\alpha}^{\mathtt c}\mathbf{{Q}}^{b\alpha}\mathsf{H}^{(10_r)a}$
\item $\displaystyle30\epsilon_{\sigma\gamma\beta cb}\mathsf{M}^{\gamma\alpha}\mathsf{M}^{c\sigma}\mathsf{H}^{(10_r)a}
    \mathsf{H}^{(210)\beta b}_{\alpha a}=10\mathcal S_{75_{_{{210}}}}\epsilon_{ab}{\mathbf U}_{\alpha}^{\mathtt c}\mathbf{{Q}}^{b\alpha}\mathsf{H}^{(10_r)a}$
\end{enumerate}

\item $\displaystyle\frac{5}{2}\epsilon_{ijklm}\mathsf{M}^{ij}\mathsf{M}^{kl}\mathsf{H}^{(10_r)n}\mathsf{H}^{(210)m}_{n}
    =-10\mathcal S_{24_{_{{210}}}}\epsilon_{ab}{\mathbf U}_{\alpha}^{\mathtt c}\mathbf{{Q}}^{b\alpha}\mathsf{H}^{(10_r)a}+\cdots$

\item $\displaystyle-\frac{3}{4}\epsilon_{ijklm}\mathsf{M}^{ij}\mathsf{M}^{kl}\mathsf{H}^{(10_r)m}\mathsf{H}^{(210)}=-6\mathcal S_{1_{_{{210}}}}\epsilon_{ab}{\mathbf U}_{\alpha}^{\mathtt c}\mathbf{{Q}}^{b\alpha}\mathsf{H}^{(10_r)a}+\cdots$

\item $\displaystyle-60\mathsf{M}^{ij}\mathsf{M}_{k}\mathsf{H}^{(10_r)}_l\mathsf{H}^{(210)kl}_{ij}
=60\mathcal S_{75_{_{{210}}}}\epsilon^{ab}{\mathbf E}^{\mathtt c}\mathbf{{L}}_{b}\mathsf{H}^{(10_r)}_a-20\mathcal S_{75_{_{{210}}}}\mathbf{{Q}}^{a\alpha}{\mathbf D}_{\alpha}^{\mathtt c}\mathsf{H}^{(10_r)}_a+\cdots$

\item $\displaystyle-20\mathsf{M}^{ij}\mathsf{M}_{i}\mathsf{H}^{(10_r)}_k\mathsf{H}^{(210)k}_{j}
=-10\mathcal S_{24_{_{{210}}}}\epsilon^{ab}{\mathbf E}^{\mathtt c}\mathbf{{L}}_{b}\mathsf{H}^{(10_r)}_a-10\mathcal S_{24_{_{{210}}}}\mathbf{{Q}}^{a\alpha}{\mathbf D}_{\alpha}^{\mathtt c}\mathsf{H}^{(10_r)}_a+\cdots$

\item $\displaystyle80\mathsf{M}^{ij}\mathsf{M}_{k}\mathsf{H}^{(10_r)}_i\mathsf{H}^{(210)k}_{j}
=-40\mathcal S_{24_{_{{210}}}}\epsilon^{ab}{\mathbf E}^{\mathtt c}\mathbf{{L}}_{b}\mathsf{H}^{(10_r)}_a+\frac{80}{3}\mathcal S_{24_{_{{210}}}}\mathbf{{Q}}^{a\alpha}{\mathbf D}_{\alpha}^{\mathtt c}\mathsf{H}^{(10_r)}_a+\cdots$
\end{enumerate}
Thus,
\begin{eqnarray}
W^{(1)}_4
&=&-i\frac{f^{(1)}}{5!M_{c}}b_r\left[\left(\frac{15}{2}\mathcal S_{75_{_{{210}}}}-10\mathcal S_{24_{_{{210}}}}-6\mathcal S_{1_{_{{210}}}}\right)\epsilon_{ab}{\mathbf U}_{\alpha}^{\mathtt c}\mathbf{{Q}}^{b\alpha}\mathsf{H}^{(10_r)a}\right.\nonumber\\
&&\left.~~~~~~~~~~~~~~~+\left(-20\mathcal S_{75_{_{{210}}}}+\frac{50}{3}\mathcal S_{24_{_{{210}}}}\right){\mathbf D}_{\alpha}^{\mathtt c}\mathbf{{Q}}^{a\alpha}\mathsf{H}^{(10_r)}_a\right.\nonumber\\
&&\left.~~~~~~~~~~~~~~~+\left(60\mathcal S_{75_{_{{210}}}}-50\mathcal S_{24_{_{{210}}}}\right)\epsilon^{ab}{\mathbf E}^{\mathtt c}\mathbf{{L}}_{b}\mathsf{H}^{(10_r)}_a+\cdots\right].
\end{eqnarray}
Using $\epsilon_{54}=-1=\epsilon^{54}$ and $\epsilon_{45}=+1=\epsilon^{45}$, we get
\begin{align*}
\displaystyle\epsilon_{ab}{\mathbf U}_{\alpha}^{\mathtt c}\mathbf{{Q}}^{b\alpha}\mathsf{H}^{(10_r)a}&=-{\mathbf U}_{\alpha}^{\mathtt c}\mathbf{{U}}^{\alpha}\mathsf{H}^{(10_r)5}+\cdots, \\
\displaystyle{\mathbf D}_{\alpha}^{\mathtt c}\mathbf{{Q}}^{a\alpha}\mathsf{H}^{(10_r)}_a&={\mathbf D}_{\alpha}^{\mathtt c}\mathbf{{D}}^{\alpha}\mathsf{H}^{(10_r)}_5+\cdots, \\
\displaystyle\epsilon^{ab}{\mathbf E}^{\mathtt c}\mathbf{{L}}_{b}\mathsf{H}^{(10_r)}_a&=-{\mathbf E}^{\mathtt c}\mathbf{{E}}\mathsf{H}^{(10_r)}_5+\cdots
\end{align*}
Further,  on using Eqs. (\ref{doublet mass eigenstates}), (\ref{Extraction of doublets}) and (\ref{Normalized doublets}) gives
\begin{align*}
\displaystyle \mathsf{H}^{(10_1)5}&\equiv{}^{({5}_{{10_1}})}\!{{\mathcal D}}^{5}=\sqrt{2}~{}^{({5}_{{10_1}})}\!{{\mathsf D}}^{5}=\sqrt{2}U_{d_{11}}\langle H_u\rangle+\cdots, \\
\displaystyle \mathsf{H}^{(10_2)5}&\equiv{}^{({5}_{{10_2}})}\!{{\mathcal D}}^{5}=\sqrt{2}~{}^{({5}_{{10_2}})}\!{{\mathsf D}}^{5}=\sqrt{2}U_{d_{21}}\langle H_u\rangle+\cdots, \\
\displaystyle \mathsf{H}^{(10_1)}_5&\equiv{}{}^{(\overline{5}_{{10_1}})}\!{{\mathcal D}}_{5}=\sqrt{2}~{}^{(\overline{5}_{{10_1}})}\!{{\mathsf D}}_{5}=\sqrt{2}V_{d_{11}}\langle H_d\rangle+\cdots, \\
\displaystyle \mathsf{H}^{(10_2)}_5&\equiv{}{}^{(\overline{5}_{{10_2}})}\!{{\mathcal D}}_{5}=\sqrt{2}~{}^{(\overline{5}_{{10_2}})}\!{{\mathsf D}}_{5}=\sqrt{2}V_{d_{21}}\langle H_d\rangle+\cdots,
\end{align*}
where, $H_d \equiv {\mathbf{H_d}}_5$ and $H_u\equiv {\mathbf{H_u}}^5$.

And finally, using (see Eqs. (\ref{Normalized singlets}) and (\ref{VEVS})),
\begin{align*}
\displaystyle\langle\mathcal S_{1_{_{{210}}}}\rangle &=4\sqrt{\frac{5}{3}}~\mathcal V_{1_{210}},\\
\displaystyle\langle\mathcal S_{24_{_{{210}}}}\rangle &=2\sqrt{\frac{3}{5}}~\mathcal V_{24_{210}},\\
\displaystyle\langle\mathcal S_{75_{_{{210}}}}\rangle &=\frac{1}{\sqrt{3}}~\mathcal V_{24_{210}},
\end{align*}
gives Eqs. (\ref{q&l masses from quartic coupling 1a})$-$(\ref{q&l masses from quartic coupling 1c}).

\item
Next we compute contribution to the Yukawas of the third generation from Eq.~(\ref{w42}).
\begin{eqnarray}
W_4^{(2)}&=&\frac{f^{(2)}}{5!M_{c}}\langle\Psi_{(+)}^*|B \Gamma_{[\lambda}\Gamma_\mu\Gamma_{\nu}\Gamma_{\rho}\Gamma_{\sigma]}|\Psi_{(+)}\rangle~
\left[-\Sigma_{\lambda\alpha\beta}\Phi_{\gamma\rho\sigma\lambda}+\Sigma_{\lambda\alpha\gamma}\Phi_{\beta\rho\sigma\lambda}
-\Sigma_{\lambda\alpha\rho}\Phi_{\beta\gamma\sigma\lambda}\right.\nonumber\\
&&\left.~~~~~~~~~~~~~~~~~~~~~~~~~~~~~~~~~~~~~~~~~~~~~~+\Sigma_{\lambda\alpha\sigma}\Phi_{\beta\gamma\rho\lambda}
+\Sigma_{\lambda\gamma\beta}\Phi_{\alpha\rho\sigma\lambda}
-\Sigma_{\lambda\rho\beta}\Phi_{\alpha\gamma\sigma\lambda}\right.\nonumber\\
&&\left.~~~~~~~~~~~~~~~~~~~~~~~~~~~~~~~~~~~~~~~~~~~~~~+\Sigma_{\lambda\sigma\beta}\Phi_{\alpha\gamma\rho\lambda}
+\Sigma_{\lambda\gamma\rho}\Phi_{\beta\alpha\sigma\lambda}
-\Sigma_{\lambda\gamma\sigma}\Phi_{\beta\alpha\rho\lambda}\right.\nonumber\\
&&\left.~~~~~~~~~~~~~~~~~~~~~~~~~~~~~~~~~~~~~~~~~~~~~~+\Sigma_{\lambda\rho\sigma}\Phi_{\beta\alpha\gamma\lambda}\right]\nonumber\\
&=&-i\frac{f^{(2)}}{5!M_{c}}\left[-\frac{15}{2}\epsilon_{ijklm}\mathsf{M}^{ij}\mathsf{M}^{no}\mathsf{H}^{(120)kl}_x\mathsf{H}^{(210)mx}_{no}
+15\epsilon_{ijklm}\mathsf{M}^{ij}\mathsf{M}^{no}\mathsf{H}^{(120)xk}_n\mathsf{H}^{(210)lm}_{xo}\right.\nonumber\\
&&\left.~~~~~~~~~~~~+5\epsilon_{ijklm}\mathsf{M}^{ij}\mathsf{M}^{no}\mathsf{H}^{(120)kl}_n\mathsf{H}^{(210)m}_{o}
+\frac{15}{2}\epsilon_{ijklm}\mathsf{M}^{ij}\mathsf{M}^{kn}\mathsf{H}^{(120)xy}_n\mathsf{H}^{(210)lm}_{xy}\right.\nonumber\\
&&\left.~~~~~~~~~~~~-\frac{5}{2}\epsilon_{ijklm}\mathsf{M}^{ij}\mathsf{M}^{kn}\mathsf{H}^{(120)lm}_x\mathsf{H}^{(210)x}_{n}
+5\epsilon_{ijklm}\mathsf{M}^{ij}\mathsf{M}^{kn}\mathsf{H}^{(120)xl}_n\mathsf{H}^{(210)m}_{x}\right.\nonumber\\
&&\left.~~~~~~~~~~~~+\frac{15}{8}\epsilon_{ijklm}\mathsf{M}^{ij}\mathsf{M}^{kl}\mathsf{H}^{(120)xy}_z\mathsf{H}^{(210)mz}_{xy}
-\frac{5}{4}\epsilon_{ijklm}\mathsf{M}^{ij}\mathsf{M}^{kl}\mathsf{H}^{(120)mx}_y\mathsf{H}^{(210)y}_{x}
\right.\nonumber\\
&&\left.~~~~~~~~~~~~+\frac{3}{16}\epsilon_{ijklm}\mathsf{M}^{ij}\mathsf{M}^{kl}\mathsf{H}^{(120)m}\mathsf{H}^{(210)}
+\frac{15}{16}\epsilon_{ijklm}\mathsf{M}^{ij}\mathsf{M}^{kl}\mathsf{H}^{(120)x}\mathsf{H}^{(210)m}_{x}\right.\nonumber\\
&&\left.~~~~~~~~~~~~-15\mathsf{M}^{ij}\mathsf{M}_{j}\mathsf{H}^{(120)x}_{yz}\mathsf{H}^{(210)yz}_{ix}
+10\mathsf{M}^{ij}\mathsf{M}_{j}\mathsf{H}^{(120)x}_{iy}\mathsf{H}^{(210)y}_{x}\right.\nonumber\\
&&\left.~~~~~~~~~~~~
-\frac{5}{2}\mathsf{M}^{ij}\mathsf{M}_{j}\mathsf{H}^{(120)}_x\mathsf{H}^{(210)x}_{i}-20\mathsf{M}^{ij}\mathsf{M}_{k}\mathsf{H}^{(120)k}_{jx}\mathsf{H}^{(210)x}_{i}
\right.\nonumber\\
&&\left.~~~~~~~~~~~~
-30\mathsf{M}^{ij}\mathsf{M}_{k}\mathsf{H}^{(120)x}_{ij}\mathsf{H}^{(210)k}_{x}
+30\mathsf{M}^{ij}\mathsf{M}_{k}\mathsf{H}^{(120)k}_{xy}\mathsf{H}^{(210)xy}_{ij}\right.\nonumber\\
&&\left.~~~~~~~~~~~~+3\mathsf{M}^{ij}\mathsf{M}_{k}\mathsf{H}^{(120)k}_{ij}\mathsf{H}^{(210)}
-10\mathsf{M}^{ij}\mathsf{M}_{k}\mathsf{H}^{(120)}_{i}\mathsf{H}^{(210)k}_{j}\right.\nonumber\\
&&\left.~~~~~~~~~~~~
-15\mathsf{M}^{ij}\mathsf{M}_{k}\mathsf{H}^{(120)}_{x}\mathsf{H}^{(210)kx}_{ij}
+\cdots\right].
\end{eqnarray}
\vskip 0.5cm
\begin{enumerate}
\item $\displaystyle-\frac{15}{2}\epsilon_{ijklm}\mathsf{M}^{ij}\mathsf{M}^{no}\mathsf{H}^{(120)kl}_x\mathsf{H}^{(210)mx}_{no}=
    -15\left[-\epsilon_{\beta\gamma\alpha ab}\mathsf{M}^{\beta\gamma}\mathsf{M}^{d\rho}\mathsf{H}^{(120)ab}_c\mathsf{H}^{(210)\alpha c}_{\rho d}\right.\\
\left.~~~~~~~~~~~~~~~~~~~~~~~~~~~~~~~~~~~~~~~~~~~~~~~~~~~~~~~~+2\epsilon_{\gamma\rho\alpha ab}\mathsf{M}^{\gamma\rho}\mathsf{M}^{d\sigma}\mathsf{H}^{(120)\alpha a}_{\beta}\mathsf{H}^{(210)b\beta}_{d\sigma}\right.\\
\left.~~~~~~~~~~~~~~~~~~~~~~~~~~~~~~~~~~~~~~~~~~~~~~~~~~~~~~~~+2\epsilon_{\gamma\lambda\alpha ba}\mathsf{M}^{\rho\sigma}\mathsf{M}^{b\lambda}\mathsf{H}^{(120)\alpha a}_{\beta}\mathsf{H}^{(210)\gamma\beta}_{\rho\sigma}
    +\cdots\right]$
\begin{enumerate}
\item $\displaystyle -\epsilon_{\beta\gamma\alpha ab}\mathsf{M}^{\beta\gamma}\mathsf{M}^{d\rho}\mathsf{H}^{(120)ab}_c\mathsf{H}^{(210)\alpha c}_{\rho d}=-\frac{2}{3}\mathcal S_{75_{_{{210}}}}\epsilon_{ab}{\mathbf U}_{\alpha}^{\mathtt c}\mathbf{{Q}}^{a\alpha}~{}^{({45}_{{120}})}\!{{\mathcal D}}^{b}$
\item $\displaystyle 2\epsilon_{\gamma\rho\alpha ab}\mathsf{M}^{\gamma\rho}\mathsf{M}^{d\sigma}\mathsf{H}^{(120)\alpha a}_{\beta}\mathsf{H}^{(210)b\beta}_{d\sigma}=\frac{2}{9}\mathcal S_{75_{_{{210}}}}\epsilon_{ab}{\mathbf U}_{\alpha}^{\mathtt c}\mathbf{{Q}}^{a\alpha}~{}^{({45}_{{120}})}\!{{\mathcal D}}^{b}$
    \item $\displaystyle 2\epsilon_{\gamma\lambda\alpha ba}\mathsf{M}^{\rho\sigma}\mathsf{M}^{b\lambda}\mathsf{H}^{(120)\alpha a}_{\beta}\mathsf{H}^{(210)\gamma\beta}_{\rho\sigma}=-\frac{4}{9}\mathcal S_{75_{_{{210}}}}\epsilon_{ab}{\mathbf U}_{\alpha}^{\mathtt c}\mathbf{{Q}}^{a\alpha}~{}^{({45}_{{120}})}\!{{\mathcal D}}^{b}$
\end{enumerate}

\item $\displaystyle15\epsilon_{ijklm}\mathsf{M}^{ij}\mathsf{M}^{no}\mathsf{H}^{(120)xk}_n\mathsf{H}^{(210)lm}_{xo}=
    15\left[2\epsilon_{\alpha\beta\rho cd}\mathsf{M}^{\alpha\beta}\mathsf{M}^{a\gamma}\mathsf{H}^{(120)bc}_a\mathsf{H}^{(210)\rho d}_{\gamma b}\right.\\
\left.~~~~~~~~~~~~~~~~~~~~~~~~~~~~~~~~~~~~~~~~~~~~~~~~~~+\epsilon_{\alpha\beta\rho cd}\mathsf{M}^{\alpha\beta}\mathsf{M}^{b\gamma}\mathsf{H}^{(120)\rho a}_{\gamma}\mathsf{H}^{(210)cd}_{ab}\right.\\
\left.~~~~~~~~~~~~~~~~~~~~~~~~~~~~~~~~~~~~~~~~~~~~~~~~~~+2\epsilon_{\alpha\beta\sigma ac}\mathsf{M}^{\alpha\beta}\mathsf{M}^{b\gamma}\mathsf{H}^{(120)\rho a}_{\gamma}\mathsf{H}^{(210)\sigma c}_{\rho b}
\right.\\
\left.~~~~~~~~~~~~~~~~~~~~~~~~~~~~~~~~~~~~~~~~~~~~~~~~~~-4\epsilon_{\alpha\rho\sigma ac}\mathsf{M}^{\beta\gamma}\mathsf{M}^{a\alpha}\mathsf{H}^{(120)\rho b}_{\beta}\mathsf{H}^{(210)\sigma c}_{\gamma b}
\right.\\
\left.~~~~~~~~~~~~~~~~~~~~~~~~~~~~~~~~~~~~~~~~~~~~~~~~~~-2\epsilon_{\alpha\sigma\lambda ab}\mathsf{M}^{\beta\gamma}\mathsf{M}^{a\alpha}\mathsf{H}^{(120)\rho b}_{\beta}\mathsf{H}^{(210)\sigma \lambda}_{\rho\gamma}
    +\cdots\right]$
\begin{enumerate}
\item $\displaystyle 2\epsilon_{\alpha\beta\rho cd}\mathsf{M}^{\alpha\beta}\mathsf{M}^{a\gamma}\mathsf{H}^{(120)bc}_a\mathsf{H}^{(210)\rho d}_{\gamma b}=-\frac{4}{3}\mathcal S_{75_{_{{210}}}}\epsilon_{ab}{\mathbf U}_{\alpha}^{\mathtt c}\mathbf{{Q}}^{a\alpha}~{}^{({45}_{{120}})}\!{{\mathcal D}}^{b}$
\item $\displaystyle \epsilon_{\alpha\beta\rho cd}\mathsf{M}^{\alpha\beta}\mathsf{M}^{b\gamma}\mathsf{H}^{(120)\rho a}_{\gamma}\mathsf{H}^{(210)cd}_{ab}=-\frac{2}{3}\mathcal S_{75_{_{{210}}}}\epsilon_{ab}{\mathbf U}_{\alpha}^{\mathtt c}\mathbf{{Q}}^{a\alpha}~{}^{({45}_{{120}})}\!{{\mathcal D}}^{b}$
    \item $\displaystyle 2\epsilon_{\alpha\beta\sigma ac}\mathsf{M}^{\alpha\beta}\mathsf{M}^{b\gamma}\mathsf{H}^{(120)\rho a}_{\gamma}\mathsf{H}^{(210)\sigma c}_{\rho b}=\frac{2}{9}\mathcal S_{75_{_{{210}}}}\epsilon_{ab}{\mathbf U}_{\alpha}^{\mathtt c}\mathbf{{Q}}^{a\alpha}~{}^{({45}_{{120}})}\!{{\mathcal D}}^{b}$
        \item $\displaystyle -4\epsilon_{\alpha\rho\sigma ac}\mathsf{M}^{\beta\gamma}\mathsf{M}^{a\alpha}\mathsf{H}^{(120)\rho b}_{\beta}\mathsf{H}^{(210)\sigma c}_{\gamma b}=\frac{4}{9}\mathcal S_{75_{_{{210}}}}\epsilon_{ab}{\mathbf U}_{\alpha}^{\mathtt c}\mathbf{{Q}}^{a\alpha}~{}^{({45}_{{120}})}\!{{\mathcal D}}^{b}$
         \item $\displaystyle -2\epsilon_{\alpha\sigma\lambda ab}\mathsf{M}^{\beta\gamma}\mathsf{M}^{a\alpha}\mathsf{H}^{(120)\rho b}_{\beta}\mathsf{H}^{(210)\sigma \lambda}_{\rho\gamma}=-\frac{4}{9}\mathcal S_{75_{_{{210}}}}\epsilon_{ab}{\mathbf U}_{\alpha}^{\mathtt c}\mathbf{{Q}}^{a\alpha}~{}^{({45}_{{120}})}\!{{\mathcal D}}^{b}$
        \end{enumerate}

\item $\displaystyle5\epsilon_{ijklm}\mathsf{M}^{ij}\mathsf{M}^{no}\mathsf{H}^{(120)kl}_n\mathsf{H}^{(210)m}_{o}=
    5\left[-2\epsilon_{\alpha\beta\rho ca}\mathsf{M}^{\alpha\beta}\mathsf{M}^{b\gamma}\mathsf{H}^{(120)\rho c}_{\gamma}\mathsf{H}^{(210)a}_{b}\right.\\
\left.~~~~~~~~~~~~~~~~~~~~~~~~~~~~~~~~~~~~~~~~~~~~~+\epsilon_{\alpha\beta\gamma bc}\mathsf{M}^{\alpha\beta}\mathsf{M}^{a\rho}\mathsf{H}^{(120)bc }_{a}\mathsf{H}^{(210)\gamma}_{\rho}\right.\\
\left.~~~~~~~~~~~~~~~~~~~~~~~~~~~~~~~~~~~~~~~~~~~~~+4\epsilon_{\alpha\sigma\beta ab}\mathsf{M}^{\rho\gamma}\mathsf{M}^{a\alpha}\mathsf{H}^{(120)\sigma b}_{\rho}\mathsf{H}^{(210)\beta}_{\gamma}
    +\cdots\right]$
\begin{enumerate}
\item $\displaystyle -2\epsilon_{\alpha\rho ca}\mathsf{M}^{\alpha\beta}\mathsf{M}^{b\gamma}\mathsf{H}^{(120)\rho c}_{\gamma}\mathsf{H}^{(210)a}_{b}=-\frac{2}{3}\mathcal S_{24_{_{{210}}}}\epsilon_{ab}{\mathbf U}_{\alpha}^{\mathtt c}\mathbf{{Q}}^{a\alpha}~{}^{({45}_{{120}})}\!{{\mathcal D}}^{b}$
\item $\displaystyle \epsilon_{\alpha\beta\gamma bc}\mathsf{M}^{\alpha\beta}\mathsf{M}^{a\rho}\mathsf{H}^{(120)bc }_{a}\mathsf{H}^{(210)\gamma}_{\rho}=-\frac{4}{3}\mathcal S_{24_{_{{210}}}}\epsilon_{ab}{\mathbf U}_{\alpha}^{\mathtt c}\mathbf{{Q}}^{a\alpha}~{}^{({45}_{{120}})}\!{{\mathcal D}}^{b}$
    \item $\displaystyle 4\epsilon_{\alpha\sigma\beta ab}\mathsf{M}^{\rho\gamma}\mathsf{M}^{a\alpha}\mathsf{H}^{(120)\sigma b}_{\rho}\mathsf{H}^{(210)\beta}_{\gamma}=\frac{8}{9}\mathcal S_{24_{_{{210}}}}\epsilon_{ab}{\mathbf U}_{\alpha}^{\mathtt c}\mathbf{{Q}}^{a\alpha}~{}^{({45}_{{120}})}\!{{\mathcal D}}^{b}$
\end{enumerate}

\item $\displaystyle\frac{15}{2}\epsilon_{ijklm}\mathsf{M}^{ij}\mathsf{M}^{kn}\mathsf{H}^{(120)xy}_n\mathsf{H}^{(210)lm}_{xy}=
   \frac{15}{2}\left[-4\epsilon_{\alpha\beta\rho ab}\mathsf{M}^{\alpha\beta}\mathsf{M}^{a\gamma}\mathsf{H}^{(120)\sigma c}_{\gamma}\mathsf{H}^{(210)b\rho}_{c\sigma}\right.\\
\left.~~~~~~~~~~~~~~~~~~~~~~~~~~~~~~~~~~~~~~~~~~~~~~~~~~~-\epsilon_{\alpha\beta\gamma ab}\mathsf{M}^{\alpha\beta}\mathsf{M}^{c\gamma}\mathsf{H}^{(120)de }_{c}\mathsf{H}^{(210)ab}_{de}\right.\\
\left.~~~~~~~~~~~~~~~~~~~~~~~~~~~~~~~~~~~~~~~~~~~~~~~~~~-8\epsilon_{\alpha\beta\rho ab}\mathsf{M}^{\beta\gamma}\mathsf{M}^{a\alpha}\mathsf{H}^{(120)\sigma c}_{\gamma}\mathsf{H}^{(210)\rho b}_{\sigma c}
    +\cdots\right]$
\begin{enumerate}
\item $\displaystyle -4\epsilon_{\alpha\beta\rho ab}\mathsf{M}^{\alpha\beta}\mathsf{M}^{a\gamma}\mathsf{H}^{(120)\sigma c}_{\gamma}\mathsf{H}^{(210)b\rho}_{c\sigma}=\frac{4}{9}\mathcal S_{75_{_{{210}}}}\epsilon_{ab}{\mathbf U}_{\alpha}^{\mathtt c}\mathbf{{Q}}^{a\alpha}~{}^{({45}_{{120}})}\!{{\mathcal D}}^{b}$
\item $\displaystyle -\epsilon_{\alpha\beta\gamma ab}\mathsf{M}^{\alpha\beta}\mathsf{M}^{c\gamma}\mathsf{H}^{(120)de }_{c}\mathsf{H}^{(210)ab}_{de}=4\mathcal S_{75_{_{{210}}}}\epsilon_{ab}{\mathbf U}_{\alpha}^{\mathtt c}\mathbf{{Q}}^{a\alpha}~{}^{({45}_{{120}})}\!{{\mathcal D}}^{b}$
    \item $\displaystyle -8\epsilon_{\alpha\beta\rho ab}\mathsf{M}^{\beta\gamma}\mathsf{M}^{a\alpha}\mathsf{H}^{(120)\sigma c}_{\gamma}\mathsf{H}^{(210)\rho b}_{\sigma c}=\frac{8}{9}\mathcal S_{75_{_{{210}}}}\epsilon_{ab}{\mathbf U}_{\alpha}^{\mathtt c}\mathbf{{Q}}^{a\alpha}~{}^{({45}_{{120}})}\!{{\mathcal D}}^{b}$
\end{enumerate}

\item $\displaystyle-\frac{5}{2}\epsilon_{ijklm}\mathsf{M}^{ij}\mathsf{M}^{kn}\mathsf{H}^{(120)lm}_x\mathsf{H}^{(210)x}_{n}=
   -\frac{5}{2}\left[-2\epsilon_{\alpha\beta\sigma ab}\mathsf{M}^{\alpha\beta}\mathsf{M}^{a\gamma}\mathsf{H}^{(120)\sigma b}_{\rho}\mathsf{H}^{(210)\rho}_{\gamma}\right.\\
\left.~~~~~~~~~~~~~~~~~~~~~~~~~~~~~~~~~~~~~~~~~~~~~~~~~~~-\epsilon_{\alpha\beta\gamma cd}\mathsf{M}^{\alpha\beta}\mathsf{M}^{a\gamma}\mathsf{H}^{(120)cd }_{b}\mathsf{H}^{(210)b}_{a}\right.\\
\left.~~~~~~~~~~~~~~~~~~~~~~~~~~~~~~~~~~~~~~~~~~~~~~~~~~-4\epsilon_{\alpha\beta\sigma ab}\mathsf{M}^{\beta\gamma}\mathsf{M}^{a\alpha}\mathsf{H}^{(120)\sigma b}_{\rho}\mathsf{H}^{(210)\rho}_{\gamma}
    +\cdots\right]$
\begin{enumerate}
\item $\displaystyle -2\epsilon_{\alpha\beta\sigma ab}\mathsf{M}^{\alpha\beta}\mathsf{M}^{a\gamma}\mathsf{H}^{(120)\sigma b}_{\rho}\mathsf{H}^{(210)\rho}_{\gamma}=-\frac{4}{9}\mathcal S_{24_{_{{210}}}}\epsilon_{ab}{\mathbf U}_{\alpha}^{\mathtt c}\mathbf{{Q}}^{a\alpha}~{}^{({45}_{{120}})}\!{{\mathcal D}}^{b}$
\item $\displaystyle -\epsilon_{\alpha\beta\gamma cd}\mathsf{M}^{\alpha\beta}\mathsf{M}^{a\gamma}\mathsf{H}^{(120)cd }_{b}\mathsf{H}^{(210)b}_{a}=-2\mathcal S_{24_{_{{210}}}}\epsilon_{ab}{\mathbf U}_{\alpha}^{\mathtt c}\mathbf{{Q}}^{a\alpha}~{}^{({45}_{{120}})}\!{{\mathcal D}}^{b}$
    \item $\displaystyle -4\epsilon_{\alpha\beta\sigma ab}\mathsf{M}^{\beta\gamma}\mathsf{M}^{a\alpha}\mathsf{H}^{(120)\sigma b}_{\rho}\mathsf{H}^{(210)\rho}_{\gamma}=-\frac{8}{9}\mathcal S_{24_{_{{210}}}}\epsilon_{ab}{\mathbf U}_{\alpha}^{\mathtt c}\mathbf{{Q}}^{a\alpha}~{}^{({45}_{{120}})}\!{{\mathcal D}}^{b}$
\end{enumerate}

\item $5\epsilon_{ijklm}\mathsf{M}^{ij}\mathsf{M}^{kn}\mathsf{H}^{(120)xl}_n\mathsf{H}^{(210)m}_{x}=
    5\left[\epsilon_{\alpha\beta\lambda ac}\mathsf{M}^{\alpha\beta}\mathsf{M}^{a\gamma}\mathsf{H}^{(120)\lambda b}_{\gamma}\mathsf{H}^{(210)c}_{b}\right.\\
\left.~~~~~~~~~~~~~~~~~~~~~~~~~~~~~~~~~~~~~~~~~~~~~+\epsilon_{\alpha\beta\lambda ab}\mathsf{M}^{\alpha\beta}\mathsf{M}^{a\gamma}\mathsf{H}^{(120)\rho b}_{\gamma}\mathsf{H}^{(210)\lambda}_{\rho}\right.\\
\left.~~~~~~~~~~~~~~~~~~~~~~~~~~~~~~~~~~~~~~~~~~~~~-\epsilon_{\alpha\beta\gamma bc}\mathsf{M}^{\alpha\beta}\mathsf{M}^{a\gamma}\mathsf{H}^{(120)db }_{a}\mathsf{H}^{(210)c}_{d}
\right.\\
\left.~~~~~~~~~~~~~~~~~~~~~~~~~~~~~~~~~~~~~~~~~~~~~+2\epsilon_{\alpha\beta\rho ac}\mathsf{M}^{\beta\gamma}\mathsf{M}^{a\alpha}\mathsf{H}^{(120)\rho b}_{\gamma}\mathsf{H}^{(210)c}_{b}
\right.\\
\left.~~~~~~~~~~~~~~~~~~~~~~~~~~~~~~~~~~~~~~~~~~~~~+2\epsilon_{\alpha\beta\lambda ab}\mathsf{M}^{\beta\gamma}\mathsf{M}^{a\alpha}\mathsf{H}^{(120)\rho b}_{\gamma}\mathsf{H}^{(210)\lambda}_{\rho}
    +\cdots\right]$
\begin{enumerate}
\item $\displaystyle \epsilon_{\alpha\beta\lambda ac}\mathsf{M}^{\alpha\beta}\mathsf{M}^{a\gamma}\mathsf{H}^{(120)\lambda b}_{\gamma}\mathsf{H}^{(210)c}_{b}=-\frac{1}{3}\mathcal S_{24_{_{{210}}}}\epsilon_{ab}{\mathbf U}_{\alpha}^{\mathtt c}\mathbf{{Q}}^{a\alpha}~{}^{({45}_{{120}})}\!{{\mathcal D}}^{b}$
\item $\displaystyle \epsilon_{\alpha\beta\lambda ab}\mathsf{M}^{\alpha\beta}\mathsf{M}^{a\gamma}\mathsf{H}^{(120)\rho b}_{\gamma}\mathsf{H}^{(210)\lambda}_{\rho}=\frac{2}{9}\mathcal S_{24_{_{{210}}}}\epsilon_{ab}{\mathbf U}_{\alpha}^{\mathtt c}\mathbf{{Q}}^{a\alpha}~{}^{({45}_{{120}})}\!{{\mathcal D}}^{b}$
    \item $\displaystyle -\epsilon_{\alpha\beta\gamma bc}\mathsf{M}^{\alpha\beta}\mathsf{M}^{a\gamma}\mathsf{H}^{(120)db }_{a}\mathsf{H}^{(210)c}_{d}=2\mathcal S_{24_{_{{210}}}}\epsilon_{ab}{\mathbf U}_{\alpha}^{\mathtt c}\mathbf{{Q}}^{a\alpha}~{}^{({45}_{{120}})}\!{{\mathcal D}}^{b}$
        \item $\displaystyle 2\epsilon_{\alpha\beta\rho ac}\mathsf{M}^{\beta\gamma}\mathsf{M}^{a\alpha}\mathsf{H}^{(120)\rho b}_{\gamma}\mathsf{H}^{(210)c}_{b}=-\frac{2}{3}\mathcal S_{24_{_{{210}}}}\epsilon_{ab}{\mathbf U}_{\alpha}^{\mathtt c}\mathbf{{Q}}^{a\alpha}~{}^{({45}_{{120}})}\!{{\mathcal D}}^{b}$
         \item $\displaystyle2\epsilon_{\alpha\beta\lambda ab}\mathsf{M}^{\beta\gamma}\mathsf{M}^{a\alpha}\mathsf{H}^{(120)\rho b}_{\gamma}\mathsf{H}^{(210)\lambda}_{\rho}=\frac{4}{9}\mathcal S_{24_{_{{210}}}}\epsilon_{ab}{\mathbf U}_{\alpha}^{\mathtt c}\mathbf{{Q}}^{a\alpha}~{}^{({45}_{{120}})}\!{{\mathcal D}}^{b}$
        \end{enumerate}

\item $\displaystyle\frac{15}{8}\epsilon_{ijklm}\mathsf{M}^{ij}\mathsf{M}^{kl}\mathsf{H}^{(120)xy}_z\mathsf{H}^{(210)mz}_{xy}=
   -15\left[\frac{1}{2}\epsilon_{\alpha\beta\gamma ab}\mathsf{M}^{\beta\gamma}\mathsf{M}^{a\alpha}\mathsf{H}^{(120)de}_{c}\mathsf{H}^{(210)bc}_{de}\right.\\
\left.~~~~~~~~~~~~~~~~~~~~~~~~~~~~~~~~~~~~~~~~~~~~~~~~~~~-\epsilon_{\alpha\beta\gamma ab}\mathsf{M}^{\beta\gamma}\mathsf{M}^{a\alpha}\mathsf{H}^{(120)\sigma c }_{\rho}\mathsf{H}^{(210)b\rho}_{c\sigma}
    +\cdots\right]$
\begin{enumerate}
\item $\displaystyle \frac{1}{2}\epsilon_{\alpha\beta\gamma ab}\mathsf{M}^{\beta\gamma}\mathsf{M}^{a\alpha}\mathsf{H}^{(120)de}_{c}\mathsf{H}^{(210)bc}_{de}=\mathcal S_{75_{_{{210}}}}\epsilon_{ab}{\mathbf U}_{\alpha}^{\mathtt c}\mathbf{{Q}}^{a\alpha}~{}^{({45}_{{120}})}\!{{\mathcal D}}^{b}$
\item $\displaystyle -\epsilon_{\alpha\beta\gamma ab}\mathsf{M}^{\beta\gamma}\mathsf{M}^{a\alpha}\mathsf{H}^{(120)\sigma c }_{\rho}\mathsf{H}^{(210)b\rho}_{c\sigma}=\frac{1}{3}\mathcal S_{75_{_{{210}}}}\epsilon_{ab}{\mathbf U}_{\alpha}^{\mathtt c}\mathbf{{Q}}^{a\alpha}~{}^{({45}_{{120}})}\!{{\mathcal D}}^{b}$
\end{enumerate}

\item $\displaystyle-\frac{5}{4}\epsilon_{ijklm}\mathsf{M}^{ij}\mathsf{M}^{kl}\mathsf{H}^{(120)mx}_y\mathsf{H}^{(210)y}_{x}=
   5\left[\epsilon_{\alpha\beta\gamma ab}\mathsf{M}^{\beta\gamma}\mathsf{M}^{a\alpha}\mathsf{H}^{(120)bc}_{d}\mathsf{H}^{(210)d}_{c}\right.\\
\left.~~~~~~~~~~~~~~~~~~~~~~~~~~~~~~~~~~~~~~~~~~~~~~~~~-\epsilon_{\alpha\beta\gamma ab}\mathsf{M}^{\beta\gamma}\mathsf{M}^{a\alpha}\mathsf{H}^{(120)\rho b }_{\sigma}\mathsf{H}^{(210)\sigma}_{\rho}
    +\cdots\right]$
\begin{enumerate}
\item $\displaystyle \epsilon_{\alpha\beta\gamma ab}\mathsf{M}^{\beta\gamma}\mathsf{M}^{a\alpha}\mathsf{H}^{(120)bc}_{d}\mathsf{H}^{(210)d}_{c}=-\mathcal S_{24_{_{{210}}}}\epsilon_{ab}{\mathbf U}_{\alpha}^{\mathtt c}\mathbf{{Q}}^{a\alpha}~{}^{({45}_{{120}})}\!{{\mathcal D}}^{b}$
\item $\displaystyle -\epsilon_{\alpha\beta\gamma ab}\mathsf{M}^{\beta\gamma}\mathsf{M}^{a\alpha}\mathsf{H}^{(120)\rho b }_{\sigma}\mathsf{H}^{(210)\sigma}_{\rho}=-\frac{2}{3}\mathcal S_{24_{_{{210}}}}\epsilon_{ab}{\mathbf U}_{\alpha}^{\mathtt c}\mathbf{{Q}}^{a\alpha}~{}^{({45}_{{120}})}\!{{\mathcal D}}^{b}$
\end{enumerate}

\item $\displaystyle\frac{3}{16}\epsilon_{ijklm}\mathsf{M}^{ij}\mathsf{M}^{kl}\mathsf{H}^{(120)m}\mathsf{H}^{(210)}
 =-\frac{3}{2}\mathcal S_{1_{_{{210}}}}\epsilon_{ab}{\mathbf U}_{\alpha}^{\mathtt c}\mathbf{{Q}}^{a\alpha}~{}^{({5}_{{120}})}\!{{\mathcal D}}^{b}$

\item $\displaystyle\frac{15}{16}\epsilon_{ijklm}\mathsf{M}^{ij}\mathsf{M}^{kl}\mathsf{H}^{(120)x}\mathsf{H}^{(210)m}_x=\frac{15}{4}\mathcal S_{24_{_{{210}}}}\epsilon_{ab}{\mathbf U}_{\alpha}^{\mathtt c}\mathbf{{Q}}^{a\alpha}~{}^{({5}_{{120}})}\!{{\mathcal D}}^{b}$

\item $\displaystyle-15\mathsf{M}^{ij}\mathsf{M}_{j}\mathsf{H}^{(120)x}_{yz}\mathsf{H}^{(210)yz}_{ix}
=-15 \left[\mathsf{M}^{ab}\mathsf{M}_{b}\mathsf{H}^{(120)c}_{de}\mathsf{H}^{(210)de}_{ac}
-2\mathsf{M}^{ab}\mathsf{M}_{b}\mathsf{H}^{(120)\gamma}_{\rho c}\mathsf{H}^{(210)\rho c}_{\gamma a}\right.\\
\left.~~~~~~~~~~~~~~~~~~~~~~~~~~~~~~~~~~~~~~~~~~~~+\mathsf{M}^{a\alpha}\mathsf{M}_{\alpha}\mathsf{H}^{(120)c}_{de}\mathsf{H}^{(210)de}_{ac}
-2\mathsf{M}^{a\alpha}\mathsf{M}_{\alpha}\mathsf{H}^{(120)\gamma}_{\rho c}\mathsf{H}^{(210)\rho c}_{\gamma a}
\right]+\cdots$
\begin{enumerate}
\item $\displaystyle \mathsf{M}^{ab}\mathsf{M}_{b}\mathsf{H}^{(120)c}_{de}\mathsf{H}^{(210)de}_{ac}
=-\mathcal S_{75_{_{{210}}}}\epsilon^{ab}{\mathbf E}^{\mathtt c}\mathbf{{L}}_{a}~{}^{(\overline{45}_{{120}})}\!{{\mathcal D}}_{b}$
\item $\displaystyle -2\mathsf{M}^{ab}\mathsf{M}_{b}\mathsf{H}^{(120)\gamma}_{\rho c}\mathsf{H}^{(210)\rho c}_{\gamma a}=-\frac{1}{3}\mathcal S_{75_{_{{210}}}}\epsilon^{ab}{\mathbf E}^{\mathtt c}\mathbf{{L}}_{a}~{}^{(\overline{45}_{{120}})}\!{{\mathcal D}}_{b}$
    \item $\displaystyle\mathsf{M}^{a\alpha}\mathsf{M}_{\alpha}\mathsf{H}^{(120)c}_{de}\mathsf{H}^{(210)de}_{ac}=\mathcal S_{75_{_{{210}}}}{\mathbf D}_{\alpha}^{\mathtt c}\mathbf{{Q}}^{a\alpha}~{}^{(\overline{45}_{{120}})}\!{{\mathcal D}}_{a}$
    \item $\displaystyle-2\mathsf{M}^{a\alpha}\mathsf{M}_{\alpha}\mathsf{H}^{(120)\gamma}_{\rho c}\mathsf{H}^{(210)\rho c}_{\gamma a}=\frac{1}{3}\mathcal S_{75_{_{{210}}}}{\mathbf D}_{\alpha}^{\mathtt c}\mathbf{{Q}}^{a\alpha}~{}^{(\overline{45}_{{120}})}\!{{\mathcal D}}_{a}$
    \end{enumerate}

\item $\displaystyle10\mathsf{M}^{ij}\mathsf{M}_{j}\mathsf{H}^{(120)x}_{iy}\mathsf{H}^{(210)y}_{x}
=10 \left[\mathsf{M}^{ab}\mathsf{M}_{b}\mathsf{H}^{(120)c}_{ab}\mathsf{H}^{(210)b}_{c}
-\mathsf{M}^{ab}\mathsf{M}_{b}\mathsf{H}^{(120)\gamma}_{\beta a}\mathsf{H}^{(210)\beta}_{\gamma}\right.\\
\left.~~~~~~~~~~~~~~~~~~~~~~~~~~~~~~~~~~~~+\mathsf{M}^{a\alpha}\mathsf{M}_{\alpha}\mathsf{H}^{(120)c}_{ab}\mathsf{H}^{(210)b}_{c}
-\mathsf{M}^{a\alpha}\mathsf{M}_{\alpha}\mathsf{H}^{(120)\gamma}_{\beta a}\mathsf{H}^{(210)\beta}_{\gamma}
\right]+\cdots$
\begin{enumerate}
\item $\displaystyle \mathsf{M}^{ab}\mathsf{M}_{b}\mathsf{H}^{(120)c}_{ab}\mathsf{H}^{(210)b}_{c}
=\frac{1}{2}\mathcal S_{24_{_{{210}}}}\epsilon^{ab}{\mathbf E}^{\mathtt c}\mathbf{{L}}_{a}~{}^{(\overline{45}_{{120}})}\!{{\mathcal D}}_{b}$
\item $\displaystyle -\mathsf{M}^{ab}\mathsf{M}_{b}\mathsf{H}^{(120)\gamma}_{\beta a}\mathsf{H}^{(210)\beta}_{\gamma}=\frac{1}{3}\mathcal S_{24_{_{{210}}}}\epsilon^{ab}{\mathbf E}^{\mathtt c}\mathbf{{L}}_{a}~{}^{(\overline{45}_{{120}})}\!{{\mathcal D}}_{b}$
    \item $\displaystyle\mathsf{M}^{a\alpha}\mathsf{M}_{\alpha}\mathsf{H}^{(120)c}_{ab}\mathsf{H}^{(210)b}_{c}=-\frac{1}{2}\mathcal S_{24_{_{{210}}}}{\mathbf D}_{\alpha}^{\mathtt c}\mathbf{{Q}}^{a\alpha}~{}^{(\overline{45}_{{120}})}\!{{\mathcal D}}_{a}$
    \item $\displaystyle-\mathsf{M}^{a\alpha}\mathsf{M}_{\alpha}\mathsf{H}^{(120)\gamma}_{\beta a}\mathsf{H}^{(210)\beta}_{\gamma}=-\frac{1}{3}\mathcal S_{24_{_{{210}}}}{\mathbf D}_{\alpha}^{\mathtt c}\mathbf{{Q}}^{a\alpha}~{}^{(\overline{45}_{{120}})}\!{{\mathcal D}}_{a}$
\end{enumerate}

\item $\displaystyle-\frac{5}{2}\mathsf{M}^{ij}\mathsf{M}_{j}\mathsf{H}^{(120)}_x\mathsf{H}^{(210)x}_{i}
=-\frac{5}{2}\left[\frac{1}{2}\mathcal S_{24_{_{{210}}}}\epsilon^{ab}{\mathbf E}^{\mathtt c}\mathbf{{L}}_{a}~{}^{(\overline{5}_{{120}})}\!{{\mathcal D}}_{b}
-\frac{1}{2}\mathcal S_{24_{_{{210}}}}{\mathbf D}_{\alpha}^{\mathtt c}\mathbf{{Q}}^{a\alpha}~{}^{(\overline{5}_{{120}})}\!{{\mathcal D}}_{a}
+\cdots\right]$

\item $\displaystyle-20\mathsf{M}^{ij}\mathsf{M}_{k}\mathsf{H}^{(120)k}_{jx}\mathsf{H}^{(210)x}_{i}
=-20\left[\mathsf{M}^{ac}\mathsf{M}_{d}\mathsf{H}^{(120)d}_{cb}\mathsf{H}^{(210)b}_{a}+
\mathsf{M}^{a\alpha}\mathsf{M}_{\beta}\mathsf{H}^{(120\beta}_{\alpha b}\mathsf{H}^{(210)b}_{a}
\right.\\
\left.~~~~~~~~~~~~~~~~~~~~~~~~~~~~~~~~~~~~~~~~~-\mathsf{M}^{a\alpha}\mathsf{M}_{\gamma}\mathsf{H}^{(120)\gamma}_{a\beta}\mathsf{H}^{(210)\beta}_{\alpha}
\right]+\cdots$
\begin{enumerate}
\item $\displaystyle \mathsf{M}^{ac}\mathsf{M}_{d}\mathsf{H}^{(120)d}_{cb}\mathsf{H}^{(210)b}_{a}
=-\mathcal S_{24_{_{{210}}}}\epsilon^{ab}{\mathbf E}^{\mathtt c}\mathbf{{L}}_{a}~{}^{(\overline{45}_{{120}})}\!{{\mathcal D}}_{b}$
    \item $\displaystyle\mathsf{M}^{a\alpha}\mathsf{M}_{\beta}\mathsf{H}^{(120\beta}_{\alpha b}\mathsf{H}^{(210)b}_{a}=-\frac{1}{6}\mathcal S_{24_{_{{210}}}}{\mathbf D}_{\alpha}^{\mathtt c}\mathbf{{Q}}^{a\alpha}~{}^{(\overline{45}_{{120}})}\!{{\mathcal D}}_{a}$
    \item $\displaystyle-\mathsf{M}^{a\alpha}\mathsf{M}_{\gamma}\mathsf{H}^{(120)\gamma}_{a\beta}\mathsf{H}^{(210)\beta}_{\alpha}=\frac{1}{9}\mathcal S_{24_{_{{210}}}}{\mathbf D}_{\alpha}^{\mathtt c}\mathbf{{Q}}^{a\alpha}~{}^{(\overline{45}_{{120}})}\!{{\mathcal D}}_{a}$
\end{enumerate}

\item $\displaystyle-30\mathsf{M}^{ij}\mathsf{M}_{k}\mathsf{H}^{(120)x}_{ij}\mathsf{H}^{(210)k}_{x}
=-30\left[\mathcal S_{24_{_{{210}}}}\epsilon^{ab}{\mathbf E}^{\mathtt c}\mathbf{{L}}_{a}~{}^{(\overline{45}_{{120}})}\!{{\mathcal D}}_{b}
-\frac{2}{9}\mathcal S_{24_{_{{210}}}}{\mathbf D}_{\alpha}^{\mathtt c}\mathbf{{Q}}^{a\alpha}~{}^{(\overline{45}_{{120}})}\!{{\mathcal D}}_{a}
+\cdots\right]$

\item $\displaystyle30\mathsf{M}^{ij}\mathsf{M}_{k}\mathsf{H}^{(120)k}_{xy}\mathsf{H}^{(210)xy}_{ij}
=30\left[-2\mathcal S_{75_{_{{210}}}}\epsilon^{ab}{\mathbf E}^{\mathtt c}\mathbf{{L}}_{a}~{}^{(\overline{45}_{{120}})}\!{{\mathcal D}}_{b}
+\frac{2}{9}\mathcal S_{75_{_{{210}}}}{\mathbf D}_{\alpha}^{\mathtt c}\mathbf{{Q}}^{a\alpha}~{}^{(\overline{45}_{{120}})}\!{{\mathcal D}}_{a}
+\cdots\right]$

\item $\displaystyle3\mathsf{M}^{ij}\mathsf{M}_{k}\mathsf{H}^{(120)k}_{ij}\mathsf{H}^{(210)}
=3\left[-2\mathcal S_{1_{_{{210}}}}\epsilon^{ab}{\mathbf E}^{\mathtt c}\mathbf{{L}}_{a}~{}^{(\overline{45}_{{120}})}\!{{\mathcal D}}_{b}
-\frac{2}{3}\mathcal S_{1_{_{{210}}}}{\mathbf D}_{\alpha}^{\mathtt c}\mathbf{{Q}}^{a\alpha}~{}^{(\overline{45}_{{120}})}\!{{\mathcal D}}_{a}
+\cdots\right]$

%

\item $\displaystyle-10\mathsf{M}^{ij}\mathsf{M}_{k}\mathsf{H}^{(120)}_{i}\mathsf{H}^{(210)k}_{j}
=-10\left[\frac{1}{2}\mathcal S_{24_{_{{210}}}}\epsilon^{ab}{\mathbf E}^{\mathtt c}\mathbf{{L}}_{a}~{}^{(\overline{5}_{{120}})}\!{{\mathcal D}}_{b}
+\frac{1}{3}\mathcal S_{24_{_{{210}}}}{\mathbf D}_{\alpha}^{\mathtt c}\mathbf{{Q}}^{a\alpha}~{}^{(\overline{5}_{{120}})}\!{{\mathcal D}}_{a}
+\cdots\right]$

\item $\displaystyle-15\mathsf{M}^{ij}\mathsf{M}_{k}\mathsf{H}^{(120)}_{x}\mathsf{H}^{(210)kx}_{ij}
=-15\left[\mathcal S_{75_{_{{210}}}}\epsilon^{ab}{\mathbf E}^{\mathtt c}\mathbf{{L}}_{a}~{}^{(\overline{5}_{{120}})}\!{{\mathcal D}}_{b}
+\frac{1}{3}\mathcal S_{75_{_{{210}}}}{\mathbf D}_{\alpha}^{\mathtt c}\mathbf{{Q}}^{a\alpha}~{}^{(\overline{5}_{{120}})}\!{{\mathcal D}}_{a}
+\cdots\right]$
\end{enumerate}
Thus,

\begin{eqnarray}
W^{(2)}_4
&=&-i\frac{f^{(2)}}{5!M_{c}}\left[\left(\frac{20}{3}\mathcal S_{75_{_{{210}}}}{}^{({45}_{{120}})}\!{{\mathcal D}}^{b}+\frac{25}{9}\mathcal S_{24_{_{{210}}}}{}^{({45}_{{120}})}\!{{\mathcal D}}^{b}+\frac{15}{4}\mathcal S_{24_{_{{210}}}}{}^{({5}_{{120}})}\!{{\mathcal D}}^{b}-\frac{3}{2}\mathcal S_{1_{_{{210}}}}{}^{({5}_{{120}})}\!{{\mathcal D}}^{b}\right)\epsilon_{ab}{\mathbf U}_{\alpha}^{\mathtt c}\mathbf{{Q}}^{a\alpha}\right.\nonumber\\
&&\left.~~~~~~~~~~~~~~+\left(-\frac{40}{3}\mathcal S_{75_{_{{210}}}}{}^{(\overline{45}_{{120}})}\!{{\mathcal D}}_a-\frac{5}{9}\mathcal S_{24_{_{{210}}}}{}^{(\overline{45}_{{120}})}\!{{\mathcal D}}_a-\frac{25}{12}\mathcal S_{24_{_{{210}}}}{}^{(\overline{5}_{{120}})}\!{{\mathcal D}}_a-2\mathcal S_{1_{_{{210}}}}{}^{(\overline{45}_{{120}})}\!{{\mathcal D}}_a\right.\right.\nonumber\\
&&\left.\left.~~~~~~~~~~~~~~-5\mathcal S_{75_{_{{210}}}}{}^{(\overline{5}_{{120}})}\!{{\mathcal D}}_a\right){\mathbf D}_{\alpha}^{\mathtt c}\mathbf{{Q}}^{a\alpha}\right.\nonumber\\
&&\left.~~~~~~~~~~~~~~+\left(-40\mathcal S_{75_{_{{210}}}}{}^{(\overline{45}_{{120}})}\!{{\mathcal D}}_b-\frac{5}{3}\mathcal S_{24_{_{{210}}}}{}^{(\overline{45}_{{120}})}\!{{\mathcal D}}_b-\frac{25}{4}\mathcal S_{24_{_{{210}}}}{}^{(\overline{5}_{{120}})}\!{{\mathcal D}}_b-6\mathcal S_{1_{_{{210}}}}{}^{(\overline{45}_{{120}})}\!{{\mathcal D}}_b\right.\right.\nonumber\\
&&\left.\left.~~~~~~~~~~~~~~-15\mathcal S_{75_{_{{210}}}}{}^{(\overline{5}_{{120}})}\!{{\mathcal D}}_b\right)\epsilon^{ab}{\mathbf E}^{\mathtt c}\mathbf{{L}}_{a}+\cdots\right].
\end{eqnarray}
 Usage of Eqs. (\ref{doublet mass eigenstates}), (\ref{Extraction of doublets}) and (\ref{Normalized doublets}) gives

\begin{align*}
\displaystyle {}^{({5}_{{120}})}\!{{\mathcal D}}^{5}=\frac{4}{\sqrt{3}}~{}^{({5}_{{120}})}\!{{\mathsf D}}^{5}=\frac{4}{\sqrt{3}}U_{d_{31}}\langle H_u\rangle+\cdots, \\
\displaystyle {}^{(\overline{5}_{{120}})}\!{{\mathcal D}}_{5}=\frac{4}{\sqrt{3}}~{}^{(\overline{5}_{{120}})}\!{{\mathsf D}}_{5}=\frac{4}{\sqrt{3}}V_{d_{31}}\langle H_d\rangle+\cdots, \\
\displaystyle {}^{({45}_{{120}})}\!{{\mathcal D}}^{5}=\frac{1}{\sqrt{2}}~{}^{({45}_{{120}})}\!{{\mathsf D}}^{5}=\frac{1}{\sqrt{2}}U_{d_{61}}\langle H_u\rangle+\cdots, \\
\displaystyle {}^{(\overline{45}_{{120}})}\!{{\mathcal D}}_{5}=\frac{1}{\sqrt{2}}~{}^{(\overline{45}_{{120}})}\!{{\mathsf D}}_{5}=\frac{1}{\sqrt{2}}V_{d_{61}}\langle H_d\rangle+\cdots, \\
\end{align*}
and finally making use of Eqs. \ref{Normalized singlets} and (\ref{VEVS}), gives Eqs. (\ref{q&l masses from quartic coupling 2a})$-$(\ref{q&l masses from quartic coupling 2c}).

\item
Finally we compute the contributions to the third generation Yukawas from Eq.~(\ref{w42}).


\begin{eqnarray}
W_4^{(3)}&=&\frac{f^{(3)}}{M_{c}}\langle\Psi_{(+)}^*|B \Gamma_{\mu}|\Psi_{(+)}\rangle\Sigma_{\rho\sigma\lambda}\Phi_{\rho\sigma\lambda\mu}\nonumber\\
&=&-\frac{3i}{64}\frac{f^{(3)}}{M_{c}}\left[\epsilon_{ijklm}\mathsf{M}^{ij}\mathsf{M}^{kl}\mathsf{H}^{(120)ab}_c\mathsf{H}^{(210)cm}_{ab}
+2\epsilon_{ijklm}\mathsf{M}^{ij}\mathsf{M}^{kl}\mathsf{H}^{(120)a\alpha}_{\beta}\mathsf{H}^{(210)\beta m}_{a\alpha}\right.\nonumber\\
&&\left.~~~~~~~~~~~~~+\frac{2}{3}\epsilon_{ijklm}\mathsf{M}^{ij}\mathsf{M}^{kl}\mathsf{H}^{(120)mb}_a\mathsf{H}^{(210)a}_{b}
+\frac{2}{3}\epsilon_{ijklm}\mathsf{M}^{ij}\mathsf{M}^{kl}\mathsf{H}^{(120)m\beta}_{\alpha}\mathsf{H}^{(210)\alpha}_{\beta}\right.\nonumber\\
&&\left.~~~~~~~~~~~~~-\frac{1}{2}\epsilon_{ijklm}\mathsf{M}^{ij}\mathsf{M}^{kl}\mathsf{H}^{(120)a}\mathsf{H}^{(210)m}_{a}
-\frac{1}{10}\epsilon_{ijklm}\mathsf{M}^{ij}\mathsf{M}^{kl}\mathsf{H}^{(120)m}\mathsf{H}^{(210)}+\cdots\right]\nonumber\\
&&-\frac{3i}{8}\frac{f^{(3)}}{M_{c}}\left[\mathsf{M}_{i}\mathsf{M}^{ij}\mathsf{H}^{(120)c}_{ab}\mathsf{H}^{(210)ab}_{cj}
+2\mathsf{M}_{i}\mathsf{M}^{ij}\mathsf{H}^{(120)\beta}_{\alpha a}\mathsf{H}^{(210)\alpha a}_{\beta j}+\frac{2}{3}\mathsf{M}_{i}\mathsf{M}^{ij}\mathsf{H}^{(120)b}_{ja}\mathsf{H}^{(210)a}_{b}\right.\nonumber\\
&&\left.~~~~~~~~~~~~~+\frac{2}{3}\mathsf{M}_{i}\mathsf{M}^{ij}\mathsf{H}^{(120)\beta}_{j\alpha}\mathsf{H}^{(210)\alpha}_{\beta}
-\frac{1}{2}\mathsf{M}_{i}\mathsf{M}^{ij}\mathsf{H}^{(120)}_{a}\mathsf{H}^{(210)a}_{j}-
\frac{1}{10}\mathsf{M}_{i}\mathsf{M}^{ij}\mathsf{H}^{(120)}_{j}\mathsf{H}^{(210)}\right.\nonumber\\
&&\left.~~~~~~~~~~~~~+\cdots\right]
\end{eqnarray}
\begin{itemize}
\item $\displaystyle\epsilon_{ijklm}\mathsf{M}^{ij}\mathsf{M}^{kl}\mathsf{H}^{(120)ab}_c\mathsf{H}^{(210)cm}_{ab}= 8\mathcal S_{75_{_{{210}}}}{\mathbf U}_{\alpha}^{\mathtt c}\mathbf{{U}}^{\alpha}~{}^{({45}_{{120}})}\!{\mathcal D}^{5}$
    \vskip 0.5cm
   \item $\displaystyle2\epsilon_{ijklm}\mathsf{M}^{ij}\mathsf{M}^{kl}\mathsf{H}^{(120)a\alpha}_{\beta}\mathsf{H}^{(210)\beta m}_{a\alpha}=\frac{8}{3}\mathcal S_{75_{_{{210}}}}{\mathbf U}_{\alpha}^{\mathtt c}\mathbf{{U}}^{\alpha}~{}^{({45}_{{120}})}\!{\mathcal D}^{5}$
       \vskip 0.5cm
   \item $\displaystyle\frac{2}{3}\epsilon_{ijklm}\mathsf{M}^{ij}\mathsf{M}^{kl}\mathsf{H}^{(120)mb}_a\mathsf{H}^{(210)a}_{b}=\frac{8}{3}\mathcal S_{24_{_{{210}}}}{\mathbf U}_{\alpha}^{\mathtt c}\mathbf{{U}}^{\alpha}~{}^{({45}_{{120}})}\!{\mathcal D}^{5}$
        \vskip 0.5cm
   \item $\displaystyle \frac{2}{3}\epsilon_{ijklm}\mathsf{M}^{ij}\mathsf{M}^{kl}\mathsf{H}^{(120)m\beta}_{\alpha}\mathsf{H}^{(210)\alpha}_{\beta}=\frac{16}{9}\mathcal S_{24_{_{{210}}}}{\mathbf U}_{\alpha}^{\mathtt c}\mathbf{{U}}^{\alpha}~{}^{({45}_{{120}})}\!{\mathcal D}^{5}$
         \vskip 0.5cm
   \item $ \displaystyle-\frac{1}{2}\epsilon_{ijklm}\mathsf{M}^{ij}\mathsf{M}^{kl}\mathsf{H}^{(120)a}\mathsf{H}^{(210)m}_{a}=-2\mathcal S_{24_{_{{210}}}}{\mathbf U}_{\alpha}^{\mathtt c}\mathbf{{U}}^{\alpha}~{}^{({5}_{{120}})}\!{\mathcal D}^{5}$
        \vskip 0.5cm
   \item $ \displaystyle-\frac{1}{10}\epsilon_{ijklm}\mathsf{M}^{ij}\mathsf{M}^{kl}\mathsf{H}^{(120)m}\mathsf{H}^{(210)}=\frac{4}{5}\mathcal S_{1_{_{{210}}}}{\mathbf U}_{\alpha}^{\mathtt c}\mathbf{{U}}^{\alpha}~{}^{({5}_{{120}})}\!{\mathcal D}^{5}$
        \vskip 0.5cm
        \item $\displaystyle\mathsf{M}_{i}\mathsf{M}^{ij}\mathsf{H}^{(120)c}_{ab}\mathsf{H}^{(210)ab}_{cj}= -\mathcal S_{75_{_{{210}}}}{\mathbf E}^{\mathtt c}\mathbf{{E}}~{}^{(\overline{45}_{{120}})}\!{\mathcal D}_{5}+\mathcal S_{75_{_{{210}}}}{\mathbf D}_{\alpha}^{\mathtt c}\mathbf{{D}}^{\alpha}~{}^{(\overline{45}_{{120}})}\!{\mathcal D}_{5}$
        \vskip 0.5cm
        \item $\displaystyle2\mathsf{M}_{i}\mathsf{M}^{ij}\mathsf{H}^{(120)\beta}_{\alpha a}\mathsf{H}^{(210)\alpha a}_{\beta j}=-\frac{1}{3}\mathcal S_{75_{_{{210}}}}{\mathbf E}^{\mathtt c}\mathbf{{E}}~{}^{(\overline{45}_{{120}})}\!{\mathcal D}_{5}+\frac{1}{3}\mathcal S_{75_{_{{210}}}}{\mathbf D}_{\alpha}^{\mathtt c}\mathbf{{D}}^{\alpha}~{}^{(\overline{45}_{{120}})}\!{\mathcal D}_{5}$
       \vskip 0.5cm
        \item $\displaystyle\frac{2}{3}\mathsf{M}_{i}\mathsf{M}^{ij}\mathsf{H}^{(120)b}_{ja}\mathsf{H}^{(210)a}_{b}=-\frac{1}{3}\mathcal S_{24_{_{{210}}}}{\mathbf E}^{\mathtt c}\mathbf{{E}}~{}^{(\overline{45}_{{120}})}\!{\mathcal D}_{5}+\frac{1}{3}\mathcal S_{24_{_{{210}}}}{\mathbf D}_{\alpha}^{\mathtt c}\mathbf{{D}}^{\alpha}~{}^{(\overline{45}_{{120}})}\!{\mathcal D}_{5}$
        \vskip 0.5cm
        \item $\displaystyle\frac{2}{3}\mathsf{M}_{i}\mathsf{M}^{ij}\mathsf{H}^{(120)\beta}_{j\alpha}\mathsf{H}^{(210)\alpha}_{\beta}=-\frac{2}{9}\mathcal S_{24_{_{{210}}}}{\mathbf E}^{\mathtt c}\mathbf{{E}}~{}^{(\overline{45}_{{120}})}\!{\mathcal D}_{5}+\frac{2}{9}\mathcal S_{24_{_{{210}}}}{\mathbf D}_{\alpha}^{\mathtt c}\mathbf{{D}}^{\alpha}~{}^{(\overline{45}_{{120}})}\!{\mathcal D}_{5}$
        \vskip 0.5cm
        \item $\displaystyle-\frac{1}{2}\mathsf{M}_{i}\mathsf{M}^{ij}\mathsf{H}^{(120)}_{a}\mathsf{H}^{(210)a}_{j}=\frac{1}{4}\mathcal S_{24_{_{{210}}}}{\mathbf E}^{\mathtt c}\mathbf{{E}}~{}^{(\overline{5}_{{120}})}\!{\mathcal D}_{5}-\frac{1}{4}\mathcal S_{24_{_{{210}}}}{\mathbf D}_{\alpha}^{\mathtt c}\mathbf{{D}}^{\alpha}~{}^{(\overline{5}_{{120}})}\!{\mathcal D}_{5}$
         \vskip 0.5cm
        \item $\displaystyle-\frac{1}{10}\mathsf{M}_{i}\mathsf{M}^{ij}\mathsf{H}^{(120)}_{j}\mathsf{H}^{(210)}=-\frac{1}{10}\mathcal S_{1_{_{{210}}}}{\mathbf E}^{\mathtt c}\mathbf{{E}}~{}^{(\overline{5}_{{120}})}\!{\mathcal D}_{5}+\frac{1}{10}\mathcal S_{1_{_{{210}}}}{\mathbf D}_{\alpha}^{\mathtt c}\mathbf{{D}}^{\alpha}~{}^{(\overline{5}_{{120}})}\!{\mathcal D}_{5}$
\end{itemize}

\end{enumerate}
\end{appendices}


\begin{thebibliography}{999}




\bibitem{georgi}
H. Georgi, in Particles and Fields (edited by C.E. Carlson), A.I.P.,
1975; H. Fritzch and P. Minkowski, Ann. Phys. {\bf 93}, 193 (1975).


\bibitem{Fritzsch:1974nn}
  H.~Fritzsch and P.~Minkowski,
  Annals Phys.\  {\bf 93}, 193 (1975).
  doi:10.1016/0003-4916(75)90211-0


\bibitem{Clark:1982ai}
  T.~E.~Clark, T.~K.~Kuo and N.~Nakagawa,
  Phys.\ Lett.\  {\bf 115B}, 26 (1982).
  doi:10.1016/0370-2693(82)90507-X


\bibitem{Aulakh:1982sw}
  C.~S.~Aulakh and R.~N.~Mohapatra,
  Phys.\ Rev.\ D {\bf 28}, 217 (1983).
  doi:10.1103/PhysRevD.28.217


\bibitem{Babu:1992ia}
  K.~S.~Babu and R.~N.~Mohapatra,
  Phys.\ Rev.\ Lett.\  {\bf 70}, 2845 (1993)
  doi:10.1103/PhysRevLett.70.2845
  [hep-ph/9209215].


\bibitem{Nath:2001uw}
  P.~Nath and R.~M.~Syed,
  Phys.\ Lett.\ B {\bf 506}, 68 (2001)
  Erratum: [Phys.\ Lett.\ B {\bf 508}, 216 (2001)]
  doi:10.1016/S0370-2693(01)00508-1, 10.1016/S0370-2693(01)00392-6
  [hep-ph/0103165].


\bibitem{Nath:2001yj}
  P.~Nath and R.~M.~Syed,
  Nucl.\ Phys.\ B {\bf 618}, 138 (2001)
  doi:10.1016/S0550-3213(01)00493-X
  [hep-th/0109116].


\bibitem{Nath:2003rc}
  P.~Nath and R.~M.~Syed,
  Nucl.\ Phys.\ B {\bf 676}, 64 (2004)
  doi:10.1016/j.nuclphysb.2003.10.018
  [hep-th/0310178].


\bibitem{Aulakh:2003kg}
  C.~S.~Aulakh, B.~Bajc, A.~Melfo, G.~Senjanovic and F.~Vissani,
  Phys.\ Lett.\ B {\bf 588}, 196 (2004)
  doi:10.1016/j.physletb.2004.03.031
  [hep-ph/0306242].


\bibitem{Bajc:2004xe}
  B.~Bajc, A.~Melfo, G.~Senjanovic and F.~Vissani,
  Phys.\ Rev.\ D {\bf 70}, 035007 (2004)
  doi:10.1103/PhysRevD.70.035007
  [hep-ph/0402122].


\bibitem{Aulakh:2004hm}
  C.~S.~Aulakh and A.~Girdhar,
  Nucl.\ Phys.\ B {\bf 711}, 275 (2005)
  doi:10.1016/j.nuclphysb.2005.01.008
  [hep-ph/0405074].


\bibitem{Aulakh:2008sn}
  C.~S.~Aulakh and S.~K.~Garg,
  Nucl.\ Phys.\ B {\bf 857}, 101 (2012)
  doi:10.1016/j.nuclphysb.2011.12.003
  [arXiv:0807.0917 [hep-ph]].


\bibitem{Nath:2006ut}
  P.~Nath and P.~Fileviez Perez,
  Phys.\ Rept.\  {\bf 441}, 191 (2007)
  doi:10.1016/j.physrep.2007.02.010
  [hep-ph/0601023].


\bibitem{Braun:2005nv}
  V.~Braun, Y.~H.~He, B.~A.~Ovrut and T.~Pantev,
  JHEP {\bf 0605}, 043 (2006)
  doi:10.1088/1126-6708/2006/05/043
  [hep-th/0512177].


\bibitem{Bouchard:2006dn}
  V.~Bouchard, M.~Cvetic and R.~Donagi,
  Nucl.\ Phys.\ B {\bf 745}, 62 (2006)
  doi:10.1016/j.nuclphysb.2006.03.032
  [hep-th/0602096].


\bibitem{Anderson:2009mh}
  L.~B.~Anderson, J.~Gray, Y.~H.~He and A.~Lukas,
  JHEP {\bf 1002}, 054 (2010)
  doi:10.1007/JHEP02(2010)054
  [arXiv:0911.1569 [hep-th]].


\bibitem{Bouchard:2005ag}
  V.~Bouchard and R.~Donagi,
  Phys.\ Lett.\ B {\bf 633}, 783 (2006)
  doi:10.1016/j.physletb.2005.12.042
  [hep-th/0512149].

   \bibitem{DW}
S. Dimopoulos and F. Wilczek. NSF-ITP-82-07 (unpublished);
K.~Babu and S.~M.~Barr,
Phys. Rev. D \textbf{50}, 3529-3536 (1994)
doi:10.1103/PhysRevD.50.3529
[arXiv:hep-ph/9402291 [hep-ph]];
D.~Lee and R.~Mohapatra,
Phys. Lett. B \textbf{324}, 376-379 (1994)
doi:10.1016/0370-2693(94)90209-7
[arXiv:hep-ph/9310371 [hep-ph]];
Y.~Chen and D.~Zhang,
JHEP \textbf{01}, 025 (2015)
doi:10.1007/JHEP01(2015)025
[arXiv:1410.5625 [hep-ph]].





\bibitem{Masiero:1982fe}
  A.~Masiero, D.~V.~Nanopoulos, K.~Tamvakis and T.~Yanagida,
  Phys.\ Lett.\  {\bf 115B}, 380 (1982).
  doi:10.1016/0370-2693(82)90522-6


\bibitem{Grinstein:1982um}
B.~Grinstein,
Nucl. Phys. B \textbf{206}, 387 (1982)
doi:10.1016/0550-3213(82)90275-9



\bibitem{Babu:2006nf}
  K.~S.~Babu, I.~Gogoladze and Z.~Tavartkiladze,
  Phys.\ Lett.\ B {\bf 650}, 49 (2007)
  doi:10.1016/j.physletb.2007.02.050
  [hep-ph/0612315].


\bibitem{Babu:2011tw}
  K.~S.~Babu, I.~Gogoladze, P.~Nath and R.~M.~Syed,
  Phys.\ Rev.\ D {\bf 85}, 075002 (2012)
  doi:10.1103/PhysRevD.85.075002
  [arXiv:1112.5387 [hep-ph]].

%


\bibitem{Du:2013nza}
  L.~Du, X.~Li and D.~X.~Zhang,
  JHEP {\bf 1404}, 027 (2014)
  doi:10.1007/JHEP04(2014)027
  [arXiv:1312.1786 [hep-ph]].


\bibitem{Ananthanarayan:1992cd}
  B.~Ananthanarayan, G.~Lazarides and Q.~Shafi,
  Phys.\ Lett.\ B {\bf 300}, 245 (1993).
  doi:10.1016/0370-2693(93)90361-K


\bibitem{Nath:2015kaa}
  P.~Nath and R.~M.~Syed,
  Phys.\ Rev.\ D {\bf 93}, no. 5, 055005 (2016)
  doi:10.1103/PhysRevD.93.055005
  [arXiv:1508.00585 [hep-ph]].


\bibitem{Nath:1996qs}
P.~Nath,
Phys. Rev. Lett. \textbf{76}, 2218-2221 (1996)
doi:10.1103/PhysRevLett.76.2218
[arXiv:hep-ph/9512415 [hep-ph]].






\bibitem{Mohapatra:1979nn}
  R.~N.~Mohapatra and B.~Sakita,
  Phys.\ Rev.\ D {\bf 21}, 1062 (1980).
  doi:10.1103/PhysRevD.21.1062


\bibitem{Nath:2005bx}
P.~Nath and R.~M.~Syed,
JHEP \textbf{02}, 022 (2006)
doi:10.1088/1126-6708/2006/02/022
[arXiv:hep-ph/0511172 [hep-ph]].






\bibitem{Aad:2012tfa}
  G.~Aad {\it et al.} [ATLAS Collaboration],
  Phys.\ Lett.\ B {\bf 716}, 1 (2012)
  doi:10.1016/j.physletb.2012.08.020
  [arXiv:1207.7214 [hep-ex]].


\bibitem{Chatrchyan:2012ufa}
  S.~Chatrchyan {\it et al.} [CMS Collaboration],
  Phys.\ Lett.\ B {\bf 716}, 30 (2012)
  doi:10.1016/j.physletb.2012.08.021
  [arXiv:1207.7235 [hep-ex]].



\bibitem{Tanabashi:2018oca}
  M.~Tanabashi {\it et al.} [Particle Data Group],
  Phys.\ Rev.\ D {\bf 98}, no. 3, 030001 (2018).
  doi:10.1103/PhysRevD.98.030001


\bibitem{sugra-uni}
A.~H.~Chamseddine, R.~Arnowitt and P.~Nath,
  Phys.\ Rev.\ Lett.\  {\bf 49} (1982) 970;
  P.~Nath, R.~L.~Arnowitt and A.~H.~Chamseddine,
  Nucl.\ Phys.\  B {\bf 227}, 121 (1983);
 L.~J.~Hall, J.~D.~Lykken and S.~Weinberg,
  Phys.\ Rev.\ D {\bf 27}, 2359 (1983).
  doi:10.1103/PhysRevD.27.2359



\bibitem{Nath:2016qzm}
  P.~Nath,
  ``Supersymmetry, Supergravity, and Unification,''
  doi:10.1017/9781139048118


\bibitem{Porod:2003um}
  W.~Porod,
  Comput.\ Phys.\ Commun.\  {\bf 153}, 275 (2003)
  doi:10.1016/S0010-4655(03)00222-4
  [hep-ph/0301101].


\bibitem{Porod:2011nf}
  W.~Porod and F.~Staub,
  Comput.\ Phys.\ Commun.\  {\bf 183}, 2458 (2012)
  doi:10.1016/j.cpc.2012.05.021
  [arXiv:1104.1573 [hep-ph]].


\bibitem{Staub:2017jnp}
  F.~Staub and W.~Porod,
  Eur.\ Phys.\ J.\ C {\bf 77}, no. 5, 338 (2017)
  doi:10.1140/epjc/s10052-017-4893-7
  [arXiv:1703.03267 [hep-ph]].


\bibitem{Babu:2005gx}
K.~Babu, I.~Gogoladze, P.~Nath and R.~M.~Syed,
Phys. Rev. D \textbf{72}, 095011 (2005)
doi:10.1103/PhysRevD.72.095011
[arXiv:hep-ph/0506312 [hep-ph]];
Phys. Rev. D \textbf{74}, 075004 (2006)
doi:10.1103/PhysRevD.74.075004
[arXiv:hep-ph/0607244 [hep-ph]].



\bibitem{Nath:2007eg}
P.~Nath and R.~M.~Syed,
Phys. Rev. D \textbf{77}, 015015 (2008)
doi:10.1103/PhysRevD.77.015015
[arXiv:0707.1332 [hep-ph]].


\bibitem{Nath:2009nf}
P.~Nath and R.~M.~Syed,
Phys. Rev. D \textbf{81}, 037701 (2010)
doi:10.1103/PhysRevD.81.037701
[arXiv:0909.2380 [hep-ph]].



\bibitem{Ajaib:2013kka}
M.~A.~Ajaib, I.~Gogoladze and Q.~Shafi,
Phys. Rev. D \textbf{88}, no.9, 095019 (2013)
doi:10.1103/PhysRevD.88.095019
[arXiv:1307.4882 [hep-ph]].


\bibitem{Dasgupta:1995js}
T.~Dasgupta, P.~Mamales and P.~Nath,
Phys. Rev. D \textbf{52}, 5366-5369 (1995)
doi:10.1103/PhysRevD.52.5366
[arXiv:hep-ph/9501325 [hep-ph]].

\bibitem{Aboubrahim:2018bil}
A.~Aboubrahim and P.~Nath,
Phys. Rev. D \textbf{98}, no.1, 015009 (2018)
doi:10.1103/PhysRevD.98.015009
[arXiv:1804.08642 [hep-ph]].







\bibitem{Belanger:2014vza}
  G.~Bélanger, F.~Boudjema, A.~Pukhov and A.~Semenov,
  Comput.\ Phys.\ Commun.\  {\bf 192}, 322 (2015)
  doi:10.1016/j.cpc.2015.03.003
  [arXiv:1407.6129 [hep-ph]].


\bibitem{Aghanim:2018eyx}
  N.~Aghanim {\it et al.} [Planck Collaboration],
  arXiv:1807.06209 [astro-ph.CO].


\bibitem{Feldman:2010wy}
  D.~Feldman, Z.~Liu, P.~Nath and G.~Peim,
  Phys.\ Rev.\ D {\bf 81}, 095017 (2010)
  doi:10.1103/PhysRevD.81.095017
  [arXiv:1004.0649 [hep-ph]].


\bibitem{Feldman:2011ms}
  D.~Feldman, P.~Fileviez Perez and P.~Nath,
  JHEP {\bf 1201}, 038 (2012)
  doi:10.1007/JHEP01(2012)038
  [arXiv:1109.2901 [hep-ph]].


\bibitem{Aboubrahim:2019mxn}
  A.~Aboubrahim and P.~Nath,
  arXiv:1909.08684 [hep-ph].


\bibitem{Baer:2018rhs}
  H.~Baer, V.~Barger, D.~Sengupta and X.~Tata,
  Eur.\ Phys.\ J.\ C {\bf 78}, no. 10, 838 (2018)
  doi:10.1140/epjc/s10052-018-6306-y
  [arXiv:1803.11210 [hep-ph]].


\bibitem{Halverson:2017deq}
  J.~Halverson, C.~Long and P.~Nath,
  Phys.\ Rev.\ D {\bf 96}, no. 5, 056025 (2017)
  doi:10.1103/PhysRevD.96.056025
  [arXiv:1703.07779 [hep-ph]].


\bibitem{Aboubrahim:2020wah}
  A.~Aboubrahim, W.~Z.~Feng and P.~Nath,
  arXiv:2003.02267 [hep-ph].


\bibitem{Kaufman:2015nda}
  B.~Kaufman, P.~Nath, B.~D.~Nelson and A.~B.~Spisak,
  Phys.\ Rev.\ D {\bf 92}, 095021 (2015)
  doi:10.1103/PhysRevD.92.095021
  [arXiv:1509.02530 [hep-ph]].


\bibitem{Aboubrahim:2017aen}
  A.~Aboubrahim, P.~Nath and A.~B.~Spisak,
  Phys.\ Rev.\ D {\bf 95}, no. 11, 115030 (2017)
  doi:10.1103/PhysRevD.95.115030
  [arXiv:1704.04669 [hep-ph]].

%


\bibitem{Aboubrahim:2018tpf}
A.~Aboubrahim and P.~Nath,
Phys. Rev. D \textbf{98}, no.9, 095024 (2018)
doi:10.1103/PhysRevD.98.095024
[arXiv:1810.12868 [hep-ph]].


\bibitem{Aboubrahim:2019vjl}
A.~Aboubrahim and P.~Nath,
Phys. Rev. D \textbf{100}, no.1, 015042 (2019)
doi:10.1103/PhysRevD.100.015042
[arXiv:1905.04601 [hep-ph]].


\bibitem{CMS:2017fij}
  CMS Collaboration [CMS Collaboration],
  CMS-PAS-SUS-16-048.


\bibitem{Sirunyan:2018iwl}
  A.~M.~Sirunyan {\it et al.} [CMS Collaboration],
  Phys.\ Lett.\ B {\bf 782}, 440 (2018)
  doi:10.1016/j.physletb.2018.05.062
  [arXiv:1801.01846 [hep-ex]].


\bibitem{Sirunyan:2019zfq}
  A.~M.~Sirunyan {\it et al.} [CMS Collaboration],
  JHEP {\bf 1908}, 150 (2019)
  doi:10.1007/JHEP08(2019)150
  [arXiv:1905.13059 [hep-ex]].


\bibitem{Aad:2019vvi}
  G.~Aad {\it et al.} [ATLAS Collaboration],
  Phys.\ Rev.\ D {\bf 101}, no. 7, 072001 (2020)
  doi:10.1103/PhysRevD.101.072001
  [arXiv:1912.08479 [hep-ex]].


\bibitem{Aad:2015eda}
  G.~Aad {\it et al.} [ATLAS Collaboration],
  Phys.\ Rev.\ D {\bf 93}, no. 5, 052002 (2016)
  doi:10.1103/PhysRevD.93.052002
  [arXiv:1509.07152 [hep-ex]].


\bibitem{Aaboud:2018jiw}
  M.~Aaboud {\it et al.} [ATLAS Collaboration],
  Eur.\ Phys.\ J.\ C {\bf 78}, no. 12, 995 (2018)
  doi:10.1140/epjc/s10052-018-6423-7
  [arXiv:1803.02762 [hep-ex]].


\bibitem{Debove:2011xj}
  J.~Debove, B.~Fuks and M.~Klasen,
  Nucl.\ Phys.\ B {\bf 849}, 64 (2011)
  doi:10.1016/j.nuclphysb.2011.03.015
  [arXiv:1102.4422 [hep-ph]].


\bibitem{Fuks:2013vua}
  B.~Fuks, M.~Klasen, D.~R.~Lamprea and M.~Rothering,
  Eur.\ Phys.\ J.\ C {\bf 73}, 2480 (2013)
  doi:10.1140/epjc/s10052-013-2480-0
  [arXiv:1304.0790 [hep-ph]].


\bibitem{Buckley:2014ana}
  A.~Buckley, J.~Ferrando, S.~Lloyd, K.~Nordström, B.~Page, M.~Rüfenacht, M.~Schönherr and G.~Watt,
  Eur.\ Phys.\ J.\ C {\bf 75}, 132 (2015)
  doi:10.1140/epjc/s10052-015-3318-8
  [arXiv:1412.7420 [hep-ph]].


\bibitem{Sjostrand:2014zea}
  T.~Sjöstrand {\it et al.},
  Comput.\ Phys.\ Commun.\  {\bf 191}, 159 (2015)
  doi:10.1016/j.cpc.2015.01.024
  [arXiv:1410.3012 [hep-ph]].


\bibitem{Cacciari:2011ma}
  M.~Cacciari, G.~P.~Salam and G.~Soyez,
  Eur.\ Phys.\ J.\ C {\bf 72}, 1896 (2012)
  doi:10.1140/epjc/s10052-012-1896-2
  [arXiv:1111.6097 [hep-ph]].


\bibitem{Cacciari:2008gp}
  M.~Cacciari, G.~P.~Salam and G.~Soyez,
  JHEP {\bf 0804}, 063 (2008)
  doi:10.1088/1126-6708/2008/04/063
  [arXiv:0802.1189 [hep-ph]].


\bibitem{deFavereau:2013fsa}
  J.~de Favereau {\it et al.} [DELPHES 3 Collaboration],
  JHEP {\bf 1402}, 057 (2014)
  doi:10.1007/JHEP02(2014)057
  [arXiv:1307.6346 [hep-ex]].


\bibitem{Antcheva:2011zz}
  I.~Antcheva {\it et al.},
  Comput.\ Phys.\ Commun.\  {\bf 182}, 1384 (2011).
  doi:10.1016/j.cpc.2011.02.008


\bibitem{CidVidal:2018eel}
  X.~Cid Vidal {\it et al.},
  CERN Yellow Rep.\ Monogr.\  {\bf 7}, 585 (2019)
  doi:10.23731/CYRM-2019-007.585
  [arXiv:1812.07831 [hep-ph]].


\bibitem{Cepeda:2019klc}
  M.~Cepeda {\it et al.},
  CERN Yellow Rep.\ Monogr.\  {\bf 7}, 221 (2019)
  doi:10.23731/CYRM-2019-007.221
  [arXiv:1902.00134 [hep-ph]].


\bibitem{Chen:2017quw}
  Z.~Y.~Chen, D.~X.~Zhang and X.~Z.~Bai,
  Int.\ J.\ Mod.\ Phys.\ A {\bf 32}, no. 36, 1750207 (2017)
  doi:10.1142/S0217751X17502074
  [arXiv:1707.00580 [hep-ph]].

     \end{thebibliography}
    \end{document}